\documentclass[12pt]{article}

  \usepackage{graphicx}
  \usepackage{epsfig}
  \usepackage{amssymb}
  \usepackage{subfigure}
    \usepackage{feynmf}
\evensidemargin - 1.truecm
\begin{document}

\begin{fmffile}{NONabFF}

\newcommand{\be}{\begin{equation}}
\newcommand{\ee}{\end{equation}}
\newcommand{\nn}{\nonumber}
\newcommand{\bea}{\begin{eqnarray}}
\newcommand{\eea}{\end{eqnarray}}
\newcommand{\bfig}{\begin{figure}}
\newcommand{\efig}{\end{figure}}
\newcommand{\bc}{\begin{center}}
\newcommand{\ec}{\end{center}}
\newcommand{\nr}{\hspace{-.4cm}}


\begin{titlepage}
\nopagebreak
{\flushright{
        \begin{minipage}{5cm}
        PITHA 04/11 \\
        Freiburg-THEP 04/10\\
        ZU-TH 11/04\\
        UCLA/04/TEP/25\\
        {\tt hep-ph/0412259}\\
        \end{minipage}        }
}
\vspace*{-1.5cm}                        
\vskip 1.5cm
\begin{center}
{\Large \bf Two-Loop QCD Corrections to the Heavy Quark \\[2mm]
Form Factors: Axial Vector Contributions}
\vskip 1.cm
{\large  W.~Bernreuther$\rm \, ^{a, \,}$\footnote{Email: 
{\tt breuther@physik.rwth-aachen.de}}},
{\large  R.~Bonciani$\rm \, ^{b, \,}$\footnote{Email: 
{\tt Roberto.Bonciani@physik.uni-freiburg.de}}},
{\large T.~Gehrmann$\rm \, ^{c, \,}$\footnote{Email: 
{\tt gehrt@physik.unizh.ch}}}, \\[2mm] 
{\large R.~Heinesch$\rm \, ^{a, \,}$\footnote{Email: 
{\tt heinesch@physik.rwth-aachen.de}}},
{\large T.~Leineweber$\rm \, ^{a, \,}$\footnote{Email: 
{\tt leineweber@physik.rwth-aachen.de}}}, 
{\large P.~Mastrolia$\rm \, ^{d, \,}$\footnote{Email: 
{\tt mastrolia@physics.ucla.edu}}}, \\[2mm] 
and {\large E.~Remiddi$\rm \, ^{e, \,}$\footnote{Email: 
{\tt Ettore.Remiddi@bo.infn.it}}}
\vskip .6cm
{\it $\rm ^a$ Institut f\"ur Theoretische Physik, RWTH Aachen,
D-52056 Aachen, Germany} 
\vskip .2cm
{\it $\rm ^b$ Fakult\"at f\"ur Mathematik und Physik, Albert-Ludwigs-Universit\"at
Freiburg, \\ D-79104 Freiburg, Germany} 
\vskip .2cm
{\it $\rm ^c$ Institut f\"ur Theoretische Physik, 
Universit\"at Z\"urich, CH-8057 Z\"urich, Switzerland}
\vskip .2cm
{\it $\rm ^d$ Department of Physics and Astronomy, UCLA,
Los Angeles, CA 90095-1547} 
\vskip .2cm
{\it $\rm ^e$ Dipartimento di Fisica dell'Universit\`a di Bologna, and
INFN, Sezione di Bologna, I-40126 Bologna, Italy} 
\end{center}
\vskip .4cm

\begin{abstract}
We consider the $Z^* Q{\bar Q}$ vertex to order $\alpha_s^2$ in the QCD
coupling for an on-shell massive quark-antiquark pair and for
arbitrary momentum transfer of the $Z$ boson. We present closed
analytic expressions for the two parity-violating
form factors of that vertex at the two-loop level in QCD, excluding the contributions from triangle
diagrams. These form factors are expressed 
in terms of 1-dimensional harmonic polylogarithms of maximum weight 4.

\vskip .3cm
\flushright{
        \begin{minipage}{12.3cm}
{\it Key words}:  Feynman diagrams, Multi-loop calculations,  Vertex diagrams,
\hspace*{18.5mm} Heavy quarks.\\
{\it PACS}: 11.15.Bt, 12.38.Bx, 14.65.Fy, 14.65.Ha, 13.88.+e
        \end{minipage}        }
\end{abstract}
\vfill
\end{titlepage}


%
\section{Introduction \label{sec_intro}}
%

This paper is the second of a series that is devoted to the computation
of the  electromagnetic and neutral current form factors of
heavy quarks $Q$ at the two-loop level in QCD \cite{bbghlmr}.
These form factors are relevant for a number of applications. For
instance, they are part of the order $\alpha_s^2$ QCD corrections
to the differential electroproduction cross sections of heavy quarks
$e^+ e^- \to \gamma^*, Z^* \to Q {\bar Q} X$ and, in particular,
part of the second order QCD corrections to the forward-backward asymmetry
$A_{fb}^Q$. As far as $b$ quarks are concerned, an order $\alpha_s^2$ 
calculation
with a non-zero $b$ quark mass  is of interest in view of
the discrepancy between experimental measurement and theoretical
expectations -- see \cite{EWWorkingGroup} and the
references  given in 
\cite{bbghlmr}
concerning the state of the theoretical predictions.  
A review of the present status of electroweak corrections to
the forward-backward asymmetry is given in \cite{Freitas:2004mn}.
At a future  linear collider,
forward-backward asymmetries will play a  prominent role in
very precise measurements of the neutral current couplings of bottom
and of top quarks \cite{tesla}. Clearly, predictions
will be required taking the  mass of the heavy quark fully into account.
\par
In \cite{bbghlmr} we presented
closed analytic expressions to order $\alpha_s^2$ of the heavy-quark
electromagnetic vertex form factors for arbitrary momentum  transfer.
Up to an overall coupling factor these are identical to the
corresponding vector, i.e., parity-conserving form factors that appear
in the amplitude of the
decay of a virtual $Z$ boson into a heavy quark-antiquark pair.
In this paper, we compute the axial vector form factors
$G_1$ and $G_2$, excluding the anomalous triangle graph contributions,
Fig.~\ref{triangle-graph}~(a) and (b), which contribute only through the
mass splitting of a quark isospin doublet in the triangle loop. These
terms deserve a separate discussion and will be given elsewhere
\cite{triangle}. \par The paper is organized as follows. In Section 2
we fix our conventions and describe how the form factors can be
obtained from the vertex amplitude by appropriate projections. In
Section 3 the renormalization constants in the scheme that we use --
which is the same as the one chosen in \cite{bbghlmr} -- are collected
for the convenience of the reader. Sections 4 and 5 contain, for
spacelike momentum transfer, the unsubtracted and renormalized axial
vector form factors at one-loop and two-loop order, respectively, for
the cases of the renormalization scale being both equal and different
from the heavy quark mass. In Section 6 the form factors are
analytically continued above the ${\bar Q} Q$ threshold, and we give
there  also their threshold and asymptotic expansions. We
conclude in Section 7.
 
%
\section{The Axial Vector Form Factors \label{sec_avff}}
%

We consider the amplitude $V^\mu_{c_1  c_2}(p_1,p_2)$ for the
decay of a virtual $Z$ boson of four-momentum $q=p_1+p_2$ into a massive quark-antiquark
pair of momenta $p_1$, $p_2$ and colors $c_1$, $c_2$. The quarks $Q,\,\bar{Q}$ are
on their mass shell, $p_1^2=p_2^2=m^2$, where $m$ denotes the pole
mass of $Q$. The squared center-of-mass energy is $S = (p_1+p_2)^2$.

A general decomposition of the vertex function $V^\mu(p_1,p_2)$
involves 6 form factors, two of which odd under a $CP$
transformation. As we consider here, besides QCD
interactions, Standard Model (SM) neutral current interactions to
lowest order, $CP$ invariance holds. This implies that $V^\mu$ depends
only on 4 form factors, and we use the decomposition
\bea
V^\mu_{c_1  c_2}(p_1,p_2) & = & \bar{u}_{c_1}(p_1)\Gamma^\mu_{c_1
  c_2}(q) v_{c_2}(p_2), \\
\Gamma^\mu_{c_1 c_2}(q) & = & (-i) \left( v_Q F_1(s)\gamma^\mu +
  v_Q\frac{1}{2m}F_2(s) i \sigma^{\mu \nu} q_\nu \right. \nn \\
& & \phantom{(-i)} + \left.a_Q G_1(s) \gamma^\mu \gamma_5 + a_Q
  \frac{1}{2m}G_2(s) \gamma_5 q^\mu \right) \delta_{c_1 c_2} \label{decomp},
\eea
where $s=S/m^2$, $\sigma^{\mu \nu} = \frac{i}{2}\left[\gamma^\mu,\gamma^\nu
\right]$, $\bar{u}_{c_1}(p_1)$, $v_{c_2}(p_2)$ are Dirac spinors and
$v_Q$, $a_Q$ are the SM vector and axial vector couplings of the massive
quark $Q$ to the Z boson:  
\be
v_Q = \frac{e}{s_w c_w} \left(
  \frac{T_3^Q}{2}-s_w^2 Q_Q\right), \quad a_Q = -
\frac{e}{s_w c_w}\frac{T_3^Q}{2},
\ee
where $s_w(c_w)$ is the sine (cosine) of the weak mixing angle,
$T_3^Q$ the third component of the weak isospin and $Q_Q$ is the charge
of the heavy quark in units of the positron charge $e>0$.

The dimensionless form factors $F_i,\, G_i$ can be obtained from
$V^\mu_{c_1c_2}(p_1,p_2)$ by appropriate projections. We consider the spinor traces
\bea
&&\mathcal{P}_1=\mathrm{Tr}\left[ \gamma_\mu \left( \not\!
    p_1+m\right) \Gamma^\mu(q) \left( \not\!
    p_2-m\right)\right], \label{proj1}\\
&&\mathcal{P}_2=\mathrm{Tr}\left[ t_\mu \left( \not\!
    p_1+m\right) \Gamma^\mu(q) \left( \not\!
    p_2-m\right)\right],\label{proj2}\\
&&\mathcal{P}_3=\mathrm{Tr}\left[ \gamma_\mu \gamma_5\left( \not\!
    p_1+m\right) \Gamma^\mu(q) \left( \not\!
    p_2-m\right)\right], \label{proj3}\\
&&\mathcal{P}_4=\mathrm{Tr}\left[ \gamma_5 q_\mu \left( \not\!
    p_1+m\right) \Gamma^\mu(q) \left( \not\!
    p_2-m\right)\right],\label{proj4}
\eea
where $t^\mu = p_2^\mu - p_1^\mu$. Since we are working in $D=4 - 2 \epsilon$ dimensions we calculate
these traces in $D$ dimensions as well. Inserting Eq.~(\ref{decomp}) into Eqs.~(\ref{proj1}-\ref{proj4})
and performing the traces one obtains:
\bea
F_1 + F_2 & = & -\frac{i}{v_Q}\frac{2\mathcal{P}_2 +
  m(4-s)\mathcal{P}_1}{4sm^3(1-\epsilon)(4-s)} \label{formfactorf1}, \\ 
F_2 & = & -\frac{i}{v_Q}\frac{ [2+(1-\epsilon)s]\mathcal{P}_2 +
  m(4-s)\mathcal{P}_1}{sm^3(1-\epsilon)(4-s)^2}  \label{formfactorf2}, \\
G_1 & = & \phantom{-}
\frac{i}{a_Q}\frac{sm\mathcal{P}_3+2\mathcal{P}_4} {4sm^3(1-\epsilon)(4-s)},
\label{formfactorg1}
 \\
G_2 & = &
-\frac{i}{a_Q}\frac{ms\mathcal{P}_3+(2(3-2\epsilon)-(1-\epsilon)s)\mathcal{P}_4}
{s^2m^3(1-\epsilon)(4-s)}.
\label{formfactorg2}
\eea
The trace of the unit matrix is kept equal to four also in $D$
dimensions. In calculating the diagrams 
considered in this paper we use an anticommuting $\gamma_5$ in $D$
dimensions. (This prescription is also used in 
the derivation of the formulae (\ref{formfactorg1}),
(\ref{formfactorg2}).)  This prescription is appropriate
as the diagrams below correspond to the  order $\alpha_S^2$ 
``non-singlet'' contributions to the matrix
element of the axial vector current, and it is well known that for
these contributions a canonical, i.e., non-anomalous Ward identity
must hold to this order. Within dimensional regularisation this is most conveniently
implemented with an anticommuting $\gamma_5$. 
In a subsequent paper \cite{triangle}, the contributions involving closed triangle loops,
Fig. 1, will be computed for a mass-split doublet of fermions in the
triangle loop. In that context, the naive anticommuting  $\gamma_5$
used here is problematic in  $D\neq 4$.
However, it will be shown there that using a different
$\gamma_5$ prescription \cite{HV,Larin} does not affect the non-anomalous
contributions presented here. \\

\bfig
\bc
\subfigure[]{
\begin{fmfgraph*}(40,25)
\fmfleft{i}
\fmfright{o1,o2}
\fmfright{o1,o2}
\fmf{dashes}{vz,i}
\fmf{phantom}{o1,v1,v2,v3,v4,vz,v5,v6,v7,v8,o2}
\fmffreeze
\fmf{plain}{vz,v3,v6,vz}
\fmf{dbl_plain}{o1,v1,v8,o2}
\fmf{gluon}{v1,v3}
\fmf{gluon}{v6,v8}
\end{fmfgraph*}
}
\subfigure[]{
\begin{fmfgraph*}(40,25)
\fmfleft{i}
\fmfright{o1,o2}
\fmfright{o1,o2}
\fmf{dashes}{vz,i}
\fmf{phantom}{o1,v1,v2,v3,v4,vz,v5,v6,v7,v8,o2}
\fmffreeze
\fmf{dbl_plain}{vz,v6,v3,vz}
\fmf{dbl_plain}{o1,v1,v8,o2}
\fmf{gluon}{v1,v3}
\fmf{gluon}{v6,v8}
\end{fmfgraph*}
}
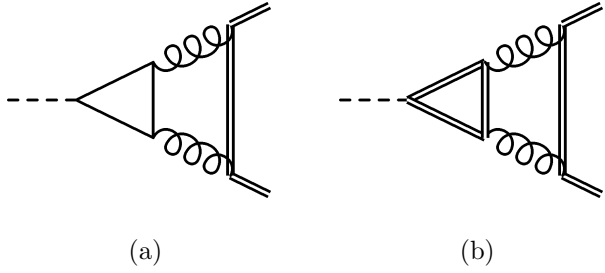
\caption{\label{triangle-graph}Triangle diagram contributions to
  $V^\mu$. Crossed diagrams are not drawn.
The external dashed line refers to an incoming
  $Z$ boson, the curly lines to gluons, the double straight lines to the
  massive quark and the simple straight lines to massless quarks.}
\ec
\efig

The formulae Eqs.~(\ref{formfactorf1}-\ref{formfactorg2}) show that with
the above projections the computation of the vector form factors
$F_{1,2}$ and the axial vector (i.e., parity-violating) form factors 
$G_{1,2}$ decouple from each other. The vector
form factors in Eqs.~(\ref{formfactorf1},\ref{formfactorf2}) were computed
in \cite{bbghlmr}. Here we determine the form factors in
Eqs.~(\ref{formfactorg1},\ref{formfactorg2}) to the second order in
the strong coupling constant $\alpha_S$, excluding the contributions
from the triangle diagrams shown in Fig.~\ref{triangle-graph}. Expanding the
renormalized form factors to the second order in
$\alpha_S$, we have:
\bea
G_1\Bigl(s,\epsilon,\frac{\mu^2}{m^2}\Bigr) = &1 \; + & \! \! \! \! \left(
  \frac{\alpha_S}{2\pi}\right)G_{1,R}^{(1l)}\Bigl(s,\epsilon,\frac{\mu^2}{m^2}\Bigr)+\left(
  \frac{\alpha_S}{2\pi}\right)^2
G_{1,R}^{(2l)}\Bigl(s,\epsilon,\frac{\mu^2}{m^2}\Bigr) \nn \\
&& \! \! \! \! + \mathcal{O}
\left(\left(\frac{\alpha_S}{2\pi}\right)^3\right)\label{g1exp},\\
G_2\Bigl(s,\epsilon,\frac{\mu^2}{m^2}\Bigr) = && \! \! \! \! \left(
  \frac{\alpha_S}{2\pi}\right)G_{2,R}^{(1l)}\Bigl(s,\epsilon,\frac{\mu^2}{m^2}\Bigr)
+ \left(  \frac{\alpha_S}{2\pi}\right)^2
G_{2,R}^{(2l)}\Bigl(s,\epsilon,\frac{\mu^2}{m^2}\Bigr)\nn \\
&& \! \! \! \! + \mathcal{O}
\left(\left(\frac{\alpha_S}{2\pi}\right)^3\right)\label{g2exp},
\eea
where the superscripts $(1l)$ and $(2l)$ denote the one- and two-loop
contributions. The subscript ``$R$'' labels the renormalization scheme
specified in the next Section. After having performed the
renormalization, the form factors still depend on the parameter
$\epsilon$, which regularizes the remaining
infrared divergencies. We keep $\alpha_S$ dimensionless also in
$D\not=4$ dimensions.

The form factors are represented as series in $\epsilon$ and expressed in terms of 1-dimensional harmonic
polylogarithms $\mbox{H}(\vec{a};x)$ up to weight 4 \cite{Polylog,Polylog3}, which are functions of the
dimensionless variable $x$ defined by
\be
x = \frac{\sqrt{-S+4m^2} - \sqrt{-S} }{\sqrt{-S+4m^2} + \sqrt{-S}} =
\frac{\sqrt{-s+4} - \sqrt{-s} }{\sqrt{-s+4} + \sqrt{-s}}.\label{xvar} 
\ee
We give our results firstly in the kinematical region in which $s$ is
spacelike ($0\leq x \leq 1$), where the form factors are real. In Section \ref{sec_analytic} we
shall perform the analytical continuation to the physical region above
threshold, $s > 4$, $-1 < x \leq 0$, and explicitly decompose the form factors into real
and imaginary parts. 

In what follows  $N_f$ denotes the
number of light quarks (which we take to be massless) running in the
loops Fig.~\ref{2-loop-graph}~(g), and $C_F =(N_c^2-1)/(2N_c)$, $C_A =
N_c$, $T_R =1/2$, where  $N_c$ is the number of colors. 

%
\section{Renormalization Scheme\label{sec_ren-scheme}}
%

As in our previous paper \cite{bbghlmr}, we use renormalized
perturbation theory with $\alpha_S = g_s^2/(4\,\pi)$ being defined as
the standard $\overline{\mathrm{MS}}$ coupling in QCD with $N_f$
massless and one massive quark, while we define the mass $m$ and the
wave-function of the heavy quark $Q$ in the on-shell (OS) scheme. 

For the renormalization procedure we need the coupling
renormalization and the gluon wave function to one-loop:
\bea
Z_g^{\overline{\mathrm{MS}}} =  1 -
C(\epsilon)\; \frac{\alpha_S}{2\pi}\;
\frac{1}{4\epsilon}\; \left(\frac{11}{3}C_A-\frac{4}{3}T_R(N_f+1)\right)\,+\,\mathcal{O}\left(\left(\frac{\alpha_S}{2\,\pi}\right)^2\right),\label{zg}
\eea
where
\bea
C(\epsilon) = (4 \pi)^{\epsilon} \, \Gamma \left( 1 + \epsilon \right) 
\,  \label{cofd}
\eea
and, in the Feynman gauge,
\bea
Z_3^{\overline{\mathrm{MS}}} = 1 +
C(\epsilon)\; \frac{\alpha_S}{2\pi}\;
\frac{1}{2\epsilon}\;
\left(\frac{5}{3}C_A-\frac{4}{3}T_R(N_f+1)\right)\, + 
\,\mathcal{O}\left(\left(\frac{\alpha_S}{2\,\pi}\right)^2\right)
\label{z3msb}.
\eea
The renormalization constants concerning the heavy quark are
defined in the on-shell scheme. Here we need
\bea
Z_m^{\mathrm{OS}} = 1 +
\frac{\alpha_S}{2\pi}\;C(\epsilon)\;C_F\;\frac{1}{2\epsilon}\;\frac{2\epsilon-3}{1-2\epsilon}\left(\frac{\mu^2}{m^2}\right)^{\epsilon}+
\mathcal{O}\left(\left(\frac{\alpha_S}{2\,\pi}\right)^2\right),
\label{zmos}
\eea
to one-loop order. Here and in the following $\mu$
denotes the mass scale of dimensional regularization/renormalization.
The wave function $Z_2^{\rm{OS}}$ is needed to two-loop
order. The latter was computed in \cite{Broadhurst,Melnikov}. Using the result of
\cite{Melnikov} and expressing it in terms of the renormalized
$\overline{\mathrm{MS}}$ coupling $\alpha_S$ we have 
\bea
Z_2^{\mathrm{OS}} = &&\nr 1 +
\frac{\alpha_S}{2\pi}\;C(\epsilon)\;C_F\;\frac{1}{2\epsilon}\;\frac{2\epsilon-3}{1-2\epsilon}\left(\frac{\mu^2}{m^2}\right)^{\epsilon}\nn
\\
&&\nr \phantom{1} + \,C^2(\epsilon) \left(\frac{\alpha_S}{2\pi}\right)^2 \;
Z_2^{(2)}+\mathcal{O}\left(\left(\frac{\alpha_S}{2\,\pi}\right)^2\right)\label{z2os}
\eea
where $Z_2^{(2)}$ is: 
\bea
Z_{2}^{(2)} = &&\nr C_{F}^2
\left(\frac{\mu^2}{m^2}\right)^{2\epsilon}\Bigg\{{\frac
  {9}{8\,\epsilon^2}} + {\frac {51}{16\,\epsilon}}+{\frac {433}{32}}-{\frac {39}{2}}\,\zeta \left( 2 \right) +24\,\zeta
 \left( 2 \right) \ln  \left( 2 \right)\nn \\
&&\nr  -6\,\zeta \left( 3 \right)\Bigg\} + C_{F} C_{A}  \Bigg\{\left(\frac{\mu^2}{m^2}\right)^{\epsilon}\bigg[{\frac {11}{12}}\,\bigg(\frac{3}{\epsilon^2}+\frac{4}{\epsilon}+8\bigg)
\bigg]\nn \\
&&\nr +\left(\frac{\mu^2}{m^2}\right)^{2\epsilon}\bigg[-{\frac
  {11}{8\,\epsilon^2}}-{\frac {101}{16\,\epsilon}} + \frac{15}{2}\,\zeta \left( 2 \right) -{\frac {803}{32}}-12\,\zeta \left( 2
 \right) \ln
  \left( 2 \right) \nn \\
&&\nr  +3\,\zeta \left( 3 \right)\bigg]\Bigg\} + C_{F} T_{R} N_f  \Bigg\{\left(\frac{\mu^2}{m^2}\right)^{\epsilon}\bigg[-\frac{1}{3}\,\bigg(\frac{3}{\epsilon^2}+\frac{4}{\epsilon}+8\bigg)
\bigg]\nn \\
&&\nr
+\left(\frac{\mu^2}{m^2}\right)^{2\epsilon}\bigg[\frac{1}{2\,\epsilon^{2}}+\frac{9}{4\,\epsilon}+2\,\zeta
\left( 2 \right) +{\frac {59}{8}}\bigg]\Bigg\} \nn\\
&&\nr + C_{F} T_{R}
\Bigg\{\left(\frac{\mu^2}{m^2}\right)^{\epsilon}\bigg[-\frac{1}{3} \,\bigg(\frac{3}{\epsilon^2}+\frac{4}{\epsilon}+8\bigg)
\bigg]\nn \\
&&\nr +\left(\frac{\mu^2}{m^2}\right)^{2\epsilon}\bigg[\frac{1}{{\epsilon}^{2}}+{
\frac {19}{12\,\epsilon}}+{\frac
  {1139}{72}}-8\,\zeta \left( 2 \right)\bigg]\Bigg\} .
\eea
Further we need the renormalization constant $Z_{\mathrm{1F}}$ for the $Q\bar{Q}$
gluon vertex to one loop. Using a Slavnov-Taylor
identitiy and Eqs.~(\ref{zg},\ref{z3msb},\ref{z2os}) we get:
\bea
Z_{\mathrm{1F}}
= &&\nr Z_g^{\overline{\mathrm{MS}}} \; Z_2^{\mathrm{OS}} \;
\sqrt{Z_3^{\overline{\mathrm{MS}}}} \nn \\
= &&\nr 1 - C(\epsilon)\; \frac{\alpha_S}{2\pi} \;
\frac{1}{2\epsilon} \; \left(C_A - C_F
  \;\frac{2\epsilon-3}{1-2\epsilon}\left(\frac{\mu^2}{m^2}\right)^{\epsilon}\right)\nn \\
&&\nr \!\!+\,\mathcal{O}\left(\left(\frac{\alpha_S}{2\,\pi}\right)^2\right)
\label{z1f}.
\eea
For the counterterm contributions to the renormalized form factors it
is convenient to define (here our notation differs slightly
  from that in \cite{bbghlmr})
\bea
\delta_{\mathrm{1F}} & = & Z_{\mathrm{1F}} \, - \, 1,\\
\delta_2 & = & Z_2^{\mathrm{OS}} \, - \, 1 = \delta_2^{(1l)} \, + \, \delta_2^{(2l)},\\
\delta_3 & = & Z_3^{\overline{\mathrm{MS}}} \, - \, 1, \\
\delta_m & = & \left(Z_2^{\mathrm{OS}}\,Z_m^{\mathrm{OS}} - 1\right) \; m,
\eea
where $\delta_2^{(1l)}$, $\delta_2^{(2l)}$  can be read off from
Eq.~(\ref{z2os}). 

In this renormalization scheme the renormalized vertex function
$\Gamma^\mu$ to order $\alpha_S^2$ is given by the contributions from
the 1-particle irreducible diagrams, Figs.~\ref{1-loop-graph}
and \ref{2-loop-graph}, which we call the unsubtracted contributions,
and the counterterms given below.

%
\begin{boldmath}
\section{One-loop Unsubtracted and UV-Renormalized Form Factors}\label{sec_1l_results}
\end{boldmath}
%

\begin{fmfgroup}
\bfig
\bc
\begin{fmfgraph*}(40,25)
\fmfleft{i}
\fmfright{o1,o2}
\fmf{dashes}{vz,i}
\fmf{dbl_plain}{o1,v1,v2,v3,v4,vz,v5,v6,v7,v8,o2}
\fmffreeze
\fmf{gluon}{v1,v8}
\fmflabel{$q$}{i}
\fmflabel{$p_1$}{o2}
\fmflabel{$p_2$}{o1}
\end{fmfgraph*}
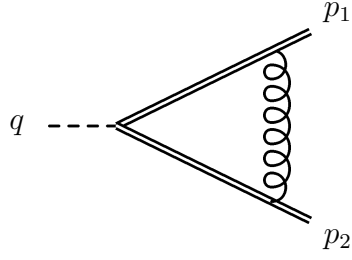
\caption{\label{1-loop-graph}One-loop QCD contribution to $V^\mu_{c_1
    c_2}(q)$.} 
\ec
\efig
\end{fmfgroup}

In this Section we give the results for $G_{1,R}^{(1l)},\,
  G_{2,R}^{(1l)}$. First we write down the contribution from the diagram
Fig.~\ref{1-loop-graph} including the terms of order $\epsilon$. They
are needed for computing the order $\alpha_S^2$ counterterms
shown in Figs.~\ref{ct-graph}~(c), (d) and (f) of Section \ref{subsec_sub_cont}
below. We obtain:
\bea
G_1^{(\mathrm{Fig.\ref{1-loop-graph}})}\Bigl(s,\epsilon,\frac{\mu^2}{m^2}\Bigr)
= && \nr \frac{\alpha_S}{2\,\pi}\;C_F \,
C(\epsilon)\,\left(\frac{\mu^2}{m^2}\right)^{\epsilon}\left( \frac{1}{\epsilon} \; a_1 +
  a_2 + \epsilon \; a_3 \right) + \mathcal{O}(\epsilon^2), \label{g11l} \\
G_2^{(\mathrm{Fig.\ref{1-loop-graph}})}\Bigl(s,\epsilon,\frac{\mu^2}{m^2}\Bigr)
= && \nr \frac{\alpha_S}{2\,\pi}\;C_F \,
C(\epsilon)\,\left(\frac{\mu^2}{m^2}\right)^{\epsilon}\left( \frac{1}{\epsilon} \; b_1 +
  b_2 + \epsilon \; b_3 \right) + \mathcal{O}(\epsilon^2), \label{g21l}  
\eea
where 
$C(\epsilon)$ is defined in Eq.~(\ref{cofd}).

The coefficients $a_i$ and $b_i$ ($i=1\dots 3$) are:
\bea
a_1 = && \nr (x^2+1)/[(x-1)(x+1)]\,H(0;x)+1/2,\\
a_2 = && \nr 1/2\,(3\,x^2-2\,x+3)/[(x-1)(x+1)]\,H(0;x)\nn\\
&& \nr-2\,(x^2+1)/[(x-1)(x+1)]\,H(-1,0;x)\nn\\
&& \nr+(x^2+1)/[(x-1)(x+1)]\,H(0,0;x)\nn\\
&& \nr-\zeta(2)\,(x^2+1)/[(x-1)(x+1)],\\
a_3 = && \nr 2\,\zeta(2)\,(x^2+1)/[(x-1)(x+1)]\,H(-1;x)\nn\\
&& \nr-(x^2+1)\,(-4+\zeta(2))/[(x-1)(x+1)]\,H(0;x)\nn\\
&& \nr-(3\,x^2-2\,x+3)/[(x-1)(x+1)]\,H(-1,0;x)\nn\\
&& \nr+4\,(x^2+1)/[(x-1)(x+1)]\,H(-1,-1,0;x)\nn\\
&& \nr-2\,(x^2+1)/[(x-1)(x+1)]\,H(0,-1,0;x)\nn\\
&& \nr+1/2\,(3\,x^2-2\,x+3)/[(x-1)(x+1)]\,H(0,0;x)\nn\\
&& \nr-2\,(x^2+1)/[(x-1)(x+1)]\,H(-1,0,0;x)\nn\\
&& \nr+(x^2+1)/[(x-1)(x+1)]\,H(0,0,0;x)\nn\\
&& \nr-1/2\,(3\,\zeta(2)\,x^2+4\,\zeta(3)\,x^2-2\,\zeta(2)\,x\nn \\
&& \nr+3\,\zeta(2)+4\,\zeta(3))/[(x-1)(x+1)],
\eea
respectively,
\bea
b_1 = && \nr 0,\\
b_2 = && \nr 2\,(3\,x^2-2\,x+3)/[(x+1)(x-1)^3]\,x\,H(0;x)-4/(x-1)^2\,x,\\
b_3 = && \nr 4\,(3\,x^2-2\,x+3)/[(x+1)(x-1)^3]\,x\,H(0;x)\nn\\
&& \nr-4\,(3\,x^2-2\,x+3)/[(x+1)(x-1)^3]\,x\,H(-1,0;x)\nn\\
&& \nr+2\,(3\,x^2-2\,x+3)/[(x+1)(x-1)^3]\,x\,H(0,0;x)\nn\\
&& \nr-2\,(3\,\zeta(2)\,x^2+4\,x^2-2\,\zeta(2)\,x-4+3\,\zeta(2))/[(x+1)(x-1)^3]\,x.
\eea
\bfig
\bc
\begin{fmfgraph*}(40,25)
\fmfleft{i}
\fmfright{o1,o2}
\fmf{dashes}{vz,i}
\fmf{dbl_plain}{o1,v1,v2,v3,v4,vz,v5,v6,v7,v8,o2}
\fmffreeze
\fmfv{decor.shape=cross,label=$\delta_2^{(1l)}$,label.angle=90.
  ,label.dist=0.1w}{vz} 
\end{fmfgraph*}
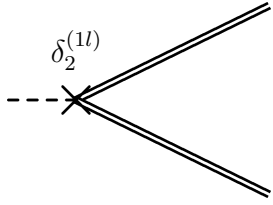
\caption{\label{1-loop-ct-graph}Subtraction term for the one-loop
  renormalization.} 
\ec
\efig
The counterterm of Fig.~\ref{1-loop-ct-graph} contributes to $G_1$:
\bea
G_1^{\mathrm{(Fig.~\ref{1-loop-ct-graph})}}\Bigl(s,\epsilon,\frac{\mu^2}{m^2}\Bigr) = \delta_2^{(1l)}\label{g11lct}
\eea
From Eqs.~(\ref{g11l}), (\ref{g21l}) and (\ref{g11lct}) we obtain the
renormalized one-loop form factors
\bea
G_{1,R}^{(1l)}\Bigl(s,\epsilon,\frac{\mu^2}{m^2}\Bigl) = && \nr C_F \,
C(\epsilon)\,\left(\frac{\mu^2}{m^2}\right)^{\epsilon}\,\left(\frac{1}{\epsilon} \;
  \tilde{a}_1 + \tilde{a}_2  + \epsilon \, \tilde{a}_3 \right) +
\mathcal{O}\left(\epsilon^2\right), \label{g1_1l_ren_decomp}\\
G_{2,R}^{(1l)}\Bigl(s,\epsilon,\frac{\mu^2}{m^2}\Bigr) = && \nr C_F \,
C(\epsilon)\,\left(\frac{\mu^2}{m^2}\right)^{\epsilon}\,\left(\frac{1}{\epsilon} \;
  \tilde{b}_1 + \tilde{b}_2  + \epsilon \, \tilde{b}_3 \right) +
\mathcal{O}\left(\epsilon^2\right), \label{g2_1l_ren_decomp}
\eea
with
\bea
\tilde{a}_1 = && \nr a_1 - 3/2,\\  
\tilde{a}_2 = && \nr a_2 - 2,\\
\tilde{a}_3 = && \nr a_3 - 4
\eea
and $\tilde{b}_i = b_i$ for $i=1\ldots3$.


\vspace{.5cm}

%
\begin{boldmath}
\section{Two-Loop Unsubtracted and Renormalized\\
Form Factors}\label{sec_2l_results}
\end{boldmath}
%

\subsection{Two-Loop Unsubtracted Form Factors \label{subsec_2l_cont}}


\bfig
\bc
%
%
\subfigure[]{
\begin{fmfgraph*}(40,25)
\fmfleft{i}
\fmfright{o1,o2}
\fmf{dashes}{vz,i}
\fmf{dbl_plain}{o1,v1,v2,v3,v4,vz,v5,v6,v7,v8,o2}
\fmffreeze
\fmf{gluon,left}{v6,v8}
\fmf{gluon}{v2,v7}
\end{fmfgraph*}
\label{2-loop-graph-1}
}
%
%
\subfigure[]{
\begin{fmfgraph*}(40,25)
\fmfleft{i}
\fmfright{o1,o2}
\fmf{dashes}{vz,i}
\fmf{dbl_plain}{o1,v1,v2,v3,v4,vz,v5,v6,v7,v8,o2}
\fmffreeze
\fmf{gluon}{v1,v8}
\fmf{gluon,left}{v5,v7}
\end{fmfgraph*}
\label{2-loop-graph-2}
}
%
%
\subfigure[]{
\begin{fmfgraph*}(40,25)
\fmfleft{i}
\fmfright{o1,o2}
\fmf{dashes}{vz,i}
\fmf{dbl_plain}{o1,v1,v2,v3,v4,vz,v5,v6,v7,v8,o2}
\fmfforce{.76w,.5h}{vtg}
\fmffreeze
\fmf{gluon,tension=0.}{vtg,v8}
\fmf{gluon,tension=0.}{v6,vtg}
\fmf{gluon,tension=0.}{v1,vtg}
\end{fmfgraph*}
\label{2-loop-graph-3}
}
%
%
\subfigure[]{
\begin{fmfgraph*}(40,25)
\fmfleft{i}
\fmfright{o1,o2}
\fmf{dashes}{vz,i}
\fmf{dbl_plain}{o1,v1,v2,v3,v4,vz,v5,v6,v7,v8,o2}
\fmffreeze
\fmf{gluon}{v1,v8}
\fmf{gluon}{v2,v7}
\end{fmfgraph*}
\label{2-loop-graph-4}
}
%
%
\subfigure[]{
\begin{fmfgraph*}(40,25)
\fmfleft{i}
\fmfright{o1,o2}
\fmf{dashes}{vz,i}
\fmf{dbl_plain}{o1,v1,v2,v3,v4,vz,v5,v6,v7,v8,o2}
\fmffreeze
\fmf{gluon}{v1,v7}
\fmf{gluon,rubout}{v2,v8}
\end{fmfgraph*}
\label{2-loop-graph-5}
}
%
%
\subfigure[]{
\begin{fmfgraph*}(40,25)
\fmfleft{i}
\fmfright{o1,o2}
\fmf{dashes}{vz,i}
\fmf{dbl_plain}{o1,v1,v2,v3,v4,vz,v5,v6,v7,v8,o2}
\fmffreeze
\fmf{gluon}{v1,vn1}
\fmf{gluon}{vn2,v8}
\fmf{dbl_plain,left,tension=0.4}{vn1,vn2,vn1}
\end{fmfgraph*}
\label{2-loop-graph-6}
}
%
%
\subfigure[]{
\begin{fmfgraph*}(40,25)
\fmfleft{i}
\fmfright{o1,o2}
\fmf{dashes}{vz,i}
\fmf{dbl_plain}{o1,v1,v2,v3,v4,vz,v5,v6,v7,v8,o2}
\fmffreeze
\fmf{gluon}{v1,vn1}
\fmf{gluon}{vn2,v8}
\fmf{plain,left,tension=0.4}{vn1,vn2,vn1}
\end{fmfgraph*}
\label{2-loop-graph-7}
}
%
%
\subfigure[]{
\begin{fmfgraph*}(40,25)
\fmfleft{i}
\fmfright{o1,o2}
\fmf{dashes}{vz,i}
\fmf{dbl_plain}{o1,v1,v2,v3,v4,vz,v5,v6,v7,v8,o2}
\fmffreeze
\fmf{gluon}{v1,vn1}
\fmf{gluon}{vn2,v8}
\fmf{gluon,left,tension=0.4}{vn1,vn2,vn1}
\end{fmfgraph*}
\label{2-loop-graph-8}
}
%
%
\subfigure[]{
\begin{fmfgraph*}(40,25)
\fmfleft{i}
\fmfright{o1,o2}
\fmf{dashes}{vz,i}
\fmf{dbl_plain}{o1,v1,v2,v3,v4,vz,v5,v6,v7,v8,o2}
\fmffreeze
\fmf{gluon}{v1,vn1}
\fmf{gluon}{vn2,v8}
\fmf{dashes,left,tension=0.4}{vn1,vn2,vn1}
\end{fmfgraph*}
\label{2-loop-graph-9}
}
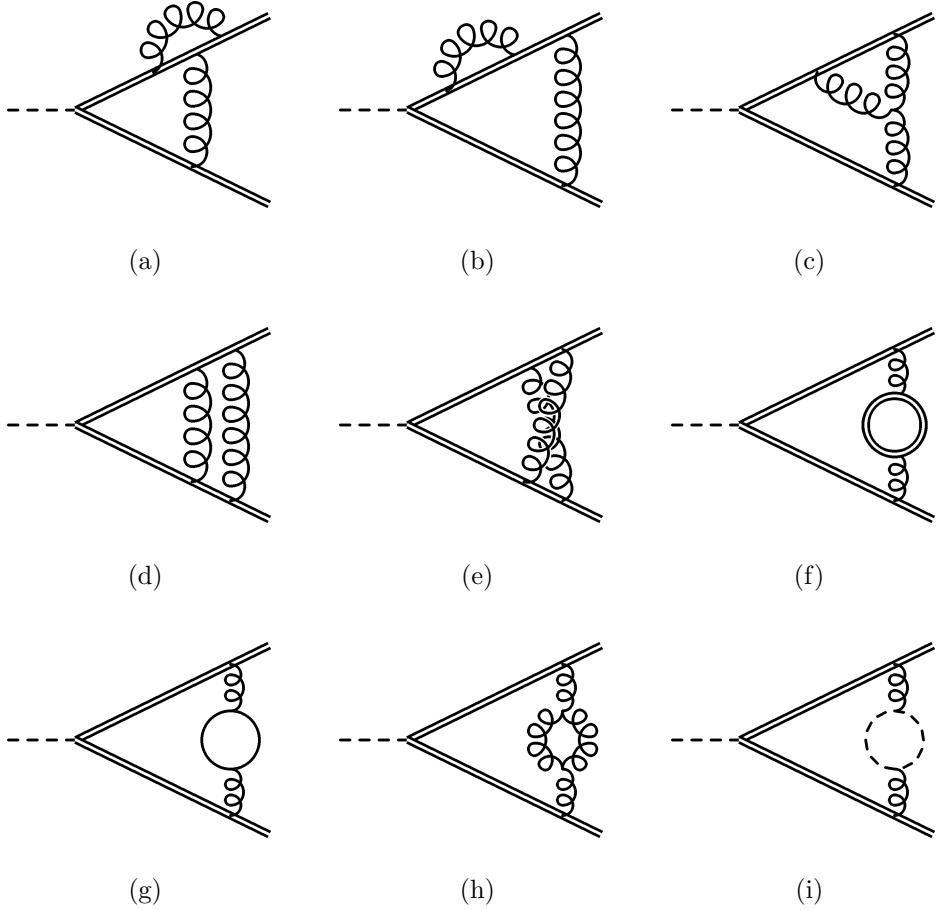
\caption{\label{2-loop-graph}Two-loop QCD contributions to $V^\mu_{c_1
    c_2}(p_1,p_2)$. The
  inner dashed line refers to ghosts. Straight lines represent massless
  quarks and the double straight line the massive quark (c.f. Fig.~\ref{triangle-graph}).}
\ec
\efig

First we compute the two-loop unsubtracted contributions represented
by the diagrams of Figs.~\ref{2-loop-graph}~(a)-(i) and the ``mirror''
diagrams to Figs.~\ref{2-loop-graph}~(a)-(c) to the
form factors $G_1,\, G_2$. 

Performing the $\gamma$ algebra we obtain, as explained in
\cite{RoPieRem1,RoPieRem2,RoPieRem3}, the form factors in
Eqs.~(\ref{formfactorg1},\ref{formfactorg2}) expressed in terms of
hundreds of different scalar integrals. These integrals are reduced to
a small set of master integrals by means of the so-called Laporta
algorithm \cite{Lap} with the help of integration-by-parts identities
\cite{Chet} and Lorentz-invariance identities \cite{Rem3}. Symmetry
relations which one can establish between different integrals are also
used during the reduction. The master integrals themselves were
evaluated with the method of differential equations
\cite{Rem3,Kot,Rem1,Rem2} in \cite{RoPieRem1,RoPieRem3}. The master
integrals, and thus the form factors are represented as series in the
regularization parameter $\epsilon$ and expressed in terms of
1-dimensional harmonic polylogarithms up to weight 4
\cite{Polylog,Polylog3}, which are functions of the dimensionless
variable $x$ defined in Eq.~(\ref{xvar}). As in the one-loop case we
take $s$ to be space-like. 

The two-loop unsubtracted form factors are found to be:
\bea
G_1^{\mathrm{(Fig.~\ref{2-loop-graph})}}\Bigl(s,\epsilon,\frac{\mu^2}{m^2}\Bigr) = && \nr  \left(\frac{\alpha_S}{2\,\pi}\right)^2\;C^2(\epsilon) \,
\left(\frac{\mu^2}{m^2}\right)^{2\epsilon}\;\Bigg\{\nn \\
&& \nr  \phantom{+}\frac{1}{\epsilon^2}\; \left(C_F\; T_R\; N_f
\;  c_1\; + \;C_F \;T_R\; c_2 \;+ \;C_F\; C_A\; c_3\; + \;C_F^2\;
c_4\right)\nn \\
&& \nr + \,\frac{1}{\epsilon}\; \left(C_F\; T_R\; N_f
\;  c_5\; + \;C_F \;T_R\; c_6 \;+ \;C_F\; C_A\; c_7\; + \;C_F^2\;
c_8\right)\nn \\
&& \nr + \, C_F\; T_R\; N_f
\;  c_9\; + \;C_F \;T_R\; c_{10} \;+ \;C_F\; C_A\; c_{11}\; + \;C_F^2\;
c_{12} \nn \\ 
&& \nr +\,\mathcal{O}(\epsilon)\Bigg\} \label{g12lu}, \\
G_2^{\mathrm{(Fig.~\ref{2-loop-graph})}}\Bigl(s,\epsilon,\frac{\mu^2}{m^2}\Bigr) = && \nr
\left(\frac{\alpha_S}{2\,\pi}\right)^2\; 
C^2(\epsilon)\,\left(\frac{\mu^2}{m^2}\right)^{2\epsilon}\;\Bigg\{\nn \\ 
&& \nr \phantom{+}\frac{1}{\epsilon^2}\;\left(C_F\; T_R\; N_f
\;  d_1\; + \;C_F \;T_R\; d_2 \;+ \;C_F\; C_A\; d_3\; + \;C_F^2\;
d_4\right)\nn \\
&& \nr +\,\frac{1}{\epsilon}\;\left(C_F\; T_R\; N_f
\;  d_5\; + \;C_F \;T_R\; d_6 \;+ \;C_F\; C_A\; d_7\; + \;C_F^2\;
d_8\right)\nn \\
&& \nr +\,C_F\; T_R\; N_f
\;  d_9\; + \;C_F \;T_R\; d_{10} \;+ \;C_F\; C_A\; d_{11}\; + \;C_F^2\;
d_{12} \nn \\
&& \nr + \,\mathcal{O}(\epsilon)\Bigg\}\label{g22lu},
\eea
where $C(\epsilon)$ is defined in Eq.~(\ref{cofd}) and 
 $N_f$,  $C_F$, $C_A$, and $T_R$ are given at the end of Section 2.
 One finds for
the $c_i$ ($i=1\dots 12$):
\bea
c_1 = && \nr \frac{1}{2}\,c_2\, = -\frac{4}{11}\,c_3\, \nn \\
 = && \nr -1/3\,(x^2+1) / [(x-1)(x+1)]\,H(0;x)-1/6,\\
c_4 = && \nr 1/2\,(x^4+20\,x^3+14\,x^2+20\,x+1)/[(x+1)^3(x-1)]\,H(0;x)\nn\\
&& \nr+(x^2+1)^2/[(x-1)^2(x+1)^2]\,H(0,0;x)\nn\\
&& \nr+1/8\,(25\,x^2+2\,x+25)/(x+1)^2,\\
c_5 = && \nr -2/9\,(7\,x^2-3\,x+7)/[(x-1)(x+1)]\,H(0;x)\nn\\
&& \nr+4/3\,(x^2+1)/[(x-1)(x+1)]\,H(-1,0;x)\nn\\
&& \nr-2/3\,(x^2+1)/[(x-1)(x+1)]\,H(0,0;x)\nn\\
&& \nr+1/36\,(-13\,x^2+24\,\zeta(2)\,x^2+13+24\,\zeta(2))/[(x-1)(x+1)],\\
c_6 = && \nr -1/3\,(3\,x^2-2\,x+3)/[(x-1)(x+1)]\,H(0;x)\nn\\
&& \nr+4/3\,(x^2+1)/[(x-1)(x+1)]\,H(-1,0;x)\nn\\
&& \nr-2/3\,(x^2+1)/[(x-1)(x+1)]\,H(0,0;x)\nn\\
&& \nr+1/12\,(-3\,x^2+8\,\zeta(2)\,x^2+3+8\,\zeta(2))/[(x-1)(x+1)],\\
c_7 = && \nr-1/18\,(27\,\zeta(2)\,x^4-83\,x^4+33\,x^3+18\,\zeta(2)\,x^2-33\,x-9\,\zeta(2)+83)\nn\\
&& \nr \hspace{1.2cm}/[(x+1)^2(x-1)^2]\,H(0;x)\nn\\
&& \nr-14/3\,(x^2+1)/[(x-1)(x+1)]\,H(-1,0;x)\nn\\
&& \nr+(x^2+1)^2/[(x+1)^2(x-1)^2]\,H(0,-1,0;x)\nn\\
&& \nr+1/6\,(23\,x^2+11)/[(x-1)(x+1)]\,H(0,0;x)\nn\\
&& \nr-2\,(x^2+1)/[(x+1)^2(x-1)^2]\,x^2\,H(0,0,0;x)\nn\\
&& \nr+(x^2+1)/[(x-1)(x+1)]\,H(1,0;x)\nn\\
&& \nr-(x^2+1)^2/[(x+1)^2(x-1)^2]\,H(0,1,0;x)\nn\\
&& \nr-1/144\,(48\,\zeta(2)\,x^4-185\,x^4+72\,\zeta(3)\,x^4+370\,x^2\nn\\
&& \nr\hspace{1.6cm}+288\,\zeta(2)\,x^2+144\,\zeta(3)\,x^2-185-336\,\zeta(2)+72\,\zeta(3))\nn\\
&& \nr\hspace{1.6cm}/[(x+1)^2(x-1)^2],\\
c_8 = && \nr-1/4\,(4\,\zeta(2)\,x^5-3\,x^5+4\,\zeta(2)\,x^4-49\,x^4+18\,x^3\nn\\
&& \nr\hspace{1.1cm}+8\,\zeta(2)\,x^3+8\,\zeta(2)\,x^2-18\,x^2+4\,\zeta(2)\,x\nn\\
&& \nr\hspace{1.1cm}+49\,x+3+4\,\zeta(2))/[(x+1)^3(x-1)^2]\,H(0;x)\nn\\
&& \nr-(x^4+20\,x^3+14\,x^2+20\,x+1)/[(x+1)^3(x-1)]\,H(-1,0;x)\nn\\
&& \nr-2\,(x^2+1)^2/[(x+1)^2(x-1)^2]\,H(0,-1,0;x)\nn\\
&& \nr+1/2\,(7\,x^5+21\,x^4+2\,x^3+14\,x^2-17\,x +5)\nn\\
&& \nr\hspace{1cm}/[(x+1)^3(x-1)^2]\,H(0,0;x)\nn\\
&& \nr-4\,(x^2+1)^2/[(x-1)^2(x+1)^2]\,H(-1,0,0;x)\nn\\
&& \nr+3\,(x^2+1)^2/[(x-1)^2(x+1)^2]\,H(0,0,0;x)\nn\\
&& \nr-1/16\,(35\,x^4+8\,\zeta(2)\,x^4+6\,x^3+160\,\zeta(2)\,x^3+112\,\zeta(2)\,x^2-6\,x\nn\\
&& \nr\hspace{1.4cm}+160\,\zeta(2)\,x-35+8\,\zeta(2))/[(x+1)^3(x-1)],\\
c_9 = && \nr -8/3\,\zeta(2)\,(x^2+1)/[(x-1)(x+1)]\,H(-1;x)\nn\\
&& \nr-1/54\,(353\,x^2-78\,x+353)/[(x-1)(x+1)]\,H(0;x)\nn\\
&& \nr+8/9\,(7\,x^2-3\,x+7)/[(x-1)(x+1)]\,H(-1,0;x)\nn\\
&& \nr-16/3\,(x^2+1)/[(x-1)(x+1)]\,H(-1,-1,0;x)\nn\\
&& \nr+8/3\,(x^2+1)/[(x-1)(x+1)]\,H(0,-1,0;x)\nn\\
&& \nr-4/9\,(7\,x^2-3\,x+7)/[(x-1)(x+1)]\,H(0,0;x)\nn\\
&& \nr+8/3\,(x^2+1)/[(x-1)(x+1)]\,H(-1,0,0;x)\nn\\
&& \nr-4/3\,(x^2+1)/[(x-1)(x+1)]\,H(0,0,0;x)\nn\\
&& \nr+1/216\,(-169\,x^2+576\,\zeta(3)\,x^2+528\,\zeta(2)\,x^2-288\,\zeta(2)\,x\nn\\
&& \nr\hspace{1.6cm}+576\,\zeta(3)+169+816\,\zeta(2))/[(x-1)(x+1)],\\ 
c_{10} = && \nr -4/3\,\zeta(2)\,(x^2+1)/[(x-1)(x+1)]\,H(-1;x)\nn\\
&& \nr-1/54\,(409\,x^4+80\,x^3+302\,x^2-216\,\zeta(2)\,x^2+80\,x+409)\nn\\
&& \nr\hspace{1.2cm}/[(x-1)(x+1)^3]\,H(0;x)\nn\\
&& \nr+2/3\,(3\,x^2-2\,x+3)/[(x-1)(x+1)]\,H(-1,0;x)\nn\\
&& \nr-8/3\,(x^2+1)/[(x-1)(x+1)]\,H(-1,-1,0;x)\nn\\
&& \nr+4/3\,(x^2+1)/[(x-1)(x+1)]\,H(0,-1,0;x)\nn\\
&& \nr+2/9\,(5\,x^5-27\,x^4+5\,x^3-23\,x^2+6\,x-14)\nn\\
&& \nr\hspace{1cm}/[(x+1)^4(x-1)]\,H(0,0;x)\nn\\
&& \nr+4/3\,(x^2+1)/[(x-1)(x+1)]\,H(-1,0,0;x)\nn\\
&& \nr-4/3\,(x^2-x+1)\,(x^2+3\,x+1)/[(x-1)(x+1)^3]\,H(0,0,0;x)\nn\\
&& \nr+1/216\,(288\,\zeta(3)\,x^5+223\,x^5+1728\,\zeta(2)\,x^5-291\,x^4-2304\,\zeta(2)\,x^4\nn\\
&& \nr\hspace{1.6cm}+864\,\zeta(3)\,x^4+1152\,\zeta(3)\,x^3+1872\,\zeta(2)\,x^3-514\,x^3\nn\\
&& \nr\hspace{1.6cm}+1152\,\zeta(3)\,x^2+514\,x^2-1008\,\zeta(2)\,x^2+291\,x+864\,\zeta(3)\,x\nn\\
&& \nr\hspace{1.6cm}+3312\,\zeta(2)\,x-1296\,\zeta(2)+288\,\zeta(3)-223)\nn\\
&& \nr\hspace{1.6cm}/[(x+1)^4(x-1)],\\
c_{11} = && \nr1/3\,(\zeta(2)\,x^4+6\,\zeta(3)\,x^4+81\,\zeta(2)\,x^3+12\,\zeta(3)\,x^2-108\,\zeta(2)\,x^2\nn\\
&& \nr\hspace{.8cm}+81\,\zeta(2)\,x+6\,\zeta(3)-37\,\zeta(2))/[(x+1)^2(x-1)^2]\,H(-1;x)\nn\\
&& \nr+\zeta(2)\,(7\,x^6+6\,x^5-29\,x^4+24\,x^3-31\,x^2+6\,x+5)\nn\\
&& \nr\hspace{1.1cm}/[(x+1)^3(x-1)^3]\,H(0,-1;x)\nn\\
&& \nr+1/216\,(648\,\zeta(2)\,x^6+4129\,x^6-1620\,\zeta(3)\,x^6+3024\,\zeta(3)\,x^5\nn\\
&& \nr\hspace{1.6cm}-12528\,\zeta(2)\,x^5-570\,x^5-4129\,x^4+18792\,\zeta(2)\,x^4\nn\\
&& \nr\hspace{1.6cm}-13716\,\zeta(3)\,x^4+25920\,\zeta(3)\,x^3-12744\,\zeta(2)\,x^3+1140\,x^3\nn\\
&& \nr\hspace{1.6cm}-13500\,\zeta(3)\,x^2+3456\,\zeta(2)\,x^2-4129\,x^2-2808\,\zeta(2)\,x\nn\\
&& \nr\hspace{1.6cm}-570\,x+3024\,\zeta(3)\,x-1404\,\zeta(3)+4129)\nn\\
&& \nr\hspace{1.6cm}/[(x+1)^3(x-1)^3]\,H(0;x)\nn\\
&& \nr-1/18\,(161\,x^4-276\,x^3-72\,\zeta(2)\,x^2+276\,x-161-72\,\zeta(2))\nn\\
&& \nr\hspace{1.2cm}/[(x-1)^2(x+1)^2]\,H(-1,0;x)\nn\\
&& \nr-10\,(x^2+1)^2/[(x-1)^2(x+1)^2]\,H(0,-1,-1,0;x)\nn\\
&& \nr-1/3\,(79\,x^4+6\,x^3-66\,x^2+6\,x-13)\nn\\
&& \nr\hspace{1cm}/[(x-1)^2(x+1)^2]\,H(0,-1,0;x)\nn\\
&& \nr-4\,(x^2+1)^2/[(x-1)^2(x+1)^2]\,H(-1,0,-1,0;x)\nn\\
&& \nr+2\,(11\,x^6+5\,x^5-15\,x^4+26\,x^3-27\,x^2+5\,x-1)\nn\\
&& \nr\hspace{.6cm}/[(x+1)^3(x-1)^3]\,H(0,0,-1,0;x)\nn\\
&& \nr+4\,(x^2+1)^2/[(x-1)^2(x+1)^2]\,H(1,0,-1,0;x)\nn\\
&& \nr-6\,(x^2+1)/[(x-1)(x+1)]\,H(1,-1,0;x)\nn\\
&& \nr+6\,(x^2+1)^2/[(x-1)^2(x+1)^2]\,H(0,1,-1,0;x)\nn\\
&& \nr+1/18\,(18\,\zeta(2)\,x^6-59\,x^6-72\,\zeta(2)\,x^5-174\,x^5+176\,x^4\nn\\
&& \nr\hspace{1.4cm}+306\,\zeta(2)\,x^4-252\,\zeta(2)\,x^3+240\,x^3+270\,\zeta(2)\,x^2\nn\\
&& \nr\hspace{1.4cm}-283\,x^2-72\,\zeta(2)\,x-66\,x+166-18\,\zeta(2))\nn\\
&& \nr\hspace{1.4cm}/[(x+1)^3(x-1)^3]\,H(0,0;x)\nn\\
&& \nr-1/3\,(64\,x^4-141\,x^3+228\,x^2-141\,x-28)\nn\\
&& \nr\hspace{1cm}/[(x-1)^2(x+1)^2]\,H(-1,0,0;x)\nn\\
&& \nr+2\,(7\,x^6-7\,x^5+33\,x^4-56\,x^3+25\,x^2-7\,x-1)\nn\\
&& \nr\hspace{.6cm}/[(x+1)^3(x-1)^3]\,H(0,-1,0,0;x)\nn\\
&& \nr+1/3\,(50\,x^6-201\,x^5+244\,x^4-177\,x^3+13\,x^2-12\,x+11)\nn\\
&& \nr\hspace{1cm}/[(x+1)^3(x-1)^3]\,H(0,0,0;x)\nn\\
&& \nr+74/3\,(x^2+1)/[(x-1)(x+1)]\,H(-1,-1,0;x)\nn\\
&& \nr+2\,(x^2+3)\,(x^2+1)/[(x-1)^2(x+1)^2]\,H(-1,0,0,0;x)\nn\\
&& \nr+2\,(x^5-5\,x^4+18\,x^3-26\,x^2+x-5)\nn\\
&& \nr\hspace{.6cm}/[(x-1)^2(x+1)^3]\,H(1,0,0,0;x)\nn\\
&& \nr-2\,(2\,x^4+5\,x^3-14\,x^2+5\,x+6)/[(x-1)^2(x+1)^2]\,H(1,0,0;x)\nn\\
&& \nr-(12\,x^5-x^4+10\,x^3-28\,x^2-2\,x-1)\nn\\
&& \nr\hspace{.4cm}/[(x+1)^3(x-1)^3]\,x\,H(0,0,0,0;x)\nn\\
&& \nr-2\,(x^6-5\,x^5+21\,x^4-34\,x^3+17\,x^2-5\,x-3)\nn\\
&& \nr\hspace{.6cm}/[(x+1)^3(x-1)^3]\,H(0,1,0,0;x)\nn\\
&& \nr+(-3\,x^5+4\,\zeta(2)\,x^5-x^4-8\,\zeta(2)\,x^4+40\,\zeta(2)\,x^3+2\,x^3\nn\\
&& \nr\hspace{.5cm}-48\,\zeta(2)\,x^2-2\,x^2+x+4\,\zeta(2)\,x+3-8\,\zeta(2))\nn
\\
&& \nr\hspace{1cm} /[(x-1)^2(x+1)^3]\,H(1,0;x)\nn\\
&& \nr-6\,(x^2+1)/[(x-1)(x+1)]\,H(-1,1,0;x)\nn\\
&& \nr+6\,(x^2+1)^2/[(x-1)^2(x+1)^2]\,H(0,-1,1,0;x)\nn\\
&& \nr+4\,(x^2+1)^2/[(x-1)^2(x+1)^2]\,H(-1,0,1,0;x)\nn\\
&& \nr-2\,(7\,x^5+7\,x^4+12\,x^3+4\,x^2+x+1)\nn\\
&& \nr\hspace{.6cm}/[(x-1)^2(x+1)^3]\,H(0,0,1,0;x)\nn\\
&& \nr-4\,(x^2+1)^2/[(x-1)^2(x+1)^2]\,H(1,0,1,0;x)\nn\\
&& \nr+2\,(x^2+1)/[(x-1)(x+1)]\,H(1,1,0;x)\nn\\
&& \nr-2\,(x^2+1)^2/[(x-1)^2(x+1)^2]\,H(0,1,1,0;x)\nn\\
&& \nr+8\,(x^2+1)/[(x-1)(x+1)^2]\,x\,H(0,1,0;x)\nn\\
&& \nr-(x^2+1)\,(\zeta(2)\,x^2+2\,\zeta(3)\,x^2-\zeta(2)\nn\\
&& \nr+2\,\zeta(3))/[(x-1)^2(x+1)^2]\,H(1;x)\nn\\
&& \nr+\zeta(2)\,(x^2+1)^2/[(x-1)^2(x+1)^2]\,H(0,1;x)\nn\\
&& \nr-1/4320\,(85920\,\zeta(2)\,x^6-77760\,\zeta(2)\,\ln(2)\,x^6+25200\,\zeta(3)\,x^6\nn\\
&& \nr\hspace{.6cm}-12925\,x^6+15336\,\zeta(2)^2\,x^6+155520\,\zeta(2)\,\ln(2)\,x^5+28512\,\zeta(2)^2\,x^5\nn\\
&& \nr\hspace{.6cm}+181440\,\zeta(3)\,x^5-279360\,\zeta(2)\,x^5+38775\,x^4-99576\,\zeta(2)^2\,x^4\nn\\
&& \nr\hspace{.6cm}-344880\,\zeta(3)\,x^4+230160\,\zeta(2)\,x^4-77760\,\zeta(2)\,\ln(2)\,x^4\nn\\
&& \nr\hspace{.6cm}+83520\,\zeta(2)\,x^3+109728\,\zeta(2)^2\,x^3+268560\,\zeta(3)\,x^2-38775\,x^2\nn\\
&& \nr\hspace{.6cm}-294720\,\zeta(2)\,x^2+77760\,\zeta(2)\,\ln(2)\,x^2-115560\,\zeta(2)^2\,x^2\nn\\
&& \nr\hspace{.6cm}+28512\,\zeta(2)^2\,x-181440\,\zeta(3)\,x+195840\,\zeta(2)\,x\nn\\
&& \nr\hspace{.6cm}-155520\,\zeta(2)\,\ln(2)\,x-648\,\zeta(2)^2+77760\,\zeta(2)\,\ln(2)+12925\nn\\
&& \nr\hspace{.6cm}-21360\,\zeta(2)+51120\,\zeta(3))/[(x+1)^3(x-1)^3],\\
c_{12} = && \nr (13\,x^5-23\,x^4+12\,x^3+24\,x^2-61\,x\nn\\
&& \nr+11)\,\zeta(2)/[(x+1)^3(x-1)^2]\,H(-1;x)\nn\\
&& \nr-2\,(7\,x^6+10\,x^5-65\,x^4+104\,x^3-75\,x^2\nn\\
&& \nr+10\,x-3)/[(x+1)^3(x-1)^3]\,H(0,-1,0,0;x)\nn\\
&& \nr-1/2\,(17\,x^6-58\,x^5-213\,x^4+224\,x^3-37\,x^2\nn\\
&& \nr-46\,x+17)/[(x+1)^3(x-1)^3]\,H(0,0,0;x)\nn\\
&& \nr-12\,(x^2+1)^2/[(x-1)^2(x+1)^2]\,H(-1,0,0,0;x)\nn\\
&& \nr+(27\,x^6-2\,x^5+27\,x^4-56\,x^3-7\,x^2-2\,x\nn\\
&& \nr-7)/[(x+1)^3(x-1)^3]\,H(0,0,0,0;x)\nn\\
&& \nr+8\,(x^5-2\,x^4+18\,x^3-14\,x^2+4\,x\nn\\
&& \nr+1)/[(x+1)^3(x-1)^2]\,H(1,0,0,0;x)\nn\\
&& \nr-2\,\zeta(2)\,(5\,x^6+6\,x^5-31\,x^4+24\,x^3-29\,x^2\nn\\
&& \nr+6\,x+7)/[(x+1)^3(x-1)^3]\,H(0,-1;x)\nn\\
&& \nr+2\,(8\,x^4-19\,x^3+46\,x^2-19\,x+8)/[(x-1)^2(x+1)^2]\,H(1,0,0;x)\nn\\
&& \nr-4\,(x^6-x^5+14\,x^4-20\,x^3+14\,x^2-x\nn\\
&& \nr+1)/[(x+1)^3(x-1)^3]\,H(0,1,0,0;x)\nn\\
&& \nr+4\,(4\,x^5+\zeta(2)\,x^5-5\,\zeta(2)\,x^4+2\,x^4+34\,\zeta(2)\,x^3-2\,x^3+2\,x^2\nn\\
&& \nr-30\,\zeta(2)\,x^2+7\,\zeta(2)\,x-2\,x-4+\zeta(2))\nn \\
&& \nr\hspace{1cm} /[(x+1)^3(x-1)^2]\,H(1,0;x)\nn\\
&& \nr-2\,(5\,x^4-10\,x^3+18\,x^2-10\,x+5)/[(x-1)^2(x+1)^2]\,H(0,1,0;x)\nn\\
&& \nr+4\,(3\,x^5+3\,x^4+6\,x^3-2\,x^2-x\nn\\
&& \nr-1)/[(x+1)^3(x-1)^2]\,H(0,0,1,0;x)\nn\\
&& \nr+4\,\zeta(3)\,(x^2+1)^2/[(x-1)^2(x+1)^2]\,H(1;x)\nn\\
&& \nr+1/160\,(1915\,x^6+4640\,\zeta(2)\,x^6+2896\,\zeta(2)^2\,x^6-640\,\zeta(3)\,x^6\nn\\
&& \nr\hspace{.6cm}-5760\,\zeta(2)\,\ln(2)\,x^6+11520\,\zeta(2)\,\ln(2)\,x^5-7680\,\zeta(3)\,x^5\nn\\
&& \nr\hspace{.6cm}-3520\,\zeta(2)\,x^5-1152\,\zeta(2)^2\,x^5-3840\,x^5+1935\,x^4+9584\,\zeta(2)^2\,x^4\nn\\
&& \nr\hspace{.6cm}-8160\,\zeta(2)\,x^4+25280\,\zeta(3)\,x^4-5760\,\zeta(2)\,\ln(2)\,x^4\nn\\
&& \nr\hspace{.6cm}-21248\,\zeta(2)^2\,x^3-1920\,\zeta(3)\,x^3-5760\,\zeta(2)\,x^3-17280\,\zeta(3)\,x^2\nn\\
&& \nr\hspace{.6cm}+5760\,\zeta(2)\,\ln(2)\,x^2+13920\,\zeta(2)\,x^2+7632\,\zeta(2)^2\,x^2-1935\,x^2\nn\\
&& \nr\hspace{.6cm}-1152\,\zeta(2)^2\,x+1920\,\zeta(3)\,x+4160\,\zeta(2)\,x-11520\,\zeta(2)\,\ln(2)\,x\nn\\
&& \nr\hspace{.6cm}+3840\,x+5760\,\zeta(2)\,\ln(2)-1915-5280\,\zeta(2)+320\,\zeta(3)\nn\\
&& \nr\hspace{.6cm}+944\,\zeta(2)^2)/[(x+1)^3(x-1)^3]\nn\\
&& \nr+4\,(x^2+1)^2/[(x-1)^2(x+1)^2]\,H(0,-1,-1,0;x)\nn\\
&& \nr+8\,(x^2+1)^2/[(x-1)^2(x+1)^2]\,H(-1,0,-1,0;x)\nn\\
&& \nr-8\,(x^2+1)^2/[(x-1)^2(x+1)^2]\,H(1,0,-1,0;x)\nn\\
&& \nr+8\,(x^2+1)^2/[(x-1)^2(x+1)^2]\,H(1,0,1,0;x)\nn\\
&& \nr+(13\,x^5-59\,x^4+72\,x^3+60\,x^2-21\,x\nn\\
&& \nr+15)/[(x+1)^3(x-1)^2]\,H(0,-1,0;x)\nn\\
&& \nr-2\,(11\,x^6-2\,x^5+33\,x^4-56\,x^3+23\,x^2\nn\\
&& \nr-2\,x+1)/[(x+1)^3(x-1)^3]\,H(0,0,-1,0;x)\nn\\
&& \nr-1/8\,(148\,\zeta(2)\,x^6-64\,\zeta(3)\,x^6-19\,x^6-160\,\zeta(3)\,x^5-224\,\zeta(2)\,x^5\nn\\
&& \nr-326\,x^5+1056\,\zeta(3)\,x^4-676\,\zeta(2)\,x^4+275\,x^4-1536\,\zeta(3)\,x^3\nn\\
&& \nr+992\,\zeta(2)\,x^3+275\,x^2-724\,\zeta(2)\,x^2+1056\,\zeta(3)\,x^2-160\,\zeta(3)\,x\nn\\
&& \nr+140\,x^3-326\,x+96\,\zeta(2)\,x-19+4\,\zeta(2)\nn\\
&& \nr-64\,\zeta(3))/[(x+1)^3(x-1)^3]\,H(0;x)\nn\\
&& \nr+4\,(-9\,x^5+\zeta(2)\,x^5-4\,x^4+\zeta(2)\,x^4+2\,\zeta(2)\,x^3+13\,x^3-13\,x^2\nn\\
&& \nr+2\,\zeta(2)\,x^2+4\,x+\zeta(2)\,x+9+\zeta(2))/[(x+1)^3(x-1)^2]\,H(-1,0;x)\nn\\
&& \nr+1/2\,(26\,\zeta(2)\,x^6+123\,x^6-8\,\zeta(2)\,x^5-112\,x^5-33\,x^4+62\,\zeta(2)\,x^4\nn\\
&& \nr+120\,x^3-160\,\zeta(2)\,x^3+42\,\zeta(2)\,x^2-135\,x^2-8\,\zeta(2)\,x+56\,x-19\nn\\
&& \nr+6\,\zeta(2))/[(x+1)^3(x-1)^3]\,H(0,0;x)\nn\\
&& \nr+2\,(x^4+20\,x^3+14\,x^2+20\,x+1)\nn \\
&& \nr\hspace{1cm} /[(x+1)^3(x-1)]\,H(-1,-1,0;x)\nn\\
&& \nr-(9\,x^5-3\,x^4+116\,x^3+128\,x^2-41\,x\nn\\
&& \nr+7)/[(x+1)^3(x-1)^2]\,H(-1,0,0;x)\nn\\
&& \nr+16\,(x^2+1)^2/[(x-1)^2(x+1)^2]\,H(-1,-1,0,0;x).
\eea
The $d_i$ ($i=1\dots 12$) are:
\bea
d_1 = && \nr d_2 = d_3 = 0,\\
d_4 = && \nr -6\,(x^2+1)/[(x+1)^3(x-1)]\,x\,H(0;x)\nn\\
&& \nr+6/(x+1)^2\,x,\\
d_5 = && \nr d_6 = -\frac{4}{11}\,d_7 = -4/3\,(3\,x^2-2\,x+3)/[(x+1)(x-1)^3]\,x\,H(0;x)\nn\\
&& \nr+8/3/(x-1)^2\,x,\\
d_8 = && \nr -(9\,x^4-60\,x^3+22\,x^2-60\,x+9)/[(x+1)^3(x-1)^3]\,x\,H(0;x)\nn\\
&& \nr+12\,(x^2+1)/[(x+1)^3(x-1)]\,x\,H(-1,0;x)\nn\\
&& \nr+2\,(x^2+1)\,(3\,x^3+11\,x^2-7\,x+9)/[(x+1)^3(x-1)^4]\,x\,H(0,0;x)\nn\\
&& \nr+2\,(-3\,x^3+3\,\zeta(2)\,x^3-25\,x^2-3\,\zeta(2)\,x^2\nn\\
&& \nr+3\,\zeta(2)\,x-25\,x-3-3\,\zeta(2))/[(x+1)^3(x-1)^2]\,x,\\
d_9 = && \nr -2/9\,(87\,x^2-50\,x+87)/[(x+1)(x-1)^3]\,x\,H(0;x)\nn\\
&& \nr+16/3\,(3\,x^2-2\,x+3)/[(x+1)(x-1)^3]\,x\,H(-1,0;x)\nn\\
&& \nr-8/3\,(3\,x^2-2\,x+3)/[(x+1)(x-1)^3]\,x\,H(0,0;x)\nn\\
&& \nr+4/9\,(31\,x^2+18\,\zeta(2)\,x^2-12\,\zeta(2)\,x\nn\\
&& \nr-31+18\,\zeta(2))/[(x+1)(x-1)^3]\,x,\\ 
d_{10} = && \nr -2/9\,(87\,x^4-128\,x^3-46\,x^2-72\,\zeta(2)\,x^2\nn\\
&& \nr-128\,x+87)/[(x+1)^3(x-1)^3]\,x\,H(0;x)\nn\\
&& \nr+8/3\,(3\,x^2-2\,x+3)/[(x+1)(x-1)^3]\,x\,H(-1,0;x)\nn\\
&& \nr-8/3\,(12\,x^4-3\,x^3+9\,x^2-5\,x+3)/[(x+1)^4(x-1)^3]\,x\,H(0,0;x)\nn\\
&& \nr+16/[(x+1)^3(x-1)^3]\,x^3\,H(0,0,0;x)\nn\\
&& \nr+8/9\,(2\,x^5+27\,\zeta(2)\,x^5-18\,x^4-57\,\zeta(2)\,x^4-20\,x^3+63\,\zeta(2)\,x^3\nn\\
&& \nr+20\,x^2-45\,\zeta(2)\,x^2+18\,x+78\,\zeta(2)\,x-2\nn\\
&& \nr-18\,\zeta(2))/[(x+1)^4(x-1)^3]\,x,\\
d_{11} = && \nr -18\,\zeta(2)\,(x^2+1)\,(x^2-4\,x+1)/[(x+1)^2(x-1)^4]\,x\,H(-1;x)\nn\\
&& \nr+24\,\zeta(2)\,(x^2+1)\,(x^2-5\,x+1)/[(x+1)^3(x-1)^5]\,x^2\,H(0,-1;x)\nn\\
&& \nr+1/18\,(1083\,x^6+324\,\zeta(2)\,x^6-238\,x^5+1008\,\zeta(3)\,x^5-3384\,\zeta(2)\,x^5\nn\\
&& \nr+3960\,\zeta(2)\,x^4-1083\,x^4-4752\,\zeta(3)\,x^4+476\,x^3+9216\,\zeta(3)\,x^3\nn\\
&& \nr-7056\,\zeta(2)\,x^3+1620\,\zeta(2)\,x^2-1083\,x^2-4752\,\zeta(3)\,x^2-238\,x\nn\\
&& \nr+1008\,\zeta(3)\,x-648\,\zeta(2)\,x+1083)/[(x+1)^3(x-1)^5]\,x\,H(0;x)\nn\\
&& \nr-4/3\,(18\,x^2-55\,x+18)/[(x+1)(x-1)^3]\,x\,H(-1,0;x)\nn\\
&& \nr-48\,(x^2-x+1)/[(x+1)^2(x-1)^4]\,x^2\,H(0,-1,0;x)\nn\\
&& \nr+8\,(5\,x^4-14\,x^3+30\,x^2-14\,x\nn\\
&& \nr+5)/[(x+1)^3(x-1)^5]\,x^2\,H(0,0,-1,0;x)\nn\\
&& \nr-2/3\,(3\,x^6+24\,\zeta(2)\,x^5+103\,x^5-126\,\zeta(2)\,x^4-27\,x^4-98\,x^3\nn\\
&& \nr+24\,\zeta(2)\,x^3+57\,x^2-126\,\zeta(2)\,x^2+24\,\zeta(2)\,x\nn\\
&& \nr-5\,x-33)/[(x+1)^3(x-1)^5]\,x\,H(0,0;x)\nn\\
&& \nr-2\,(9\,x^4-92\,x^3+130\,x^2-92\,x\nn\\
&& \nr+9)/[(x+1)^2(x-1)^4]\,x\,H(-1,0,0;x)\nn\\
&& \nr-8\,(7\,x^4-29\,x^3+62\,x^2-29\,x\nn\\
&& \nr+7)/[(x+1)^3(x-1)^5]\,x^2\,H(0,-1,0,0;x)\nn\\
&& \nr+2\,(9\,x^4-112\,x^3+146\,x^2-196\,x\nn\\
&& \nr+9)/[(x+1)^3(x-1)^5]\,x^3\,H(0,0,0;x)\nn\\
&& \nr+4\,(x^4-5\,x^3+38\,x^2-5\,x+1)\nn \\
&& \nr\hspace{1cm} /[(x+1)^3(x-1)^5]\,x^2\,H(0,0,0,0;x)\nn\\
&& \nr-16\,(3\,x^2-8\,x+3)/[(x+1)^3(x-1)^3]\,x^2\,H(1,0,0,0;x)\nn\\
&& \nr-24\,(3\,x^2-2\,x+3)/[(x+1)^2(x-1)^4]\,x^2\,H(1,0,0;x)\nn\\
&& \nr+8\,(5\,x^4-14\,x^3+42\,x^2-14\,x\nn\\
&& \nr+5)/[(x+1)^3(x-1)^5]\,x^2\,H(0,1,0,0;x)\nn\\
&& \nr-4\,(x^4+12\,\zeta(2)\,x^3-2\,x^2-32\,\zeta(2)\,x^2+12\,\zeta(2)\,x\nn\\
&& \nr+1)/[(x+1)^3(x-1)^3]\,x\,H(1,0;x)\nn\\
&& \nr-16/[(x+1)^2(x-1)^2]\,x^2\,H(0,1,0;x)\nn\\
&& \nr-32/[(x+1)^3(x-1)^3]\,x^3\,H(0,0,1,0;x)\nn\\
&& \nr+2/45\,(270\,\zeta(3)\,x^6-945\,\zeta(2)\,x^6-935\,x^6+540\,\zeta(2)\,\ln(2)\,x^6\nn\\
&& \nr-3690\,\zeta(3)\,x^5-594\,\zeta(2)^2\,x^5-2160\,\zeta(2)\,\ln(2)\,x^5+5415\,\zeta(2)\,x^5\nn\\
&& \nr+2805\,x^4+2439\,\zeta(2)^2\,x^4+2700\,\zeta(2)\,\ln(2)\,x^4+5490\,\zeta(3)\,x^4\nn\\
&& \nr-3465\,\zeta(2)\,x^4-2010\,\zeta(2)\,x^3-1584\,\zeta(2)^2\,x^3-2700\,\zeta(2)\,\ln(2)\,x^2\nn\\
&& \nr+2439\,\zeta(2)^2\,x^2-2805\,x^2-5490\,\zeta(3)\,x^2+4185\,\zeta(2)\,x^2\nn\\
&& \nr-594\,\zeta(2)^2\,x+3690\,\zeta(3)\,x-3405\,\zeta(2)\,x+2160\,\zeta(2)\,\ln(2)\,x+935\nn\\
&& \nr-540\,\zeta(2)\,\ln(2)+225\,\zeta(2)-270\,\zeta(3))/[(x+1)^3(x-1)^5]\,x,\\
d_{12} = && \nr 24\,(x^2+1)\,(x^3-3\,x^2-6\,x+2)\,\zeta(2)/[(x+1)^3(x-1)^4]\,x\,H(-1;x)\nn\\
&& \nr-48\,\zeta(2)\,(x^2+1)\,(x^2-5\,x+1)/[(x+1)^3(x-1)^5]\,x^2\,H(0,-1;x)\nn\\
&& \nr-1/2\,(84\,\zeta(2)\,x^6+45\,x^6-310\,x^5-32\,\zeta(3)\,x^5-168\,\zeta(2)\,x^5\nn\\
&& \nr+467\,x^4-1116\,\zeta(2)\,x^4+1056\,\zeta(3)\,x^4-1664\,\zeta(3)\,x^3-404\,x^3\nn\\
&& \nr+544\,\zeta(2)\,x^3+467\,x^2-564\,\zeta(2)\,x^2+1056\,\zeta(3)\,x^2+104\,\zeta(2)\,x\nn\\
&& \nr-32\,\zeta(3)\,x-310\,x+45-36\,\zeta(2))/[(x+1)^3(x-1)^5]\,x\,H(0;x)\nn\\
&& \nr-4\,(13\,x^2+46\,x+13)/[(x+1)^3(x-1)]\,x\,H(-1,0;x)\nn\\
&& \nr-24\,(x^2+1)/[(x+1)^3(x-1)]\,x\,H(-1,-1,0;x)\nn\\
&& \nr+4\,(9\,x^5-29\,x^4+114\,x^3+90\,x^2-11\,x\nn\\
&& \nr+3)/[(x+1)^3(x-1)^4]\,x\,H(0,-1,0;x)\nn\\
&& \nr-16\,(x^4+19\,x^3-28\,x^2+19\,x\nn\\
&& \nr+1)/[(x+1)^3(x-1)^5]\,x^2\,H(0,0,-1,0;x)\nn\\
&& \nr+2\,(63\,x^6-78\,x^5+48\,\zeta(2)\,x^4+13\,x^4-216\,\zeta(2)\,x^3+188\,x^3\nn\\
&& \nr+48\,\zeta(2)\,x^2-215\,x^2+50\,x-21)/[(x+1)^3(x-1)^5]\,x\,H(0,0;x)\nn\\
&& \nr-8\,(6\,x^5+4\,x^4+85\,x^3+97\,x^2-5\,x\nn\\
&& \nr+9)/[(x+1)^3(x-1)^4]\,x\,H(-1,0,0;x)\nn\\
&& \nr-16\,(x^4-37\,x^3+54\,x^2-37\,x\nn\\
&& \nr+1)/[(x+1)^3(x-1)^5]\,x^2\,H(0,-1,0,0;x)\nn\\
&& \nr-2\,(3\,x^6-40\,x^5-329\,x^4+88\,x^3-7\,x^2-24\,x\nn\\
&& \nr+21)/[(x+1)^3(x-1)^5]\,x\,H(0,0,0;x)\nn\\
&& \nr-8\,(x^4-5\,x^3+38\,x^2-5\,x+1)\nn \\
&& \nr\hspace{1cm} /[(x+1)^3(x-1)^5]\,x^2\,H(0,0,0,0;x)\nn\\
&& \nr-32\,(x^2-17\,x+1)/[(x+1)^3(x-1)^3]\,x^2\,H(1,0,0,0;x)\nn\\
&& \nr+8\,(3\,x^4-7\,x^3+64\,x^2-7\,x+3)/[(x+1)^2(x-1)^4]\,x\,H(1,0,0;x)\nn\\
&& \nr-16\,(x^4+19\,x^3-16\,x^2+19\,x\nn\\
&& \nr+1)/[(x+1)^3(x-1)^5]\,x^2\,H(0,1,0,0;x)\nn\\
&& \nr+32\,(x^4-\zeta(2)\,x^3+x^3+17\,\zeta(2)\,x^2-\zeta(2)\,x\nn\\
&& \nr+x+1)/[(x+1)^3(x-1)^3]\,x\,H(1,0;x)\nn\\
&& \nr-8\,(3\,x^4-6\,x^3+14\,x^2-6\,x+3)/[(x+1)^2(x-1)^4]\,x\,H(0,1,0;x)\nn\\
&& \nr+64/[(x+1)^3(x-1)^3]\,x^3\,H(0,0,1,0;x)\nn\\
&& \nr-1/5\,(75\,x^6+240\,\zeta(2)\,\ln(2)\,x^6-230\,\zeta(2)\,x^6-120\,\zeta(3)\,x^6\nn\\
&& \nr+560\,\zeta(3)\,x^5-960\,\zeta(2)\,\ln(2)\,x^5+320\,x^5+8\,\zeta(2)^2\,x^5\nn\\
&& \nr+340\,\zeta(2)\,x^5-865\,x^4-3760\,\zeta(3)\,x^4+1200\,\zeta(2)\,\ln(2)\,x^4\nn\\
&& \nr-1160\,\zeta(2)^2\,x^4+790\,\zeta(2)\,x^4+480\,\zeta(3)\,x^3+1400\,\zeta(2)\,x^3\nn\\
&& \nr+3240\,\zeta(2)^2\,x^3-1200\,\zeta(2)\,\ln(2)\,x^2+2920\,\zeta(3)\,x^2+865\,x^2\nn\\
&& \nr-1160\,\zeta(2)^2\,x^2-2330\,\zeta(2)\,x^2-140\,\zeta(2)\,x-320\,x+8\,\zeta(2)^2\,x\nn\\
&& \nr-80\,\zeta(3)\,x+960\,\zeta(2)\,\ln(2)\,x-75+170\,\zeta(2)\nn\\
&& \nr -240\,\zeta(2)\,\ln(2))/[(x+1)^3(x-1)^5]\,x, 
\eea
with $\zeta(n)$ being Riemann's $\zeta$-function. 

%
\subsection{Contributions from the Counterterms}\label{subsec_sub_cont}
%
\bfig
\bc
\subfigure[]{
\begin{fmfgraph*}(40,25)
\fmfleft{i}
\fmfright{o1,o2}
\fmf{dashes}{vz,i}
\fmf{dbl_plain}{o1,v1,v2,v3,v4,vz,v5,v6,v7,v8,o2}
\fmffreeze
\fmf{gluon}{v1,v8}
\fmfv{decor.shape=cross,label=$\delta_2^{(1l)}\not\!p-\delta_m$,label.angle=90.,label.dist=0.15w}{v6}
\end{fmfgraph*}
\label{ct-graph-1}
}
\subfigure[]{
\begin{fmfgraph*}(40,25)
\fmfleft{i}
\fmfright{o1,o2}
\fmf{dashes}{vz,i}
\fmf{dbl_plain}{o1,v1,v2,v3,v4,vz,v5,v6,v7,v8,o2}
\fmffreeze
\fmf{gluon}{v1,v8}
\fmfv{decor.shape=cross,label=$\delta_2^{(1l)}\not\!p-\delta_m$,label.angle=-90,label.dist=0.15w}{v3}
\end{fmfgraph*}
\label{ct-graph-2}
}
\subfigure[]{
\begin{fmfgraph*}(40,25)
\fmfleft{i}
\fmfright{o1,o2}
\fmf{dashes}{vz,i}
\fmf{dbl_plain}{o1,v1,v2,v3,v4,vz,v5,v6,v7,v8,o2}
\fmffreeze
\fmfv{decor.shape=cross}{v8}
\fmfv{label=$\delta_{1{\mathrm F}}$,label.angle=135,label.dist=0.1w}{v8}
\fmf{gluon}{v1,v8}
\end{fmfgraph*}
\label{ct-graph-3}
}
\subfigure[]{
\begin{fmfgraph*}(40,25)
\fmfleft{i}
\fmfright{o1,o2}
\fmf{dashes}{vz,i}
\fmf{dbl_plain}{o1,v1,v2,v3,v4,vz,v5,v6,v7,v8,o2}
\fmffreeze
\fmfv{decor.shape=cross}{v1}
\fmfv{label=$\delta_{1{\mathrm F}}$,label.angle=-135,label.dist=0.1w}{v1}
\fmf{gluon}{v1,v8}
\end{fmfgraph*}
\label{ct-graph-4}
}
\subfigure[]{
\begin{fmfgraph*}(40,25)
\fmfleft{i}
\fmfright{o1,o2}
\fmf{dashes}{vz,i}
\fmf{dbl_plain}{o1,v1,v2,v3,v4,vz,v5,v6,v7,v8,o2}
\fmfforce{.76w,.5h}{vn}
\fmffreeze
\fmfv{decor.shape=cross,label=$\delta_3$,label.dist=.1w}{vn}
\fmf{gluon}{v1,vn}
\fmf{gluon}{vn,v8}
\end{fmfgraph*}
\label{ct-graph-5}
}
\subfigure[]{
\begin{fmfgraph*}(40,25)
\fmfleft{i}
\fmfright{o1,o2}
\fmf{dashes}{vz,i}
\fmf{dbl_plain}{o1,v1,v2,v3,v4,vz,v5,v6,v7,v8,o2}
\fmffreeze
\fmfv{decor.shape=cross,label=$\delta_2^{(1l)}$,label.angle=90.,label.dist=0.1w}{vz}
\fmf{gluon}{v1,v8}
\end{fmfgraph*}
\label{ct-graph-6}
}
\subfigure[]{
\begin{fmfgraph*}(40,25)
\fmfleft{i}
\fmfright{o1,o2}
\fmf{dashes}{vz,i}
\fmf{dbl_plain}{o1,v1,v2,v3,v4,vz,v5,v6,v7,v8,o2}
\fmffreeze
\fmfv{decor.shape=cross,label=$\delta_2^{(2l)}$,label.angle=90.,label.dist=0.1w}{vz}
\end{fmfgraph*}
\label{ct-graph-7}
}
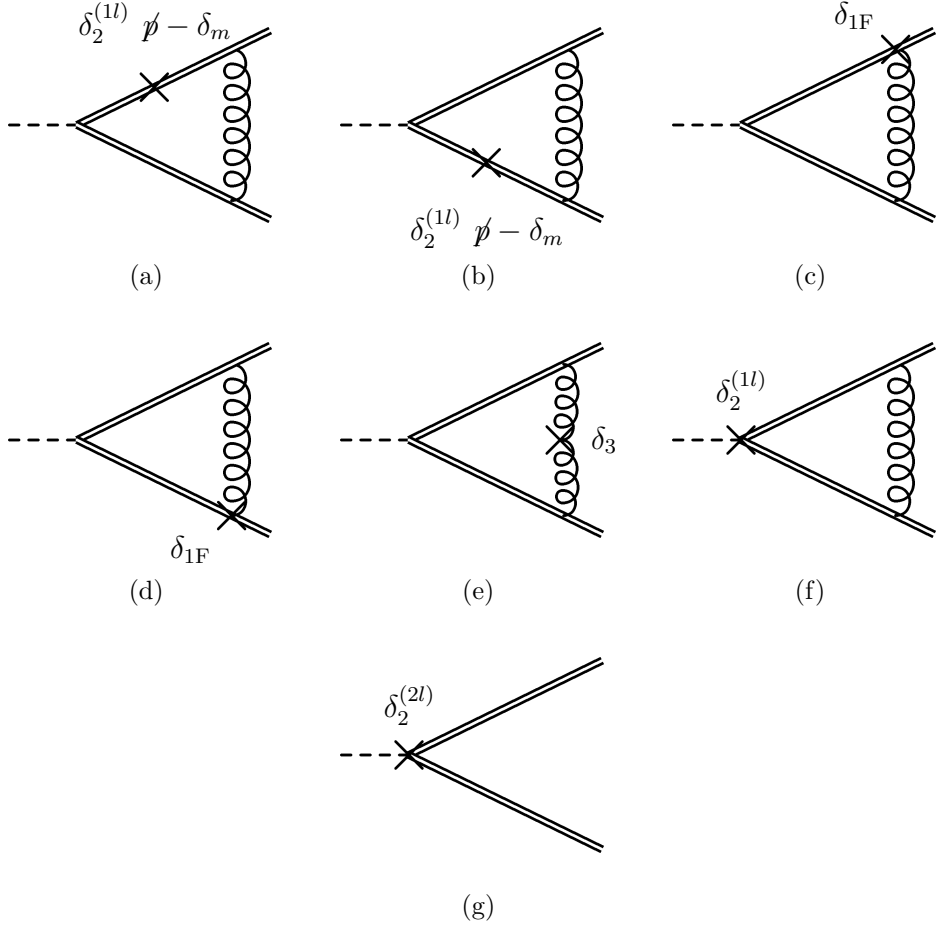
\caption{\label{ct-graph}Subtraction terms for the two-loop
  renormalization. Note that the diagrams (a)-(b) and (c)-(d)
  yield the same contribution, respectively.}
\ec
\efig


The terms in Fig.~\ref{ct-graph}~(a)-(f) are
defined as one-loop vertex amplitudes in $D$ dimensions including the
corresponding one-loop
renormalization insertions. They read explicitly:

\begin{fmfgroup}
\fmfset{curly_len}{2mm}

$\bullet$ Figs.~\ref{ct-graph}~(a) and \ref{ct-graph}~(b):
\bea
\parbox{25mm}{
\begin{fmfgraph*}(24,15)
\fmfleft{i}
\fmfright{o1,o2}
\fmf{dashes}{vz,i}
\fmf{dbl_plain}{o1,v1,v2,v3,v4,vz,v5,v6,v7,v8,o2}
\fmffreeze
\fmf{gluon}{v1,v8}
\fmfv{decor.shape=cross,decor.size=4mm}{v6}
\end{fmfgraph*}
}
& = &
\parbox{25mm}{
\begin{fmfgraph*}(24,15)
\fmfleft{i}
\fmfright{o1,o2}
\fmf{dashes}{vz,i}
\fmf{dbl_plain}{o1,v1,v2,v3,v4,vz,v5,v6,v7,v8,o2}
\fmffreeze
\fmf{gluon}{v1,v8}
\fmfv{decor.shape=cross,decor.size=4mm}{v3}
\end{fmfgraph*}
} \nn \\
& \stackrel{\mathrm{def.}}{=} & \frac{\alpha_S}{2\pi}\,C_F \,
C(\epsilon)\,\left(\frac{\mu^2}{m^2}\right)^{\epsilon}\nn \\
&&\times\int \mathfrak{D}^Dk \frac{\mathcal{N}_1}{\left[\left(p_1+k\right)^2 -
    m^2\right]^2\left[\left(p_2-k\right)^2 -m^2\right]k^2},
\eea
with
\bea
\mathcal{N}_1 & = &\gamma^\rho\left(\not\!k_{\phantom{1}} + \not\!p_1 + m\right)\left(\delta_2^{\left(1l\right)}
\left(\not\!k_{\phantom{1}} + \not\!p_1\right) - \delta_m\right)\left(\not\!k_{\phantom{1}} + \not\!p_1 + m\right)\nn \\
&&\times\left(v^Q\gamma^\mu + a^Q\gamma^\mu\gamma_5\right) \left(\not\!k_{\phantom{1}} - \not\!p_2 +
m\right)\gamma_\rho \label{ct_n1}  
\eea
and
\be
\mathfrak{D}^Dk = \frac{1}{C(\epsilon)} \left( \frac{m^2}{\mu^2}  \right)^\epsilon \frac{d^Dk}{(2\pi)^{D-2}}.
\ee
The contributions from these two diagrams to the axial vector form factors are:
\bea
G_1^{(a,b)}\Bigl(s,\epsilon,\frac{\mu^2}{m^2}\Bigr) = && \nr  \left(\frac{\alpha_S}{2\,\pi}\right)^2\;C_F^2 \,
C^2(\epsilon)\,\left(\frac{\mu^2}{m^2}\right)^{2\,\epsilon}\,\Bigg\{\,
\nn \\
&& \nr \phantom{+}\frac{1}{\epsilon^2}\, \bigg[3/2\,(x-1)\,(x^2+x+1)/(x+1)^3\,H(0;x)\nn\\
&& \nr\hspace{1cm} -3/4\,(x-1)^2/(x+1)^2\bigg]\nn \\
&& \nr + \, \frac{1}{\epsilon} \, \bigg[
1/4\,(17\,x^4+4\,x^3+6\,x^2+4\,x+17)\nn\\
&& \nr\hspace{1.5cm}/[(x-1)(x+1)^3]\,H(0;x)\nn\\
&& \nr-3\,(x-1)\,(x^2+x+1)/(x+1)^3\,H(-1,0;x)\nn\\
&& \nr+3/2\,(x-1)\,(x^2+x+1)/(x+1)^3\,H(0,0;x)\nn\\
&& \nr-1/2\,(x^2+x+1)\,(3\,\zeta(2)\,x-4\,x-3\,\zeta(2)-4)/(x+1)^3\bigg]\nn \\
&& \nr +3\,(x-1)\,(x^2+x+1)\,\zeta(2)/(x+1)^3\,H(-1;x)\nn\\
&& \nr-1/2\,(-26\,x^4+3\,\zeta(2)\,x^4-6\,x^3-3\,\zeta(2)\,x^3+8\,x^2-3\,\zeta(2)\,x\nn\\
&& \nr-6\,x+3\,\zeta(2)-26)/[(x-1)(x+1)^3]\,H(0;x)\nn\\
&& \nr-1/2\,(17\,x^4+4\,x^3+6\,x^2+4\,x+17)\nn\\
&& \nr\hspace{1cm}/[(x-1)(x+1)^3]\,H(-1,0;x)\nn\\
&& \nr+6\,(x-1)\,(x^2+x+1)/(x+1)^3\,H(-1,-1,0;x)\nn\\
&& \nr-3\,(x-1)\,(x^2+x+1)/(x+1)^3\,H(0,-1,0;x)\nn\\
&& \nr+1/4\,(17\,x^4+4\,x^3+6\,x^2+4\,x+17)\nn\\
&& \nr\hspace{1cm}/[(x-1)(x+1)^3]\,H(0,0;x)\nn\\
&& \nr-3\,(x-1)\,(x^2+x+1)/(x+1)^3\,H(-1,0,0;x)\nn\\
&& \nr+3/2\,(x-1)\,(x^2+x+1)/(x+1)^3\,H(0,0,0;x)\nn\\
&& \nr-1/4\,(17\,\zeta(2)\,x^4+12\,\zeta(3)\,x^4+16\,x^4-16\,x^3-12\,\zeta(3)\,x^3\nn\\
&& \nr\hspace{.6cm}+4\,\zeta(2)\,x^3+6\,\zeta(2)\,x^2+4\,\zeta(2)\,x+16\,x-12\,\zeta(3)\,x+17\,\zeta(2)\nn\\
&& \nr\hspace{.6cm}+12\,\zeta(3)-16)/[(x-1)(x+1)^3]\Bigg\},\\
G_2^{(a,b)}\Bigl(s,\epsilon,\frac{\mu^2}{m^2}\Bigr) = && \nr\left(\frac{\alpha_S}{2\,\pi}\right)^2\;
C_F^2\,C^2(\epsilon)\,\left(\frac{\mu^2}{m^2}\right)^{2\,\epsilon}\,\Bigg\{
\nn \\
&& \nr \phantom{+}\frac{1}{\epsilon^2}\bigg[ 3\,(x^2+1)/[(x-1)(x+1)^3]\,x\,H(0;x)-3/(x+1)^2\,x\bigg] \nn \\
&& \nr +\frac{1}{\epsilon}\bigg[ (13\,x^2+6\,x+13)/[(x-1)(x+1)^3]\,x\,H(0;x)\nn\\
&& \nr-6\,(x^2+1)/[(x-1)(x+1)^3]\,x\,H(-1,0;x)\nn\\
&& \nr+3\,(x^2+1)/[(x-1)(x+1)^3]\,x\,H(0,0;x)\nn\\
&& \nr-(3\,\zeta(2)\,x^2+4\,x^2-4+3\,\zeta(2))/[(x-1)(x+1)^3]\,x\bigg] \nn \\
&& \nr + 6\,(x^2+1)\,\zeta(2)/[(x-1)(x+1)^3]\,x\,H(-1;x)\nn\\
&& \nr-(-41\,x^4+3\,\zeta(2)\,x^4+20\,x^3-6\,\zeta(2)\,x^3+6\,\zeta(2)\,x^2-6\,x^2\nn\\
&& \nr-6\,\zeta(2)\,x+20\,x-41+3\,\zeta(2))/[(x-1)^3(x+1)^3]\,x\,H(0;x)\nn\\
&& \nr-2\,(13\,x^2+6\,x+13)/[(x-1)(x+1)^3]\,x\,H(-1,0;x)\nn\\
&& \nr+12\,(x^2+1)/[(x-1)(x+1)^3]\,x\,H(-1,-1,0;x)\nn\\
&& \nr-6\,(x^2+1)/[(x-1)(x+1)^3]\,x\,H(0,-1,0;x)\nn\\
&& \nr+(13\,x^2+6\,x+13)/[(x-1)(x+1)^3]\,x\,H(0,0;x)\nn\\
&& \nr-6\,(x^2+1)/[(x-1)(x+1)^3]\,x\,H(-1,0,0;x)\nn\\
&& \nr+3\,(x^2+1)/[(x-1)(x+1)^3]\,x\,H(0,0,0;x)\nn\\
&& \nr-(13\,\zeta(2)\,x^3+6\,\zeta(3)\,x^3+14\,x^3-7\,\zeta(2)\,x^2-6\,\zeta(3)\,x^2\nn\\
&& \nr\hspace{.6cm}+10\,x^2+6\,\zeta(3)\,x+7\,\zeta(2)\,x+10\,x-13\,\zeta(2)+14-6\,\zeta(3))\nn\\
&& \nr\hspace{.6cm}\hspace{.4cm}/[(x+1)^3(x-1)^2]\,x\Bigg\}. 
\eea

$\bullet$ Figs.~\ref{ct-graph}~(c) and \ref{ct-graph}~(d):
\bea
\parbox{25mm}{
\begin{fmfgraph*}(24,15)
\fmfleft{i}
\fmfright{o1,o2}
\fmf{dashes}{vz,i}
\fmf{dbl_plain}{o1,v1,v2,v3,v4,vz,v5,v6,v7,v8,o2}
\fmffreeze
\fmfv{decor.shape=cross,decor.size=4mm}{v8}
\fmf{gluon}{v1,v8}
\end{fmfgraph*}
}
& = &
\parbox{25mm}{
\begin{fmfgraph*}(24,15)
\fmfleft{i}
\fmfright{o1,o2}
\fmf{dashes}{vz,i}
\fmf{dbl_plain}{o1,v1,v2,v3,v4,vz,v5,v6,v7,v8,o2}
\fmffreeze
\fmfv{decor.shape=cross,decor.size=4mm}{v1}
\fmf{gluon}{v1,v8}
\end{fmfgraph*}
}
\stackrel{\mathrm{def.}}{=}
\delta_{\mathrm{1F}}\times
\parbox{25mm}{
\begin{fmfgraph*}(24,15)
\fmfleft{i}
\fmfright{o1,o2}
\fmf{dashes}{vz,i}
\fmf{dbl_plain}{o1,v1,v2,v3,v4,vz,v5,v6,v7,v8,o2}
\fmffreeze
\fmf{gluon}{v1,v8}
\end{fmfgraph*}
} \nn \\
&=&
 \delta_{1\mathbf{F}}\times \frac{\alpha_S}{2\pi}\,C_F \,
 C(\epsilon)\left(\frac{\mu^2}{m^2}\right)^{\epsilon}\,\nn \\
&& \times\int \mathfrak{D}^Dk
\frac{\mathcal{N}_2}{\left[\left(p_1+k\right)^2 -
    m^2\right]\left[\left(p_2-k\right)^2 -m^2\right]k^2}, 
\eea
with
\be\label{ren2loopn2}
\mathcal{N}_2 =
-\gamma^\rho\left(\not\!k_{\phantom{1}}+\not\!p_1+m\right)
\left(v^Q\gamma^\mu +
  a^Q\gamma^\mu\gamma_5\right)\left(\not\!k_{\phantom{1}}-\not\!p_2 +
  m\right)\gamma_\rho. 
\ee
These diagrams yield 
\bea
G_1^{(c,d)}\Bigl(s,\epsilon,\frac{\mu^2}{m^2}\Bigr) = && \nr \left(\frac{\alpha_S}{2\,\pi}\right)^2\;C_F \, C_A
\,C^2(\epsilon)\,\left(\frac{\mu^2}{m^2}\right)^{\epsilon}\,\Bigg\{\nn \\
&& \nr \phantom{+} \frac{1}{\epsilon^2}\bigg[ -1/2\,(x^2+1)/[(x-1)(x+1)]\,H(0;x)-1/4\bigg] \nn \\
&& \nr +  \frac{1}{\epsilon} \bigg[ -1/4\,(3\,x^2-2\,x+3)/[(x-1)(x+1)]\,H(0;x)\nn\\
&& \nr+(x^2+1)/[(x-1)(x+1)]\,H(-1,0;x)\nn\\
&& \nr-1/2\,(x^2+1)/[(x-1)(x+1)]\,H(0,0;x)\nn\\
&& \nr+1/2\,\zeta(2)\,(x^2+1)/[(x-1)(x+1)]\bigg] \nn \\
&& \nr-\zeta(2)\,(x^2+1)/[(x-1)(x+1)]\,H(-1;x)\nn\\
&& \nr+1/2\,(x^2+1)\,(\zeta(2)-4)/[(x-1)(x+1)]\,H(0;x)\nn\\
&& \nr+1/2\,(3\,x^2-2\,x+3)/[(x-1)(x+1)]\,H(-1,0;x)\nn\\
&& \nr-2\,(x^2+1)/[(x-1)(x+1)]\,H(-1,-1,0;x)\nn\\
&& \nr+(x^2+1)/[(x-1)(x+1)]\,H(0,-1,0;x)\nn\\
&& \nr-1/4\,(3\,x^2-2\,x+3)/[(x-1)(x+1)]\,H(0,0;x)\nn\\
&& \nr+(x^2+1)/[(x-1)(x+1)]\,H(-1,0,0;x)\nn\\
&& \nr-1/2\,(x^2+1)/[(x-1)(x+1)]\,H(0,0,0;x)\nn\\
&& \nr+1/4\,(3\,\zeta(2)\,x^2+4\,\zeta(3)\,x^2-2\,\zeta(2)\,x+4\,\zeta(3)+3\,\zeta(2))\nn\\
&& \nr\hspace{.4cm}/[(x-1)(x+1)]\Bigg\}\nn \\
+ && \nr \left(\frac{\alpha_S}{2\,\pi}\right)^2\;C_F^2
\,C^2(\epsilon)\,\left(\frac{\mu^2}{m^2}\right)^{2\,\epsilon}\,\Bigg\{
\nn \\
&& \nr \phantom{+}\frac{1}{\epsilon^2}\bigg[ -3/2\,(x^2+1)/[(x-1)(x+1)]\,H(0;x)-3/4\bigg] \nn \\
&& \nr +  \frac{1}{\epsilon} \bigg[ -1/4\,(17\,x^2-6\,x+17)/[(x-1)(x+1)]\,H(0;x)\nn\\
&& \nr+3\,(x^2+1)/[(x-1)(x+1)]\,H(-1,0;x)\nn\\
&& \nr-3/2\,(x^2+1)/[(x-1)(x+1)]\,H(0,0;x)\nn\\
&& \nr+1/2\,(-2\,x^2+3\,\zeta(2)\,x^2+2+3\,\zeta(2))/[(x-1)(x+1)]\bigg] \nn \\
&& \nr -3\,\zeta(2)\,(x^2+1)/[(x-1)(x+1)]\,H(-1;x)\nn\\
&& \nr+1/2\,(3\,\zeta(2)\,x^2-26\,x^2+4\,x+3\,\zeta(2)-26)\nn\\
&& \nr\hspace{1cm}/[(x-1)(x+1)]\,H(0;x)\nn\\
&& \nr+1/2\,(17\,x^2-6\,x+17)/[(x-1)(x+1)]\,H(-1,0;x)\nn\\
&& \nr-6\,(x^2+1)/[(x-1)(x+1)]\,H(-1,-1,0;x)\nn\\
&& \nr+3\,(x^2+1)/[(x-1)(x+1)]\,H(0,-1,0;x)\nn\\
&& \nr-1/4\,(17\,x^2-6\,x+17)/[(x-1)(x+1)]\,H(0,0;x)\nn\\
&& \nr+3\,(x^2+1)/[(x-1)(x+1)]\,H(-1,0,0;x)\nn\\
&& \nr-3/2\,(x^2+1)/[(x-1)(x+1)]\,H(0,0,0;x)\nn\\
&& \nr+1/4\,(17\,\zeta(2)\,x^2-8\,x^2+12\,\zeta(3)\,x^2-6\,\zeta(2)\,x+8\nn\\
&& \nr\hspace{.6cm}+17\,\zeta(2)+12\,\zeta(3))/[(x-1)(x+1)]\Bigg\},\\
G_2^{(c,d)}\Bigl(s,\epsilon,\frac{\mu^2}{m^2}\Bigr) = && \nr\left(\frac{\alpha_S}{2\,\pi}\right)^2\;
C_F\,C_A\,C^2(\epsilon)\,\left(\frac{\mu^2}{m^2}\right)^{\epsilon}\,
\Bigg\{ \nn \\
&& \nr \phantom{+}\frac{1}{\epsilon}\bigg[ -(3\,x^2-2\,x+3)/[(x+1)(x-1)^3]\,x\,H(0;x)\nn\\
&& \nr+2/(x-1)^2\,x\bigg] \nn \\
&& \nr -2\,(3\,x^2-2\,x+3)/[(x+1)(x-1)^3]\,x\,H(0;x)\nn\\
&& \nr+2\,(3\,x^2-2\,x+3)/[(x+1)(x-1)^3]\,x\,H(-1,0;x)\nn\\
&& \nr-(3\,x^2-2\,x+3)/[(x+1)(x-1)^3]\,x\,H(0,0;x)\nn\\
&& \nr+(3\,\zeta(2)\,x^2+4\,x^2-2\,\zeta(2)\,x+3\,\zeta(2)-4)\nn\\
&& \nr\hspace{.4cm}/[(x+1)(x-1)^3]\,x\Bigg\}\nn \\
+ && \nr \left(\frac{\alpha_S}{2\,\pi}\right)^2\;C_F^2
\,C^2(\epsilon)\,\left(\frac{\mu^2}{m^2}\right)^{2\,\epsilon}\,\Bigg\{
\nn \\
&& \nr \phantom{+}\frac{1}{\epsilon} \bigg[ -3\,(3\,x^2-2\,x+3)/[(x+1)(x-1)^3]\,x\,H(0;x)\nn\\
&& \nr+6/(x-1)^2\,x\bigg] \nn \\
&& \nr -10\,(3\,x^2-2\,x+3)/[(x+1)(x-1)^3]\,x\,H(0;x)\nn\\
&& \nr+6\,(3\,x^2-2\,x+3)/[(x+1)(x-1)^3]\,x\,H(-1,0;x)\nn\\
&& \nr-3\,(3\,x^2-2\,x+3)/[(x+1)(x-1)^3]\,x\,H(0,0;x)\nn\\
&& \nr+(20\,x^2+9\,\zeta(2)\,x^2-6\,\zeta(2)\,x-20+9\,\zeta(2))\nn\\
&& \nr\hspace{.4cm}/[(x+1)(x-1)^3]\,x\Bigg\}.
\eea

$\bullet$ Fig.~\ref{ct-graph}~(e):
\bea
\parbox{25mm}{
\begin{fmfgraph*}(24,15)
\fmfleft{i}
\fmfright{o1,o2}
\fmf{dashes}{vz,i}
\fmf{dbl_plain}{o1,v1,v2,v3,v4,vz,v5,v6,v7,v8,o2}
\fmfforce{.76w,.5h}{vn}
\fmffreeze
\fmfv{decor.shape=cross,decor.size=4mm,label=$\delta_3$,label.dist=.1w}{vn}
\fmf{gluon}{v1,vn}
\fmf{gluon}{vn,v8}
\end{fmfgraph*}
}
&\stackrel{\mathrm{def.}}{=}&
\delta_3 \times \frac{\alpha_S}{2\pi}\,C_F \,
C(\epsilon)\,\left(\frac{\mu^2}{m^2}\right)^{\epsilon}\nn \\
&&\times\int \mathfrak{D}^Dk \frac{\mathcal{N}_3}{\left[\left(p_1+k\right)^2 -
    m^2\right]\left[\left(p_2-k\right)^2 -m^2\right]\left(k^2\right)^2},
\eea
with
\bea
\mathcal{N}_3 &=& \gamma^\rho\left(\not\!k_{\phantom{1}}+\not\!p_1+m\right)
\left(v^Q\gamma^\mu +
  a^Q\gamma^\mu\gamma_5\right)\left(\not\!k_{\phantom{1}}-\not\!p_2 +
  m\right)\gamma^\sigma \nn \\
&&\times\left(k^2 g_{\rho \sigma}-k_\rho k_\sigma \right).
\eea
This leads to the contributions
\bea
G_1^{(e)}\Bigl(s,\epsilon,\frac{\mu^2}{m^2}\Bigr) = && \nr \left(\frac{\alpha_S}{2\,\pi}\right)^2\;C_F\, T_R \, N_f
\,C^2(\epsilon)\,\left(\frac{\mu^2}{m^2}\right)^{\epsilon}\,\Bigg\{\nn
\\
&& \nr \phantom{+}\frac{1}{\epsilon^2}\bigg[ 2/3\,(x^2+1)/[(x-1)(x+1)]\,H(0;x)+1/3\bigg] \nn \\
&& \nr + \frac{1}{\epsilon}\bigg[ 1/3\,(3\,x^2-2\,x+3)/[(x-1)(x+1)]\,H(0;x)\nn\\
&& \nr-4/3\,(x^2+1)/[(x-1)(x+1)]\,H(-1,0;x)\nn\\
&& \nr+2/3\,(x^2+1)/[(x-1)(x+1)]\,H(0,0;x)\nn\\
&& \nr-2/3\,\zeta(2)\,(x^2+1)/[(x-1)(x+1)]\bigg] \nn \\
&& \nr + 4/3\,\zeta(2)\,(x^2+1)/[(x-1)(x+1)]\,H(-1;x)\nn\\
&& \nr-2/3\,(x^2+1)\,(-4+\zeta(2))/[(x-1)(x+1)]\,H(0;x)\nn\\
&& \nr-2/3\,(3\,x^2-2\,x+3)/[(x-1)(x+1)]\,H(-1,0;x)\nn\\
&& \nr+8/3\,(x^2+1)/[(x-1)(x+1)]\,H(-1,-1,0;x)\nn\\
&& \nr-4/3\,(x^2+1)/[(x-1)(x+1)]\,H(0,-1,0;x)\nn\\
&& \nr+1/3\,(3\,x^2-2\,x+3)/[(x-1)(x+1)]\,H(0,0;x)\nn\\
&& \nr-4/3\,(x^2+1)/[(x-1)(x+1)]\,H(-1,0,0;x)\nn\\
&& \nr+2/3\,(x^2+1)/[(x-1)(x+1)]\,H(0,0,0;x)\nn\\
&& \nr-1/3\,(3\,\zeta(2)\,x^2+4\,\zeta(3)\,x^2-2\,\zeta(2)\,x+4\,\zeta(3)+3\,\zeta(2))\nn\\
&& \nr\hspace{1cm}/[(x-1)(x+1)]\Bigg\}\nn \\
+ && \nr \left(\frac{\alpha_S}{2\,\pi}\right)^2\;C_F \, T_R
\,C^2(\epsilon)\,\left(\frac{\mu^2}{m^2}\right)^{\epsilon}\,\Bigg\{
\nn \\
&& \nr \phantom{+}\frac{1}{\epsilon^2}\bigg[ 2/3\,(x^2+1)/[(x-1)(x+1)]\,H(0;x)+1/3\bigg]\nn \\
&& \nr + \frac{1}{\epsilon}\bigg[ 1/3\,(3\,x^2-2\,x+3)/[(x-1)(x+1)]\,H(0;x)\nn\\
&& \nr-4/3\,(x^2+1)/[(x-1)(x+1)]\,H(-1,0;x)\nn\\
&& \nr+2/3\,(x^2+1)/[(x-1)(x+1)]\,H(0,0;x)\nn\\
&& \nr-2/3\,\zeta(2)\,(x^2+1)/[(x-1)(x+1)]\bigg] \nn \\
&& \nr + 4/3\,\zeta(2)\,(x^2+1)/[(x-1)(x+1)]\,H(-1;x)\nn\\
&& \nr-2/3\,(x^2+1)\,(-4+\zeta(2))/[(x-1)(x+1)]\,H(0;x)\nn\\
&& \nr-2/3\,(3\,x^2-2\,x+3)/[(x-1)(x+1)]\,H(-1,0;x)\nn\\
&& \nr+8/3\,(x^2+1)/[(x-1)(x+1)]\,H(-1,-1,0;x)\nn\\
&& \nr-4/3\,(x^2+1)/[(x-1)(x+1)]\,H(0,-1,0;x)\nn\\
&& \nr+1/3\,(3\,x^2-2\,x+3)/[(x-1)(x+1)]\,H(0,0;x)\nn\\
&& \nr-4/3\,(x^2+1)/[(x-1)(x+1)]\,H(-1,0,0;x)\nn\\
&& \nr+2/3\,(x^2+1)/[(x-1)(x+1)]\,H(0,0,0;x)\nn\\
&& \nr-1/3\,(3\,\zeta(2)\,x^2+4\,\zeta(3)\,x^2-2\,\zeta(2)\,x+4\,\zeta(3)+3\,\zeta(2))\nn\\
&& \nr\hspace{1cm}/[(x-1)(x+1)]\Bigg\} \nn \\
+ && \nr \left(\frac{\alpha_S}{2\,\pi}\right)^2\;C_F\, C_A
\,C^2(\epsilon)\,\left(\frac{\mu^2}{m^2}\right)^{\epsilon}\,\Bigg\{
\nn \\
&& \nr \phantom{+}\frac{1}{\epsilon^2}\bigg[ -5/6\,(x^2+1)/[(x-1)(x+1)]\,H(0;x)-5/12\bigg]\nn \\
&& \nr + \frac{1}{\epsilon} \bigg[ -5/12\,(3\,x^2-2\,x+3)/[(x-1)(x+1)]\,H(0;x)\nn\\
&& \nr+5/3\,(x^2+1)/[(x-1)(x+1)]\,H(-1,0;x)\nn\\
&& \nr-5/6\,(x^2+1)/[(x-1)(x+1)]\,H(0,0;x)\nn\\
&& \nr+5/6\,\zeta(2)\,(x^2+1)/[(x-1)(x+1)]\bigg]\nn \\
&& \nr + -5/3\,\zeta(2)\,(x^2+1)/[(x-1)(x+1)]\,H(-1;x)\nn\\
&& \nr+5/6\,(x^2+1)\,(-4+\zeta(2))/[(x-1)(x+1)]\,H(0;x)\nn\\
&& \nr+5/6\,(3\,x^2-2\,x+3)/[(x-1)(x+1)]\,H(-1,0;x)\nn\\
&& \nr-10/3\,(x^2+1)/[(x-1)(x+1)]\,H(-1,-1,0;x)\nn\\
&& \nr+5/3\,(x^2+1)/[(x-1)(x+1)]\,H(0,-1,0;x)\nn\\
&& \nr-5/12\,(3\,x^2-2\,x+3)/[(x-1)(x+1)]\,H(0,0;x)\nn\\
&& \nr+5/3\,(x^2+1)/[(x-1)(x+1)]\,H(-1,0,0;x)\nn\\
&& \nr-5/6\,(x^2+1)/[(x-1)(x+1)]\,H(0,0,0;x)\nn\\
&& \nr+5/12\,(3\,\zeta(2)\,x^2+4\,\zeta(3)\,x^2-2\,\zeta(2)\,x+4\,\zeta(3)+3\,\zeta(2))\nn\\
&& \nr\hspace{1.2cm}/[(x-1)(x+1)]\Bigg\} ,\\
G_2^{(e)}\Bigl(s,\epsilon,\frac{\mu^2}{m^2}\Bigr) = && \nr \left(\frac{\alpha_S}{2\,\pi}\right)^2\;C_F\, T_R \, N_f
\,C^2(\epsilon)\,\left(\frac{\mu^2}{m^2}\right)^{\epsilon}\,\Bigg\{
\nn \\
&& \nr \phantom{+}\frac{1}{\epsilon}\bigg[ 4/3\,(3\,x^2-2\,x+3)/[(x+1)(x-1)^3]\,x\,H(0;x)\nn\\
&& \nr-8/3/(x-1)^2\,x\bigg] \nn \\
&& \nr + 8/3\,(3\,x^2-2\,x+3)/[(x+1)(x-1)^3]\,x\,H(0;x)\nn\\
&& \nr-8/3\,(3\,x^2-2\,x+3)/[(x+1)(x-1)^3]\,x\,H(-1,0;x)\nn\\
&& \nr+4/3\,(3\,x^2-2\,x+3)/[(x+1)(x-1)^3]\,x\,H(0,0;x)\nn\\
&& \nr-4/3\,(4\,x^2+3\,\zeta(2)\,x^2-2\,\zeta(2)\,x-4+3\,\zeta(2))\nn\\
&& \nr\hspace{1cm}/[(x+1)(x-1)^3]\,x\Bigg\}\nn \\
+ && \nr \left(\frac{\alpha_S}{2\,\pi}\right)^2\;C_F \, T_R
\,C^2(\epsilon)\,\left(\frac{\mu^2}{m^2}\right)^{\epsilon}\,\Bigg\{
\nn \\
&& \nr \phantom{+}\frac{1}{\epsilon}\bigg[ 4/3\,(3\,x^2-2\,x+3)/[(x+1)(x-1)^3]\,x\,H(0;x)\nn\\
&& \nr-8/3/(x-1)^2\,x\bigg] \nn \\
&& \nr + 8/3\,(3\,x^2-2\,x+3)/[(x+1)(x-1)^3]\,x\,H(0;x)\nn\\
&& \nr-8/3\,(3\,x^2-2\,x+3)/[(x+1)(x-1)^3]\,x\,H(-1,0;x)\nn\\
&& \nr+4/3\,(3\,x^2-2\,x+3)/[(x+1)(x-1)^3]\,x\,H(0,0;x)\nn\\
&& \nr-4/3\,(4\,x^2+3\,\zeta(2)\,x^2-2\,\zeta(2)\,x-4+3\,\zeta(2))\nn\\
&& \nr\hspace{1cm}/[(x+1)(x-1)^3]\,x\Bigg\} \nn \\
+ && \nr \left(\frac{\alpha_S}{2\,\pi}\right)^2\;C_F\, C_A
\,C^2(\epsilon)\,\left(\frac{\mu^2}{m^2}\right)^{\epsilon}\,\Bigg\{
\nn \\
&& \nr \phantom{+}\frac{1}{\epsilon} \bigg[ -5/3\,(3\,x^2-2\,x+3)/[(x+1)(x-1)^3]\,x\,H(0;x)\nn\\
&& \nr+10/3/(x-1)^2\,x\bigg]\nn \\
&& \nr -10/3\,(3\,x^2-2\,x+3)/[(x+1)(x-1)^3]\,x\,H(0;x)\nn\\
&& \nr+10/3\,(3\,x^2-2\,x+3)/[(x+1)(x-1)^3]\,x\,H(-1,0;x)\nn\\
&& \nr-5/3\,(3\,x^2-2\,x+3)/[(x+1)(x-1)^3]\,x\,H(0,0;x)\nn\\
&& \nr+5/3\,(4\,x^2+3\,\zeta(2)\,x^2-2\,\zeta(2)\,x-4+3\,\zeta(2))\nn\\
&&\nr\hspace{1cm}/[(x+1)(x-1)^3]\,x\Bigg\}.
\eea

$\bullet$ Fig.~\ref{ct-graph}~(f):
\bea
\parbox{25mm}{
\begin{fmfgraph*}(24,15)
\fmfleft{i}
\fmfright{o1,o2}
\fmf{dashes}{vz,i}
\fmf{dbl_plain}{o1,v1,v2,v3,v4,vz,v5,v6,v7,v8,o2}
\fmffreeze
\fmfv{decor.shape=cross,decor.size=4mm,label=$\delta_2^{(1l)}$,label.angle=90.,label.dist=0.1w}{vz}
\fmf{gluon}{v1,v8}
\end{fmfgraph*}
}
&=&
\delta_2^{(1l)}\times
\parbox{25mm}{
\begin{fmfgraph*}(24,15)
\fmfleft{i}
\fmfright{o1,o2}
\fmf{dashes}{vz,i}
\fmf{dbl_plain}{o1,v1,v2,v3,v4,vz,v5,v6,v7,v8,o2}
\fmffreeze
\fmf{gluon}{v1,v8}
\end{fmfgraph*}
} \nn \\
&\stackrel{\mathrm{def.}}{=}&\delta_2^{(1l)}\times
\frac{\alpha_S}{2\pi}\,C_F\, C(\epsilon)\,
\left(\frac{\mu^2}{m^2}\right)^{\epsilon}\nn \\
&&\times\int \mathfrak{D}^Dk
\frac{\mathcal{N}_2}{\left[\left(p_1+k\right)^2 -
    m^2\right]\left[\left(p_2-k\right)^2 -m^2\right]k^2}, 
\eea
with $\mathcal{N}_2$ as defined in Eq.(\ref{ren2loopn2}).
They contribute
\bea
G_1^{(f)}\Bigl(s,\epsilon,\frac{\mu^2}{m^2}\Bigr) = && \nr\left(\frac{\alpha_S}{2\,\pi}\right)^2\;
C_F^2\,C^2(\epsilon)\,\left(\frac{\mu^2}{m^2}\right)^{2\,\epsilon}\,
\Bigg\{ \nn \\
&& \nr \phantom{+}\frac{1}{\epsilon^2}\bigg[ -3/2\,(x^2+1)/[(x-1)(x+1)]\,H(0;x)-3/4\bigg]\nn \\
&& \nr + \frac{1}{\epsilon}\bigg[ -1/4\,(17\,x^2-6\,x+17)/[(x-1)(x+1)]\,H(0;x)\nn\\
&& \nr+3\,(x^2+1)/[(x-1)(x+1)]\,H(-1,0;x)\nn\\
&& \nr-3/2\,(x^2+1)/[(x-1)(x+1)]\,H(0,0;x)\nn\\
&& \nr+1/2\,(-2\,x^2+3\,\zeta(2)\,x^2+2+3\,\zeta(2))/[(x-1)(x+1)]\bigg] \nn \\
&& \nr -3\,\zeta(2)\,(x^2+1)/[(x-1)(x+1)]\,H(-1;x)\nn\\
&& \nr+1/2\,(-26\,x^2+3\,\zeta(2)\,x^2+4\,x+3\,\zeta(2)-26)\nn\\
&& \nr\hspace{1cm}/[(x-1)(x+1)]\,H(0;x)\nn\\
&& \nr+1/2\,(17\,x^2-6\,x+17)/[(x-1)(x+1)]\,H(-1,0;x)\nn\\
&& \nr-6\,(x^2+1)/[(x-1)(x+1)]\,H(-1,-1,0;x)\nn\\
&& \nr+3\,(x^2+1)/[(x-1)(x+1)]\,H(0,-1,0;x)\nn\\
&& \nr-1/4\,(17\,x^2-6\,x+17)/[(x-1)(x+1)]\,H(0,0;x)\nn\\
&& \nr+3\,(x^2+1)/[(x-1)(x+1)]\,H(-1,0,0;x)\nn\\
&& \nr-3/2\,(x^2+1)/[(x-1)(x+1)]\,H(0,0,0;x)\nn\\
&& \nr+1/4\,(17\,\zeta(2)\,x^2-8\,x^2+12\,\zeta(3)\,x^2-6\,\zeta(2)\,x\nn\\
&& \nr\hspace{.6cm}+12\,\zeta(3)+8+17\,\zeta(2))/[(x-1)(x+1)]\Bigg\},\\
G_2^{(f)}\Bigl(s,\epsilon,\frac{\mu^2}{m^2}\Bigr) = && \nr\left(\frac{\alpha_S}{2\,\pi}\right)^2\;
C_F^2\,C^2(\epsilon)\,\left(\frac{\mu^2}{m^2}\right)^{2\,\epsilon}\,
\Bigg\{\nn \\
&& \nr \phantom{+} \frac{1}{\epsilon}\bigg[ -3\,(3\,x^2-2\,x+3)/[(x+1)(x-1)^3]\,x\,H(0;x)\nn\\
&& \nr+6/(x-1)^2\,x\bigg]\nn \\
&& \nr -10\,(3\,x^2-2\,x+3)/[(x+1)(x-1)^3]\,x\,H(0;x)\nn\\
&& \nr+6\,(3\,x^2-2\,x+3)/[(x+1)(x-1)^3]\,x\,H(-1,0;x)\nn\\
&& \nr-3\,(3\,x^2-2\,x+3)/[(x+1)(x-1)^3]\,x\,H(0,0;x)\nn\\
&& \nr+(9\,\zeta(2)\,x^2+20\,x^2-6\,\zeta(2)\,x-20+9\,\zeta(2))\nn\\
&& \nr\hspace{.4cm}/[(x+1)(x-1)^3]\,x\Bigg\}.
\eea

$\bullet$ Fig.~\ref{ct-graph}~(g) is defined as the two-loop renormalization
constant $\delta_2^{(2l)}$ times the Born level amplitude and reads:
\bea
\parbox{25mm}{
\begin{fmfgraph*}(24,15)
\fmfleft{i}
\fmfright{o1,o2}
\fmf{dashes}{vz,i}
\fmf{dbl_plain}{o1,v1,v2,v3,v4,vz,v5,v6,v7,v8,o2}
\fmffreeze
\fmfv{decor.shape=cross,label=$\delta_2^{(2l)}$,label.angle=90.,label.dist=0.1w}{vz}
\end{fmfgraph*}
}
&\stackrel{\mathrm{def.}}{=}&
\delta_2^{(2l)}\times
\parbox{25mm}{
\begin{fmfgraph*}(24,15)
\fmfleft{i}
\fmfright{o1,o2}
\fmf{dashes}{vz,i}
\fmf{dbl_plain}{o1,v1,v2,v3,v4,vz,v5,v6,v7,v8,o2}
\fmffreeze
\end{fmfgraph*}
} \nn \\
&=& \delta_2^{(2l)} \times (-i)\left(v^Q \gamma^{\mu} + a^Q \gamma^\mu
  \gamma_5\right).
\eea
Its contribution to $G_1$ is:
\bea
G_1^{(g)}\Bigl(s,\epsilon,\frac{\mu^2}{m^2}\Bigr) =  \,\delta_2^{(2l)}.
\eea

\end{fmfgroup}

%
\subsection{UV-Renormalized Two-Loop Form Factors\label{sec_2lren}}
%

Adding the terms given in Sections \ref{subsec_2l_cont} and \ref{subsec_sub_cont} we
obtain the UV-renormalized
axial vector form factors to second order in QCD. They still contain
terms proportional to $\epsilon^{-2}$ and $\epsilon^{-1}$ due to
infrared and collinear singularities in the loops. In this Subsection  we
put the renormalization scale $\mu$ equal to the on-shell mass
$m$ (The logarithms that are present if $\mu \not=m$ are given in
Section \ref{munotm}.). Then we get:
\bea
G_{1,R}^{(2l)}(s,\epsilon) & = &  C^2(\epsilon) \, \Bigg\{\nn \\
&&
\frac{1}{\epsilon^2}\; \left(C_F\; T_R\; N_f
\;  \tilde{c}_1\; + \;C_F \;T_R\; \tilde{c}_2 \;+ \;C_F\; C_A\; \tilde{c}_3\; + \;C_F^2\;
\tilde{c}_4\right)\nn \\
& + & \frac{1}{\epsilon}\; \left(C_F\; T_R\; N_f
\;  \tilde{c}_5\; + \;C_F \;T_R\; \tilde{c}_6 \;+ \;C_F\; C_A\; \tilde{c}_7\; + \;C_F^2\;
\tilde{c}_8\right)\nn \\
& + &  \left(C_F\; T_R\; N_f
\;  \tilde{c}_9\; + \;C_F \;T_R\; \tilde{c}_{10} \;+ \;C_F\; C_A\; \tilde{c}_{11}\; + \;C_F^2\;
\tilde{c}_{12}\right) \nn \\ 
& + & \mathcal{O}(\epsilon) \Bigg\} \label{g1_2l_ren_decomp} ,\\
G_{2,R}^{(2l)}(s,\epsilon) & = & C^2(\epsilon)\,\Bigg\{\nn \\
&& \frac{1}{\epsilon^2}\;\left(C_F\; T_R\; N_f
\;  \tilde{d}_1\; + \;C_F \;T_R\; \tilde{d}_2 \;+ \;C_F\; C_A\; \tilde{d}_3\; + \;C_F^2\;
\tilde{d}_4\right)\nn \\
& + & \frac{1}{\epsilon}\;\left(C_F\; T_R\; N_f
\;  \tilde{d}_5\; + \;C_F \;T_R\; \tilde{d}_6 \;+ \;C_F\; C_A\; \tilde{d}_7\; + \;C_F^2\;
\tilde{d}_8\right)\nn \\
& + & \left(C_F\; T_R\; N_f
\;  \tilde{d}_9\; + \;C_F \;T_R\; \tilde{d}_{10} \;+ \;C_F\; C_A\; \tilde{d}_{11}\; + \;C_F^2\;
\tilde{d}_{12}\right) \nn \\
& + &  \mathcal{O}(\epsilon)\Bigg\} \label{g2_2l_ren_decomp} ,  
\eea
where:
\bea
\tilde{c}_1 = && \nr -\frac{4}{11}\,\tilde{c}_3 =
-\frac{3}{5}\,\tilde{c}_5 = 1/3\,(x^2+1)/[(x-1)(x+1)]\,H(0;x)-1/3,\\
\tilde{c}_2 = && \nr \tilde{c}_6 = 0,\\
\tilde{c}_4 = && \nr -(x^2+1)/[(x-1)(x+1)]\,H(0;x)\nn\\
&& \nr+(x^2+1)^2/[(x+1)^2(x-1)^2]\,H(0,0;x)+1/2,\\
\tilde{c}_7 = && \nr
-1/36\,(x^2+1)\,(-67\,x^2+54\,\zeta(2)\,x^2+67-18\,\zeta(2))\nn\\
&& \nr\hspace{1.2cm}/[(x+1)^2(x-1)^2]\,H(0;x)\nn\\
&& \nr-(x^2+1)/[(x-1)(x+1)]\,H(-1,0;x)\nn\\
&& \nr+(x^2+1)^2/[(x+1)^2(x-1)^2]\,H(0,-1,0;x)\nn\\
&& \nr+2/[(x-1)(x+1)]\,x^2\,H(0,0;x)\nn\\
&& \nr-2\,(x^2+1)/[(x+1)^2(x-1)^2]\,x^2\,H(0,0,0;x)\nn\\
&& \nr+(x^2+1)/[(x-1)(x+1)]\,H(1,0;x)\nn\\
&& \nr-(x^2+1)^2/[(x+1)^2(x-1)^2]\,H(0,1,0;x)\nn\\
&& \nr+1/36\,(54\,\zeta(2)\,x^4-49\,x^4-18\,\zeta(3)\,x^4-36\,\zeta(3)\,x^2+98\,x^2\nn\\
&& \nr\hspace{.6cm}-72\,\zeta(2)\,x^2-49+18\,\zeta(2)-18\,\zeta(3))/[(x+1)^2(x-1)^2],\\
\tilde{c}_8 = && \nr
-1/2\,(2\,\zeta(2)\,x^4+7\,x^4-2\,x^3+4\,\zeta(2)\,x^2+2\,x+2\,\zeta(2)-7)\nn\\
&& \nr\hspace{1cm}/[(x+1)^2(x-1)^2]\,H(0;x)\nn\\
&& \nr+2\,(x^2+1)/[(x-1)(x+1)]\,H(-1,0;x)\nn\\
&& \nr-2\,(x^2+1)^2/[(x+1)^2(x-1)^2]\,H(0,-1,0;x)\nn\\
&& \nr+2\,(x^2+1)\,(x^2-x+2)/[(x+1)^2(x-1)^2]\,H(0,0;x)\nn\\
&& \nr-4\,(x^2+1)^2/[(x+1)^2(x-1)^2]\,H(-1,0,0;x)\nn\\
&& \nr+3\,(x^2+1)^2/[(x+1)^2(x-1)^2]\,H(0,0,0;x)\nn\\
&& \nr+(2\,x^2+\zeta(2)\,x^2-2+\zeta(2))/[(x-1)(x+1)],\\
\tilde{c}_9 = && \nr -4/3\,\zeta(2)\,(x^2+1)/[(x-1)(x+1)]\,H(-1;x)\nn\\
&& \nr-1/54\,(36\,\zeta(2)\,x^2+209\,x^2-78\,x+36\,\zeta(2)+209)\nn\\
&& \nr\hspace{1.2cm}/[(x-1)(x+1)]\,H(0;x)\nn\\
&& \nr+2/9\,(19\,x^2-6\,x+19)/[(x-1)(x+1)]\,H(-1,0;x)\nn\\
&& \nr-8/3\,(x^2+1)/[(x-1)(x+1)]\,H(-1,-1,0;x)\nn\\
&& \nr+4/3\,(x^2+1)/[(x-1)(x+1)]\,H(0,-1,0;x)\nn\\
&& \nr-1/9\,(19\,x^2-6\,x+19)/[(x-1)(x+1)]\,H(0,0;x)\nn\\
&& \nr+4/3\,(x^2+1)/[(x-1)(x+1)]\,H(-1,0,0;x)\nn\\
&& \nr-2/3\,(x^2+1)/[(x-1)(x+1)]\,H(0,0,0;x)\nn\\
&& \nr+1/27\,(93\,\zeta(2)\,x^2+106\,x^2+36\,\zeta(3)\,x^2-18\,\zeta(2)\,x\nn\\
&& \nr\hspace{.6cm}+36\,\zeta(3)-106+21\,\zeta(2))/[(x-1)(x+1)],\\
\tilde{c}_{10} = && \nr
-1/54\,(36\,\zeta(2)\,x^4+265\,x^4+72\,\zeta(2)\,x^3-208\,x^3-144\,\zeta(2)\,x^2\nn\\
&& \nr\hspace{1.2cm}+14\,x^2+72\,\zeta(2)\,x-208\,x+36\,\zeta(2)+265)\nn\\
&& \nr\hspace{1.2cm}/[(x+1)^3(x-1)]\,H(0;x)\nn\\
&& \nr+1/9\,(19\,x^4-14\,x^3+14\,x^2-14\,x+19)/(x+1)^4\,H(0,0;x)\nn\\
&& \nr-2/3\,(x^4+2\,x^3-4\,x^2+2\,x+1)/[(x+1)^3(x-1)]\,H(0,0,0;x)\nn\\
&& \nr-1/27\,(-383\,x^4+27\,\zeta(2)\,x^4+1026\,\zeta(2)\,x^3-1412\,x^3-2058\,x^2\nn\\
&& \nr\hspace{.6cm}+1278\,\zeta(2)\,x^2+1026\,\zeta(2)\,x-1412\,x-383+27\,\zeta(2))/(x+1)^4,\\
\tilde{c}_{11} = && \nr -2/3\,(34\,x^4+3\,x^3-33\,x^2+3\,x-1)\nn\\
&& \nr\hspace{1cm}/[(x+1)^2(x-1)^2]\,H(0,-1,0;x)\nn\\
&& \nr-4\,(x^2+1)^2/[(x+1)^2(x-1)^2]\,H(-1,0,-1,0;x)\nn\\
&& \nr+2\,(11\,x^6+5\,x^5-15\,x^4+26\,x^3-27\,x^2+5\,x-1)\nn\\
&& \nr\hspace{.6cm}/[(x+1)^3(x-1)^3]\,H(0,0,-1,0;x)\nn\\
&& \nr+4\,(x^2+1)^2/[(x+1)^2(x-1)^2]\,H(1,0,-1,0;x)\nn\\
&& \nr-6\,(x^2+1)/[(x-1)(x+1)]\,H(1,-1,0;x)\nn\\
&& \nr+6\,(x^2+1)^2/[(x+1)^2(x-1)^2]\,H(0,1,-1,0;x)\nn\\
&& \nr+1/36\,(36\,\zeta(2)\,x^6-217\,x^6-144\,\zeta(2)\,x^5-282\,x^5+612\,\zeta(2)\,x^4\nn\\
&& \nr\hspace{1.2cm}+451\,x^4-504\,\zeta(2)\,x^3+348\,x^3-467\,x^2+540\,\zeta(2)\,x^2-66\,x\nn\\
&& \nr\hspace{1.2cm}-144\,\zeta(2)\,x-36\,\zeta(2)+233)/[(x+1)^3(x-1)^3]\,H(0,0;x)\nn\\
&& \nr-1/3\,(53\,x^4-141\,x^3+228\,x^2-141\,x-17)\nn\\
&& \nr\hspace{1cm}/[(x+1)^2(x-1)^2]\,H(-1,0,0;x)\nn\\
&& \nr+2\,(7\,x^6-7\,x^5+33\,x^4-56\,x^3+25\,x^2-7\,x-1)\nn\\
&& \nr\hspace{.6cm}/[(x+1)^3(x-1)^3]\,H(0,-1,0,0;x)\nn\\
&& \nr+1/6\,(89\,x^6-402\,x^5+499\,x^4-354\,x^3+37\,x^2-24\,x+11)\nn\\
&& \nr\hspace{1cm}/[(x+1)^3(x-1)^3]\,H(0,0,0;x)\nn\\
&& \nr+2\,(x^2+3)\,(x^2+1)/[(x+1)^2(x-1)^2]\,H(-1,0,0,0;x)\nn\\
&& \nr+2\,(x^5-5\,x^4+18\,x^3-26\,x^2+x-5)\nn\\
&& \nr\hspace{.6cm}/[(x+1)^3(x-1)^2]\,H(1,0,0,0;x)\nn\\
&& \nr-2\,(2\,x^4+5\,x^3-14\,x^2+5\,x+6)/[(x+1)^2(x-1)^2]\,H(1,0,0;x)\nn\\
&& \nr-(12\,x^5-x^4+10\,x^3-28\,x^2-2\,x-1)\nn\\
&& \nr\hspace{.4cm}/[(x+1)^3(x-1)^3]\,x\,H(0,0,0,0;x)\nn\\
&& \nr-2\,(x^6-5\,x^5+21\,x^4-34\,x^3+17\,x^2-5\,x-3)\nn\\
&& \nr\hspace{.6cm}/[(x+1)^3(x-1)^3]\,H(0,1,0,0;x)\nn\\
&& \nr+(-3\,x^5+4\,\zeta(2)\,x^5-x^4-8\,\zeta(2)\,x^4+40\,\zeta(2)\,x^3+2\,x^3-48\,\zeta(2)\,x^2\nn\\
&& \nr\hspace{.5cm}-2\,x^2+x+4\,\zeta(2)\,x+3-8\,\zeta(2))/[(x+1)^3(x-1)^2]\,H(1,0;x)\nn\\
&& \nr-6\,(x^2+1)/[(x-1)(x+1)]\,H(-1,1,0;x)\nn\\
&& \nr+6\,(x^2+1)^2/[(x+1)^2(x-1)^2]\,H(0,-1,1,0;x)\nn\\
&& \nr+4\,(x^2+1)^2/[(x+1)^2(x-1)^2]\,H(-1,0,1,0;x)\nn\\
&& \nr-2\,(7\,x^5+7\,x^4+12\,x^3+4\,x^2+x+1)\nn\\
&& \nr\hspace{.6cm}/[(x+1)^3(x-1)^2]\,H(0,0,1,0;x)\nn\\
&& \nr-4\,(x^2+1)^2/[(x+1)^2(x-1)^2]\,H(1,0,1,0;x)\nn\\
&& \nr+2\,(x^2+1)/[(x-1)(x+1)]\,H(1,1,0;x)\nn\\
&& \nr-2\,(x^2+1)^2/[(x+1)^2(x-1)^2]\,H(0,1,1,0;x)\nn\\
&& \nr+8\,(x^2+1)/[(x+1)^2(x-1)]\,x\,H(0,1,0;x)\nn\\
&& \nr-(x^2+1)\,(\zeta(2)\,x^2+2\,\zeta(3)\,x^2-\zeta(2)+2\,\zeta(3))\nn\\
&& \nr\hspace{.3cm}/[(x+1)^2(x-1)^2]\,H(1;x)\nn\\
&& \nr+1/540\,(-1917\,\zeta(2)^2\,x^6+3240\,\zeta(2)\,\ln(2)\,x^6-7975\,x^6\nn\\
&& \nr\hspace{1.6cm}-5205\,\zeta(2)\,x^6+450\,\zeta(3)\,x^6-3564\,\zeta(2)^2\,x^5\nn\\
&& \nr\hspace{1.6cm}-22680\,\zeta(3)\,x^5+33930\,\zeta(2)\,x^5-19440\,\zeta(2)\,\ln(2)\,x^5\nn\\
&& \nr\hspace{1.6cm}+23925\,x^4+29160\,\zeta(2)\,\ln(2)\,x^4 \nn \\
&& \nr\hspace{1.6cm}+36270\,\zeta(3)\,x^4+12447\,\zeta(2)^2\,x^4-42405\,\zeta(2)\,x^4\nn\\
&& \nr\hspace{1.6cm}-13716\,\zeta(2)^2\,x^3-8460\,\zeta(2)\,x^3+14445\,\zeta(2)^2\,x^2\nn\\
&& \nr\hspace{1.6cm}-29160\,\zeta(2)\,\ln(2)\,x^2+47505\,\zeta(2)\,x^2-30690\,\zeta(3)\,x^2\nn\\
&& \nr\hspace{1.6cm}-23925\,x^2-25470\,\zeta(2)\,x+22680\,\zeta(3)\,x\nn\\
&& \nr\hspace{1.6cm}+19440\,\zeta(2)\,\ln(2)\,x-3564\,\zeta(2)^2\,x-6030\,\zeta(3)\nn\\
&& \nr\hspace{1.6cm}-3240\,\zeta(2)\,\ln(2)+7975+81\,\zeta(2)^2+105\,\zeta(2))\nn
\\
&& \nr\hspace{2.6cm} /[(x+1)^3(x-1)^3]\nn\\
&& \nr+\zeta(2)\,(x^2+1)^2/[(x+1)^2(x-1)^2]\,H(0,1;x)\nn\\
&& \nr+52/3\,(x^2+1)/[(x-1)(x+1)]\,H(-1,-1,0;x)\nn\\
&& \nr-1/3\,(10\,\zeta(2)\,x^4-6\,\zeta(3)\,x^4-81\,\zeta(2)\,x^3-12\,\zeta(3)\,x^2+108\,\zeta(2)\,x^2\nn\\
&& \nr\hspace{1.2cm}-81\,\zeta(2)\,x+26\,\zeta(2)-6\,\zeta(3))/[(x+1)^2(x-1)^2]\,H(-1;x)\nn\\
&& \nr+1/216\,(1044\,\zeta(2)\,x^6-1620\,\zeta(3)\,x^6+2545\,x^6+3024\,\zeta(3)\,x^5\nn\\
&& \nr\hspace{1.6cm}-570\,x^5-12528\,\zeta(2)\,x^5-13716\,\zeta(3)\,x^4-2545\,x^4\nn\\
&& \nr\hspace{1.6cm}+18396\,\zeta(2)\,x^4-12744\,\zeta(2)\,x^3+1140\,x^3+25920\,\zeta(3)\,x^3\nn\\
&& \nr\hspace{1.6cm}+3060\,\zeta(2)\,x^2-2545\,x^2-13500\,\zeta(3)\,x^2-570\,x\nn\\
&& \nr\hspace{1.6cm}-2808\,\zeta(2)\,x+3024\,\zeta(3)\,x-1404\,\zeta(3)\nn\\
&& \nr\hspace{1.6cm}+396\,\zeta(2)+2545)/[(x+1)^3(x-1)^3]\,H(0;x)\nn\\
&& \nr+\zeta(2)\,(7\,x^6+6\,x^5-29\,x^4+24\,x^3-31\,x^2+6\,x+5)\nn\\
&& \nr\hspace{1.1cm}/[(x+1)^3(x-1)^3]\,H(0,-1;x)\nn\\
&& \nr-1/9\,(31\,x^4-105\,x^3-36\,\zeta(2)\,x^2+105\,x-36\,\zeta(2)-31)\nn\\
&& \nr\hspace{1cm}/[(x+1)^2(x-1)^2]\,H(-1,0;x)\nn\\
&& \nr-10\,(x^2+1)^2/[(x+1)^2(x-1)^2]\,H(0,-1,-1,0;x),\\
\tilde{c}_{12} = && \nr 8\,(x^2+1)^2/[(x+1)^2(x-1)^2]\,H(-1,0,-1,0;x)\nn\\
&& \nr-8\,(x^2+1)^2/[(x+1)^2(x-1)^2]\,H(1,0,-1,0;x)\nn\\
&& \nr+8\,(x^2+1)^2/[(x+1)^2(x-1)^2]\,H(1,0,1,0;x)\nn\\
&& \nr+2\,\zeta(2)\,(5\,x^4-27\,x^3+36\,x^2-27\,x+7)\nn \\
&& \nr\hspace{1cm} /[(x+1)^2(x-1)^2]\,H(-1;x)\nn\\
&& \nr-2\,\zeta(2)\,(5\,x^6+6\,x^5-31\,x^4+24\,x^3-29\,x^2+6\,x+7)\nn\\
&& \nr\hspace{1.4cm}/[(x+1)^3(x-1)^3]\,H(0,-1;x)\nn\\
&& \nr-1/8\,(85\,x^6-64\,\zeta(3)\,x^6+136\,\zeta(2)\,x^6-160\,\zeta(3)\,x^5-6\,x^5\nn\\
&& \nr\hspace{1.2cm}-296\,\zeta(2)\,x^5+1056\,\zeta(3)\,x^4-85\,x^4-568\,\zeta(2)\,x^4\nn\\
&& \nr\hspace{1.2cm}+944\,\zeta(2)\,x^3+12\,x^3-1536\,\zeta(3)\,x^3-85\,x^2\nn\\
&& \nr\hspace{1.2cm}-616\,\zeta(2)\,x^2+1056\,\zeta(3)\,x^2+24\,\zeta(2)\,x\nn\\
&& \nr\hspace{1.2cm}-160\,\zeta(3)\,x-6\,x-64\,\zeta(3)+85-8\,\zeta(2))\nn\\
&& \nr\hspace{1.2cm}/[(x+1)^3(x-1)^3]\,H(0;x)\nn\\
&& \nr+1/2\,(-55\,x^4+8\,\zeta(2)\,x^4+82\,x^3+16\,\zeta(2)\,x^2-82\,x+8\,\zeta(2)+55)\nn\\
&& \nr\hspace{1cm}/[(x+1)^2(x-1)^2]\,H(-1,0;x)\nn\\
&& \nr+2\,(8\,x^4-27\,x^3+60\,x^2-27\,x+6)\nn\\
&& \nr\hspace{.6cm}/[(x+1)^2(x-1)^2]\,H(0,-1,0;x)\nn\\
&& \nr-2\,(11\,x^6-2\,x^5+33\,x^4-56\,x^3+23\,x^2-2\,x+1)\nn\\
&& \nr\hspace{.6cm}/[(x+1)^3(x-1)^3]\,H(0,0,-1,0;x)\nn\\
&& \nr-4\,(x^2+1)/[(x-1)(x+1)]\,H(-1,-1,0;x)\nn\\
&& \nr+1/4\,(229\,x^6+52\,\zeta(2)\,x^6-266\,x^5-16\,\zeta(2)\,x^5+124\,\zeta(2)\,x^4\nn\\
&& \nr\hspace{1.1cm}+15\,x^4-320\,\zeta(2)\,x^3+196\,x^3+84\,\zeta(2)\,x^2-189\,x^2+70\,x\nn\\
&& \nr\hspace{1.1cm}-16\,\zeta(2)\,x-55+12\,\zeta(2))/[(x+1)^3(x-1)^3]\,H(0,0;x)\nn\\
&& \nr-2\,(3\,x^4-15\,x^3+76\,x^2-15\,x+5)\nn\\
&& \nr\hspace{.6cm}/[(x+1)^2(x-1)^2]\,H(-1,0,0;x)\nn\\
&& \nr+16\,(x^2+1)^2/[(x+1)^2(x-1)^2]\,H(-1,-1,0,0;x)\nn\\
&& \nr-2\,(7\,x^6+10\,x^5-65\,x^4+104\,x^3-75\,x^2+10\,x-3)\nn\\
&& \nr\hspace{.6cm}/[(x+1)^3(x-1)^3]\,H(0,-1,0,0;x)\nn\\
&& \nr-2\,(5\,x^6-10\,x^5-60\,x^4+59\,x^3-16\,x^2-7\,x+5)\nn\\
&& \nr\hspace{.6cm}/[(x+1)^3(x-1)^3]\,H(0,0,0;x)\nn\\
&& \nr-12\,(x^2+1)^2/[(x+1)^2(x-1)^2]\,H(-1,0,0,0;x)\nn\\
&& \nr+(27\,x^6-2\,x^5+27\,x^4-56\,x^3-7\,x^2-2\,x-7)\nn\\
&& \nr\hspace{.3cm}/[(x+1)^3(x-1)^3]\,H(0,0,0,0;x)\nn\\
&& \nr+8\,(x^5-2\,x^4+18\,x^3-14\,x^2+4\,x+1)\nn\\
&& \nr\hspace{.6cm}/[(x+1)^3(x-1)^2]\,H(1,0,0,0;x)\nn\\
&& \nr+2\,(8\,x^4-19\,x^3+46\,x^2-19\,x+8)/[(x+1)^2(x-1)^2]\,H(1,0,0;x)\nn\\
&& \nr-4\,(x^6-x^5+14\,x^4-20\,x^3+14\,x^2-x+1)\nn\\
&& \nr\hspace{.6cm}/[(x+1)^3(x-1)^3]\,H(0,1,0,0;x)\nn\\
&& \nr+4\,(\zeta(2)\,x^5+4\,x^5+2\,x^4-5\,\zeta(2)\,x^4+34\,\zeta(2)\,x^3-2\,x^3+2\,x^2\nn\\
&& \nr\hspace{.8cm}-30\,\zeta(2)\,x^2-2\,x+7\,\zeta(2)\,x-4+\zeta(2))\nn\\
&& \nr\hspace{.8cm}/[(x+1)^3(x-1)^2]\,H(1,0;x)\nn\\
&& \nr-2\,(5\,x^4-10\,x^3+18\,x^2-10\,x+5)/[(x+1)^2(x-1)^2]\,H(0,1,0;x)\nn\\
&& \nr+4\,(3\,x^5+3\,x^4+6\,x^3-2\,x^2-x-1)\nn\\
&& \nr\hspace{.5cm}/[(x+1)^3(x-1)^2]\,H(0,0,1,0;x)\nn\\
&& \nr+4\,\zeta(3)\,(x^2+1)^2/[(x+1)^2(x-1)^2]\,H(1;x)\nn\\
&& \nr-1/20\,(-362\,\zeta(2)^2\,x^6+240\,\zeta(2)\,\ln(2)\,x^6-230\,x^6-275\,\zeta(2)\,x^6\nn\\
&& \nr\hspace{1.4cm}+140\,\zeta(3)\,x^6+144\,\zeta(2)^2\,x^5+600\,\zeta(3)\,x^5+230\,\zeta(2)\,x^5\nn\\
&& \nr\hspace{1.4cm}-1440\,\zeta(2)\,\ln(2)\,x^5+690\,x^4+2160\,\zeta(2)\,\ln(2)\,x^4\nn\\
&& \nr\hspace{1.4cm}-2980\,\zeta(3)\,x^4-1198\,\zeta(2)^2\,x^4+255\,\zeta(2)\,x^4\nn\\
&& \nr\hspace{1.4cm}+2656\,\zeta(2)^2\,x^3+500\,\zeta(2)\,x^3-954\,\zeta(2)^2\,x^2\nn\\
&& \nr\hspace{1.4cm}-2160\,\zeta(2)\,\ln(2)\,x^2-165\,\zeta(2)\,x^2+3060\,\zeta(3)\,x^2\nn\\
&& \nr\hspace{1.4cm}-690\,x^2-730\,\zeta(2)\,x-600\,\zeta(3)\,x+1440\,\zeta(2)\,\ln(2)\,x\nn\\
&& \nr\hspace{1.4cm}+144\,\zeta(2)^2\,x-220\,\zeta(3)-240\,\zeta(2)\,\ln(2)+230\nn\\
&& \nr\hspace{1.4cm}-118\,\zeta(2)^2+185\,\zeta(2))/[(x+1)^3(x-1)^3]\nn\\
&& \nr+4\,(x^2+1)^2/[(x+1)^2(x-1)^2]\,H(0,-1,-1,0;x), 
\eea
\bea
\tilde{d}_j = && \nr 0 \quad \mbox{for} \quad j=1\dots7,\\
\tilde{d}_8 = && \nr -2\,(5\,x^2-2\,x+5)/[(x-1)^3(x+1)]\,x\,H(0;x)\nn\\
&& \nr+4\,(3\,x^2-2\,x+3)\,(x^2+1)/[(x+1)^2(x-1)^4]\,x\,H(0,0;x)\nn\\
&& \nr+4/(x-1)^2\,x,\\
\tilde{d}_9 = && \nr -2/9\,(51\,x^2-26\,x+51)/[(x+1)(x-1)^3]\,x\,H(0;x)\nn\\
&& \nr+8/3\,(3\,x^2-2\,x+3)/[(x+1)(x-1)^3]\,x\,H(-1,0;x)\nn\\
&& \nr-4/3\,(3\,x^2-2\,x+3)/[(x+1)(x-1)^3]\,x\,H(0,0;x)\nn\\
&& \nr+4/9\,(19\,x^2+9\,\zeta(2)\,x^2-6\,\zeta(2)\,x-19+9\,\zeta(2)) \nn \\
&& \nr \hspace{1cm}/[(x+1)(x-1)^3]\,x,\\
\tilde{d}_{10} = && \nr
-2/9\,(51\,x^4-176\,x^3-70\,x^2-72\,\zeta(2)\,x^2-176\,x+51)\nn\\
&& \nr\hspace{1cm}/[(x+1)^3(x-1)^3]\,x\,H(0;x)\nn\\
&& \nr+4/3\,(3\,x^4-14\,x^3-2\,x^2-14\,x+3)/[(x+1)^4(x-1)^2]\,x\,H(0,0;x)\nn\\
&& \nr+16/[(x+1)^3(x-1)^3]\,x^3\,H(0,0,0;x)\nn\\
&& \nr+4/9\,(-8\,x^4+45\,\zeta(2)\,x^4-90\,\zeta(2)\,x^3-80\,x^3-144\,x^2+18\,\zeta(2)\,x^2\nn\\
&& \nr\hspace{1.1cm}\:-90\,\zeta(2)\,x-80\,x-8+45\,\zeta(2))/[(x+1)^4(x-1)^2]\,x,\\
\tilde{d}_{11} = && \nr -18\,\zeta(2)\,(x^2-4\,x+1)\,(x^2+1)/[(x+1)^2(x-1)^4]\,x\,H(-1;x)\nn\\
&& \nr+24\,\zeta(2)\,(x^2+1)\,(x^2-5\,x+1)/[(x+1)^3(x-1)^5]\,x^2\,H(0,-1;x)\nn\\
&& \nr+1/18\,(324\,\zeta(2)\,x^6+687\,x^6+1008\,\zeta(3)\,x^5-3384\,\zeta(2)\,x^5+26\,x^5\nn\\
&& \nr\hspace{1.4cm}-4752\,\zeta(3)\,x^4+3960\,\zeta(2)\,x^4-687\,x^4-7056\,\zeta(2)\,x^3\nn\\
&& \nr\hspace{1.4cm}+9216\,\zeta(3)\,x^3-52\,x^3-687\,x^2+1620\,\zeta(2)\,x^2\nn\\
&& \nr\hspace{1.4cm}-4752\,\zeta(3)\,x^2+1008\,\zeta(3)\,x+26\,x-648\,\zeta(2)\,x+687)\nn\\
&& \nr\hspace{1.4cm}/[(x+1)^3(x-1)^5]\,x\,H(0;x)\nn\\
&& \nr-2/3\,(3\,x^2-88\,x+3)/[(x+1)(x-1)^3]\,x\,H(-1,0;x)\nn\\
&& \nr-48\,(x^2-x+1)/[(x+1)^2(x-1)^4]\,x^2\,H(0,-1,0;x)\nn\\
&& \nr+8\,(5\,x^4-14\,x^3+30\,x^2-14\,x+5)\nn\\
&& \nr\hspace{.6cm}/[(x+1)^3(x-1)^5]\,x^2\,H(0,0,-1,0;x)\nn\\
&& \nr-1/3\,(39\,x^6+48\,\zeta(2)\,x^5+184\,x^5-252\,\zeta(2)\,x^4-87\,x^4+48\,\zeta(2)\,x^3\nn\\
&& \nr\hspace{1.1cm}-152\,x^3+81\,x^2-252\,\zeta(2)\,x^2-32\,x+48\,\zeta(2)\,x-33)\nn\\
&& \nr\hspace{1.1cm}/[(x+1)^3(x-1)^5]\,x\,H(0,0;x)\nn\\
&& \nr-2\,(9\,x^4-92\,x^3+130\,x^2-92\,x+9)\nn\\
&& \nr\hspace{.6cm}/[(x+1)^2(x-1)^4]\,x\,H(-1,0,0;x)\nn\\
&& \nr-8\,(7\,x^4-29\,x^3+62\,x^2-29\,x+7)\nn\\
&& \nr\hspace{.6cm}/[(x+1)^3(x-1)^5]\,x^2\,H(0,-1,0,0;x)\nn\\
&& \nr+2\,(9\,x^4-112\,x^3+146\,x^2-196\,x+9)\nn\\
&& \nr\hspace{.6cm}/[(x+1)^3(x-1)^5]\,x^3\,H(0,0,0;x)\nn\\
&& \nr+4\,(x^4-5\,x^3+38\,x^2-5\,x+1)\nn \\
&& \nr\hspace{1cm} /[(x+1)^3(x-1)^5]\,x^2\,H(0,0,0,0;x)\nn\\
&& \nr-16\,(3\,x^2-8\,x+3)/[(x+1)^3(x-1)^3]\,x^2\,H(1,0,0,0;x)\nn\\
&& \nr-24\,(3\,x^2-2\,x+3)/[(x+1)^2(x-1)^4]\,x^2\,H(1,0,0;x)\nn\\
&& \nr+8\,(5\,x^4-14\,x^3+42\,x^2-14\,x+5)\nn\\
&& \nr\hspace{.6cm}/[(x+1)^3(x-1)^5]\,x^2\,H(0,1,0,0;x)\nn\\
&& \nr-4\,(x^4+12\,\zeta(2)\,x^3-32\,\zeta(2)\,x^2-2\,x^2+12\,\zeta(2)\,x+1)\nn\\
&& \nr\hspace{.6cm}/[(x+1)^3(x-1)^3]\,x\,H(1,0;x)\nn\\
&& \nr-16/[(x+1)^2(x-1)^2]\,x^2\,H(0,1,0;x)\nn\\
&& \nr-32/[(x+1)^3(x-1)^3]\,x^3\,H(0,0,1,0;x)\nn\\
&& \nr+1/45\,(1080\,\zeta(2)\,\ln(2)\,x^6+540\,\zeta(3)\,x^6-1395\,\zeta(2)\,x^6-1210\,x^6\nn\\
&& \nr\hspace{1.4cm}-7380\,\zeta(3)\,x^5-4320\,\zeta(2)\,\ln(2)\,x^5-1188\,\zeta(2)^2\,x^5\nn\\
&& \nr\hspace{1.4cm}+10500\,\zeta(2)\,x^5+3630\,x^4+10980\,\zeta(3)\,x^4+4878\,\zeta(2)^2\,x^4\nn\\
&& \nr\hspace{1.4cm}-7425\,\zeta(2)\,x^4+5400\,\zeta(2)\,\ln(2)\,x^4-3360\,\zeta(2)\,x^3\nn\\
&& \nr\hspace{1.4cm}-3168\,\zeta(2)^2\,x^3+4878\,\zeta(2)^2\,x^2-10980\,\zeta(3)\,x^2\nn\\
&& \nr\hspace{1.4cm}+7875\,\zeta(2)\,x^2-3630\,x^2-5400\,\zeta(2)\,\ln(2)\,x^2\nn\\
&& \nr\hspace{1.4cm}+4320\,\zeta(2)\,\ln(2)\,x-1188\,\zeta(2)^2\,x-7140\,\zeta(2)\,x\nn\\
&& \nr\hspace{1.4cm}+7380\,\zeta(3)\,x-540\,\zeta(3)-1080\,\zeta(2)\,\ln(2)\nn\\
&& \nr\hspace{1.4cm}+1210+945\,\zeta(2))/[(x+1)^3(x-1)^5]\,x,\\
\tilde{d}_{12} = && \nr 36\,\zeta(2)\,(x^2-4\,x+1)\,(x^2+1)/[(x+1)^2(x-1)^4]\,x\,H(-1;x)\nn\\
&& \nr-48\,\zeta(2)\,(x^2+1)\,(x^2-5\,x+1)/[(x+1)^3(x-1)^5]\,x^2\,H(0,-1;x)\nn\\
&& \nr-1/2\,(61\,x^6+96\,\zeta(2)\,x^6-22\,x^5-216\,\zeta(2)\,x^5-32\,\zeta(3)\,x^5\nn\\
&& \nr\hspace{1.2cm}-1032\,\zeta(2)\,x^4-61\,x^4+1056\,\zeta(3)\,x^4+448\,\zeta(2)\,x^3+44\,x^3\nn\\
&& \nr\hspace{1.2cm}-1664\,\zeta(3)\,x^3+1056\,\zeta(3)\,x^2-480\,\zeta(2)\,x^2-61\,x^2\nn\\
&& \nr\hspace{1.2cm}-32\,\zeta(3)\,x-22\,x+56\,\zeta(2)\,x+61-24\,\zeta(2))\nn\\
&& \nr\hspace{1.2cm}/[(x+1)^3(x-1)^5]\,x\,H(0;x)\nn\\
&& \nr-2\,(25\,x^2-86\,x+25)/[(x-1)^3(x+1)]\,x\,H(-1,0;x)\nn\\
&& \nr+8\,(3\,x^4-13\,x^3+64\,x^2-13\,x+3)\nn\\
&& \nr\hspace{.6cm}/[(x+1)^2(x-1)^4]\,x\,H(0,-1,0;x)\nn\\
&& \nr-16\,(x^4+19\,x^3-28\,x^2+19\,x+1)\nn\\
&& \nr\hspace{.8cm}/[(x+1)^3(x-1)^5]\,x^2\,H(0,0,-1,0;x)\nn\\
&& \nr+(125\,x^6-230\,x^5+187\,x^4+96\,\zeta(2)\,x^4-432\,\zeta(2)\,x^3+204\,x^3\nn\\
&& \nr\hspace{.5cm}+96\,\zeta(2)\,x^2-269\,x^2+26\,x-43)/[(x+1)^3(x-1)^5]\,x\,H(0,0;x)\nn\\
&& \nr-4\,(15\,x^4-16\,x^3+198\,x^2-16\,x+15)\nn\\
&& \nr\hspace{.6cm}/[(x+1)^2(x-1)^4]\,x\,H(-1,0,0;x)\nn\\
&& \nr-16\,(x^4-37\,x^3+54\,x^2-37\,x+1)\nn\\
&& \nr\hspace{.8cm}/[(x+1)^3(x-1)^5]\,x^2\,H(0,-1,0,0;x)\nn\\
&& \nr+4\,(14\,x^5+175\,x^4-56\,x^3+14\,x^2+6\,x-9)\nn\\
&& \nr\hspace{.6cm}/[(x+1)^3(x-1)^5]\,x\,H(0,0,0;x)\nn\\
&& \nr-8\,(x^4-5\,x^3+38\,x^2-5\,x+1)\nn \\
&& \nr\hspace{1cm} /[(x+1)^3(x-1)^5]\,x^2\,H(0,0,0,0;x)\nn\\
&& \nr-32\,(x^2-17\,x+1)/[(x+1)^3(x-1)^3]\,x^2\,H(1,0,0,0;x)\nn\\
&& \nr+8\,(3\,x^4-7\,x^3+64\,x^2-7\,x+3)/[(x+1)^2(x-1)^4]\,x\,H(1,0,0;x)\nn\\
&& \nr-16\,(x^4+19\,x^3-16\,x^2+19\,x+1)\nn\\
&& \nr\hspace{.8cm}/[(x+1)^3(x-1)^5]\,x^2\,H(0,1,0,0;x)\nn\\
&& \nr+32\,(x^4-\zeta(2)\,x^3+x^3+17\,\zeta(2)\,x^2-\zeta(2)\,x+x+1)\nn\\
&& \nr\hspace{.8cm}/[(x+1)^3(x-1)^3]\,x\,H(1,0;x)\nn\\
&& \nr-8\,(3\,x^4-6\,x^3+14\,x^2-6\,x+3)/[(x+1)^2(x-1)^4]\,x\,H(0,1,0;x)\nn\\
&& \nr+64/[(x+1)^3(x-1)^3]\,x^3\,H(0,0,1,0;x)\nn\\
&& \nr-1/5\,(240\,\zeta(2)\,\ln(2)\,x^6-60\,\zeta(3)\,x^6-235\,\zeta(2)\,x^6-85\,x^6\nn\\
&& \nr\hspace{1.2cm}+320\,\zeta(3)\,x^5-960\,\zeta(2)\,\ln(2)\,x^5+8\,\zeta(2)^2\,x^5\nn\\
&& \nr\hspace{1.2cm}-30\,\zeta(2)\,x^5+255\,x^4-3340\,\zeta(3)\,x^4-1160\,\zeta(2)^2\,x^4\nn\\
&& \nr\hspace{1.2cm}+1595\,\zeta(2)\,x^4+1200\,\zeta(2)\,\ln(2)\,x^4+540\,\zeta(2)\,x^3\nn\\
&& \nr\hspace{1.2cm}+3240\,\zeta(2)^2\,x^3-1160\,\zeta(2)^2\,x^2+3340\,\zeta(3)\,x^2-1525\,\zeta(2)\,x^2\nn\\
&& \nr\hspace{1.2cm}-255\,x^2-1200\,\zeta(2)\,\ln(2)\,x^2+960\,\zeta(2)\,\ln(2)\,x+8\,\zeta(2)^2\,x\nn\\
&& \nr\hspace{1.2cm}-510\,\zeta(2)\,x-320\,\zeta(3)\,x+60\,\zeta(3)-240\,\zeta(2)\,\ln(2)\nn\\
&& \nr\hspace{1.2cm}+85+165\,\zeta(2))/[(x+1)^3(x-1)^5]\,x. 
\eea

%
\boldmath
\subsection{Form Factors for $\mu \not{\! \! =} \, m$ \label{munotm}}
\unboldmath
%

In this Section we give the expressions for the renormalized
axial vector two-loop form factors
for the case of $\mu \not{\! \! =} \, m$.

At the one-loop level 
we do not have an explicit dependence on the logarithm of the ratio of the 
renormalization scale and the mass of the heavy quark, 
because an overall factor $(\mu^2/m^2)^{\epsilon}$ can be taken
out, see Eqs.~(\ref{g1_1l_ren_decomp},\ref{g2_1l_ren_decomp}).

At the two-loop level, such a dependence results from the coupling constant 
renormalization, first appearing at this level.
Factoring an overall $(\mu^2/m^2)^{2 \epsilon}$, we have:
\bea
G_{i,R}^{(2l)}\Bigl(\epsilon,s,\frac{\mu^2}{m^2}\Bigr) &=& 
C^2(\epsilon) \, \left( \frac{\mu^2}{m^2} \right)^{2 \epsilon} 
\Biggl\{
G_{i,R}^{(2l)}(\epsilon,s) 
+ {\mathcal O}_{i}^{(2l)}(\epsilon,s) \ln{\left( \frac{\mu^2}{m^2} \right)}
\nn\\
& & \hspace*{30mm}
+ {\mathcal P}_{i}^{(2l)}(s) \ln^2{\left( \frac{\mu^2}{m^2} \right)}
\Biggr\} 
\, ,
\label{G1G2_2l_munotm}
\eea
where the functions $G_{i,R}^{(2l)}(\epsilon,s)$ are given in 
Eqs.~(\ref{g1_2l_ren_decomp},\ref{g2_2l_ren_decomp}) and the functions 
${\mathcal O}_{i}^{(2l)}(\epsilon,s)$ and ${\mathcal P}_{i}^{(2l)}(s)$ 
can be obtained either from the renormalization group \cite{bbghlmr}
or by explicit expansion. We find:  
\bea
{\mathcal O}_{1}^{(2l)}(\epsilon,s) = && - C_F\,[\,T_R\,(N_f+1)\,-11/4 \,N_c\,]\nn\\
&& \times \Bigg\{ \phantom{+}\,\frac{2}{3\,\epsilon} \, \bigg\{\,(x^2+1)/[(x-1)(x+1)]\,H(0;x)-1\,\bigg\}\nn \\
&& \hspace{.7cm}+ 1/3\,(3\,x^2-2\,x+3)/[(x-1)(x+1)]\,H(0;x)\nn\\
&& \hspace{.7cm}- 4/3\,(x^2+1)/[(x-1)(x+1)]\,H(-1,0;x)\nn \\
&& \hspace{.7cm}+ 2/3\,(x^2+1)/[(x-1)(x+1)]\,H(0,0;x)\nn \\
&& \hspace{.7cm}- 2/3\,(x^2\,\zeta(2)+2\,x^2+\zeta(2)-2)/[(x-1)(x+1)]\Bigg\},\\
{\mathcal P}_{1}^{(2l)}(s) = && C_F\,[\,T_R\,(N_f+1)\,-11/4 \,N_c\,]\nn\\
&&\times 1/3\;\{(x^2+1)/[(x-1)\,(x+1)]\,H(0;x)-1\},\\
{\mathcal O}_{2}^{(2l)}(\epsilon,s) = && - C_F\,[\,T_R\,(N_f+1)\,-11/4 \,N_c\,]\nn\\
&& \times 4/3\,\{(3\,x^2-2\,x+3)/[(x-1)^3 (x+1)]\,x\,H(0;x)\nn \\
&& \hspace{1.3cm}-2\,(x-1)^2\,x\},\\
{\mathcal P}_{2}^{(2l)}(s) = && 0 . 
\eea

%
\section{Analytical Continuation above Threshold\label{sec_analytic}}
%

The results for the renormalized form factors can be analytically
continued into the timelike region $s>0$ and in particular above the
physical threshold $s>4$ by using the substitution \cite{bbghlmr}
\be
x\rightarrow-y+i\epsilon,
\ee
with
\be
y=\frac{\sqrt{s}-\sqrt{s-4}}{\sqrt{s}+\sqrt{s-4}}.
\ee
For $s>4$ the form factors become complex due to
absorptive parts. We write:
\bea
G_1(s+i\epsilon,\epsilon) & = & \Re \, G_1(s,\epsilon)+ i\pi \, \Im \,
G_1(s,\epsilon),\\
G_1(s+i\epsilon,\epsilon) & = & \Re \, G_1(s,\epsilon)+ i\pi \, \Im \,
G_1(s,\epsilon).
\eea
These imaginary parts arise from the analytical continuation of the
harmonic polylogarithms with rightmost index 0.
In the following two Subsections we shall give the real and imaginary
parts of $G_{1,2}$ at one and two loops, putting $\mu = m$.

%
\begin{boldmath}
\subsection{One-Loop Form Factors above Threshold \label{subsec_1l_anacon}}
\end{boldmath}
%
The analytical continuation of the coefficients
$\tilde{a}_i$ and $\tilde{b}_i$ in
Eqs.~(\ref{g1_1l_ren_decomp},\ref{g2_1l_ren_decomp}) is:
\bea
\Re \, \tilde{a}_1  = && \nr (y^2+1)/[(y-1)\,(y+1)]\,H(0;y)-1,\\
\Re \, \tilde{a}_2  = && \nr 1/2\,(3\,y^2+2\,y+3)/[(y-1)\,(y+1)]\,H(0;y) \nn \\
&& \nr+(y^2+1)/[(y-1)\,(y+1)]\,H(0,0;y) \nn \\
&& \nr+2\,(y^2+1)/[(y-1)\,(y+1)]\,H(1,0;y) \nn \\
&& \nr-2\,(y^2+2\,\zeta(2)\,y^2-1+2\,\zeta(2))/[(y-1)\,(y+1)],\\  
\Re \, \tilde{a}_3  = && \nr -4\,(y^2+1)\,(-1+\zeta(2))/[(y-1)\,(y+1)]\,H(0;y) \nn \\
&& \nr+1/2\,(3\,y^2+2\,y+3)/[(y-1)\,(y+1)]\,H(0,0;y) \nn \\
&& \nr+(y^2+1)/[(y-1)\,(y+1)]\,H(0,0,0;y) \nn \\
&& \nr+2\,(y^2+1)/[(y-1)\,(y+1)]\,H(1,0,0;y) \nn \\
&& \nr+(3\,y^2+2\,y+3)/[(y-1)\,(y+1)]\,H(1,0;y) \nn \\
&& \nr+2\,(y^2+1)/[(y-1)\,(y+1)]\,H(0,1,0;y) \nn \\
&& \nr+4\,(y^2+1)/[(y-1)\,(y+1)]\,H(1,1,0;y) \nn \\
&& \nr-8\,\zeta(2)\,(y^2+1)/[(y-1)\,(y+1)]\,H(1;y) \nn \\
&& \nr-2\,(\zeta(3)\,y^2+3\,\zeta(2)\,y^2+2\,y^2+2\,\zeta(2)\,y-2+3\,\zeta(2)+\zeta(3))\nn\\
&& \nr\hspace{.7cm}/[(y-1)\,(y+1)],\\
\Im \, \tilde{a}_1 = && \nr (y^2+1)/[(y-1)\,(y+1)],\\
\Im \, \tilde{a}_2 = && \nr (y^2+1)/[(y-1)\,(y+1)]\,H(0;y) \nn \\
&& \nr+2\,(y^2+1)/[(y-1)\,(y+1)]\,H(1;y) \nn \\
&& \nr+1/2\,(3\,y^2+2\,y+3)/[(y-1)\,(y+1)],\\
\Im \, \tilde{a}_3 = && \nr 1/2\,(3\,y^2+2\,y+3)/[(y-1)\,(y+1)]\,H(0;y) \nn \\
&& \nr+(y^2+1)/[(y-1)\,(y+1)]\,H(0,0;y) \nn \\
&& \nr+2\,(y^2+1)/[(y-1)\,(y+1)]\,H(1,0;y) \nn \\
&& \nr+(3\,y^2+2\,y+3)/[(y-1)\,(y+1)]\,H(1;y) \nn \\
&& \nr+2\,(y^2+1)/[(y-1)\,(y+1)]\,H(0,1;y) \nn \\
&& \nr+4\,(y^2+1)/[(y-1)\,(y+1)]\,H(1,1;y) \nn \\
&& \nr-2\,(y^2+1)\,(\zeta(2)-2)/[(y-1)\,(y+1)],\\
\Re \, \tilde{b}_2  = && \nr
-2\,(3\,y^2+2\,y+3)/[(y-1)\,(y+1)^3]\,y\,H(0;y) +4/(y+1)^2\,y,\\
\Re \, \tilde{b}_3  = && \nr -4\,(3\,y^2+2\,y+3)/[(y-1)\,(y+1)^3]\,y\,H(0;y) \nn \\
&& \nr-2\,(3\,y^2+2\,y+3)/[(y-1)\,(y+1)^3]\,y\,H(0,0;y) \nn \\
&& \nr-4\,(3\,y^2+2\,y+3)/[(y-1)\,(y+1)^3]\,y\,H(1,0;y) \nn \\
&& \nr+8\,(y^2+3\,\zeta(2)\,y^2+2\,\zeta(2)\,y+3\,\zeta(2)-1)\nn\\
&& \nr\hspace{.7cm}/[(y-1)\,(y+1)^3]\,y,\\
\Im \, \tilde{b}_2  = && \nr -2\,(3\,y^2+2\,y+3)/[(y-1)\,(y+1)^3]\,y,\\
\Im \, \tilde{b}_3  = && \nr -2\,(3\,y^2+2\,y+3)/[(y-1)\,(y+1)^3]\,y\,H(0;y) \nn \\
&& \nr-4\,(3\,y^2+2\,y+3)/[(y-1)\,(y+1)^3]\,y\,H(1;y) \nn \\
&& \nr-4\,(3\,y^2+2\,y+3)/[(y-1)\,(y+1)^3]\,y.
\eea
The above expressions for the singular term and the term of order 
$\epsilon^0$ of $G_1$ agree with those of \cite{Jersak:1981sp}.
%
\begin{boldmath}
\subsection{Two-Loop Form Factors above Threshold \label{subsec_2l_anacon}}
\end{boldmath}
%

The real and imaginary parts of the coefficients $\tilde{c}_i$ and $\tilde{d}_i$
in Eq.~(\ref{g1_2l_ren_decomp}) are:
\bea
\Re \, \tilde{c}_{1}  = && \nr -\frac{4}{11}\, \Re \, \tilde{c}_3 =-\frac{3}{5}\, \Re \, \tilde{c}_5=
\nn \\
&& \nr 1/3\,(y^2+1)/[(y-1)(y+1)]\,H(0;y)-1/3,\\
\Re \, \tilde{c}_{2}  = && \nr \Re \, \tilde{c}_{6} = 0,\\
\Re \, \tilde{c}_{4}  = && \nr -(y^2+1)/[(y-1)(y+1)]\,H(0;y)\nn\\
&& \nr+(y^2+1)^2/[(y-1)^2(y+1)^2]\,H(0,0;y)\nn\\
&& \nr-1/2\,(-y^4+6\,\zeta(2)\,y^4+2\,y^2+12\,\zeta(2)\,y^2-1+6\,\zeta(2))\nn
\\
&& \nr\hspace{1cm} /[(y-1)^2(y+1)^2],\\
\Re \, \tilde{c}_{7}  = && \nr
1/36\,(y^2+1)\,(67\,y^2+162\,\zeta(2)\,y^2-67+18\,\zeta(2))\nn \\
&& \nr\hspace{1cm} /[(y-1)^2(y+1)^2]\,H(0;y)\nn\\
&& \nr-(y^2+1)/[(y+1)(y-1)]\,H(-1,0;y)\nn\\
&& \nr+(y^2+1)^2/[(y-1)^2(y+1)^2]\,H(0,-1,0;y)\nn\\
&& \nr+2/[(y+1)(y-1)]\,y^2\,H(0,0;y)\nn\\
&& \nr-2\,(y^2+1)/[(y-1)^2(y+1)^2]\,y^2\,H(0,0,0;y)\nn\\
&& \nr+(y^2+1)/[(y+1)(y-1)]\,H(1,0;y)\nn\\
&& \nr-(y^2+1)^2/[(y-1)^2(y+1)^2]\,H(0,1,0;y)\nn\\
&& \nr-1/36\,(49\,y^4+18\,\zeta(3)\,y^4+162\,\zeta(2)\,y^4-144\,\zeta(2)\,y^2+36\,\zeta(3)\,y^2\nn
\\
&& \nr\hspace{1.4cm} -98\,y^2-18\,\zeta(2)+18\,\zeta(3)+49)/[(y-1)^2(y+1)^2],\\
\Re \, \tilde{c}_{8}  = && \nr
-1/2\,(20\,\zeta(2)\,y^4+7\,y^4+2\,y^3+40\,\zeta(2)\,y^2-2\,y-7+20\,\zeta(2))\nn
\\
&& \nr\hspace{1cm} /[(y+1)^2(y-1)^2]\,H(0;y)\nn\\
&& \nr+2\,(y^2+1)\,(y^2+y+2)/[(y+1)^2(y-1)^2]\,H(0,0;y)\nn\\
&& \nr+3\,(y^2+1)^2/[(y+1)^2(y-1)^2]\,H(0,0,0;y)\nn\\
&& \nr+4\,(y^2+1)^2/[(y+1)^2(y-1)^2]\,H(1,0,0;y)\nn\\
&& \nr-2\,(y^2+1)/[(y+1)(y-1)]\,H(1,0;y)\nn\\
&& \nr+2\,(y^2+1)^2/[(y-1)^2(y+1)^2]\,H(0,1,0;y)\nn\\
&& \nr-12\,\zeta(2)\,(y^2+1)^2/[(y+1)^2(y-1)^2]\,H(1;y)\nn\\
&& \nr-(5\,\zeta(2)\,y^4-2\,y^4+6\,\zeta(2)\,y^3+18\,\zeta(2)\,y^2+4\,y^2+6\,\zeta(2)\,y\nn
\\
&& \nr\hspace{0.6cm} -2+13\,\zeta(2))/[(y+1)^2(y-1)^2],\\
\Re \, \tilde{c}_{9}  = && \nr
1/54\,(72\,\zeta(2)\,y^2-209\,y^2-78\,y+72\,\zeta(2)-209)\nn \\
&& \nr\hspace{1cm} /[(y-1)(y+1)]\,H(0;y)\nn\\
&& \nr-1/9\,(19\,y^2+6\,y+19)/[(y-1)(y+1)]\,H(0,0;y)\nn\\
&& \nr-2/3\,(y^2+1)/[(y-1)(y+1)]\,H(0,0,0;y)\nn\\
&& \nr-4/3\,(y^2+1)/[(y-1)(y+1)]\,H(1,0,0;y)\nn\\
&& \nr-2/9\,(19\,y^2+6\,y+19)/[(y-1)(y+1)]\,H(1,0;y)\nn\\
&& \nr-4/3\,(y^2+1)/[(y-1)(y+1)]\,H(0,1,0;y)\nn\\
&& \nr-8/3\,(y^2+1)/[(y-1)(y+1)]\,H(1,1,0;y)\nn\\
&& \nr+16/3\,\zeta(2)\,(y^2+1)/[(y-1)(y+1)]\,H(1;y)\nn\\
&& \nr+2/27\,(18\,\zeta(3)\,y^2+53\,y^2+132\,\zeta(2)\,y^2+36\,\zeta(2)\,y-53\nn
\\
&& \nr\hspace{1.4cm} +96\,\zeta(2)+18\,\zeta(3))/[(y-1)(y+1)],\\
\Re \, \tilde{c}_{10}  = && \nr
1/54\,(-265\,y^4+72\,\zeta(2)\,y^4-208\,y^3-144\,\zeta(2)\,y^3-14\,y^2\nn
\\
&& \nr\hspace{1cm}
-288\,\zeta(2)\,y^2-208\,y-144\,\zeta(2)\,y-265+72\,\zeta(2))\nn \\
&& \nr\hspace{2cm} /[(y-1)^3(y+1)]\,H(0;y)\nn\\
&& \nr+1/9\,(19\,y^4+14\,y^3+14\,y^2+14\,y+19)/(y-1)^4\,H(0,0;y)\nn\\
&& \nr-2/3\,(y^4-2\,y^3-4\,y^2-2\,y+1)/[(y-1)^3(y+1)]\,H(0,0,0;y)\nn\\
&& \nr-1/27\,(-383\,y^2+198\,\zeta(2)\,y^2-504\,\zeta(2)\,y+646\,y\nn \\
&& \nr\hspace{1.4cm} -383+198\,\zeta(2))/(y-1)^2,\\
\Re \, \tilde{c}_{11}  = && \nr
(-11\,\zeta(2)\,y^4+2\,\zeta(3)\,y^4+30\,\zeta(2)\,y^3+84\,\zeta(2)\,y^2+4\,\zeta(3)\,y^2\nn
\\
&& \nr\hspace{1cm} +30\,\zeta(2)\,y-37\,\zeta(2)+2\,\zeta(3))/[(y-1)^2(y+1)^2]\,H(-1;y)\nn\\
&& \nr-\zeta(2)\,(7\,y^6+30\,y^5+127\,y^4+204\,y^3+101\,y^2+30\,y-19)\nn
\\
&& \nr\hspace{1cm} /[(y-1)^3(y+1)^3]\,H(0,-1;y)\nn\\
&& \nr-1/216\,(8568\,\zeta(2)\,y^6+1620\,\zeta(3)\,y^6-2545\,y^6-570\,y^5\nn
\\
&& \nr\hspace{1.4cm}
+3024\,\zeta(3)\,y^5+30888\,\zeta(2)\,y^5+35496\,\zeta(2)\,y^4+2545\,y^4\nn
\\
&& \nr\hspace{1.4cm}
+13716\,\zeta(3)\,y^4+1140\,y^3+25920\,\zeta(3)\,y^3+25488\,\zeta(2)\,y^3\nn
\\
&& \nr\hspace{1.4cm}
+2545\,y^2+936\,\zeta(2)\,y^2+13500\,\zeta(3)\,y^2-216\,\zeta(2)\,y\nn
\\
&& \nr\hspace{1.4cm}
-570\,y+3024\,\zeta(3)\,y+792\,\zeta(2)-2545+1404\,\zeta(3))\nn \\
&& \nr\hspace{2.4cm} /[(y-1)^3(y+1)^3]\,H(0;y)\nn\\
&& \nr+(3\,y^5+2\,\zeta(2)\,y^5+22\,\zeta(2)\,y^4-y^4-2\,y^3+68\,\zeta(2)\,y^3-2\,y^2\nn
\\
&& \nr\hspace{1cm} +108\,\zeta(2)\,y^2+2\,\zeta(2)\,y-y+22\,\zeta(2)+3)\nn
\\
&& \nr\hspace{2cm} /[(y-1)^3(y+1)^2]\,H(-1,0;y)\nn\\
&& \nr+2\,(y^2+1)/[(y+1)(y-1)]\,H(-1,-1,0;y)\nn\\
&& \nr-2\,(y^2+1)^2/[(y-1)^2(y+1)^2]\,H(0,-1,-1,0;y)\nn\\
&& \nr-4\,(y^2+1)^2/[(y-1)^2(y+1)^2]\,H(-1,0,-1,0;y)\nn\\
&& \nr+2\,(7\,y^5-7\,y^4+12\,y^3-4\,y^2+y-1)\nn \\
&& \nr\hspace{1cm} /[(y-1)^3(y+1)^2]\,H(0,0,-1,0;y)\nn\\
&& \nr+4\,(y^2+1)^2/[(y-1)^2(y+1)^2]\,H(1,0,-1,0;y)\nn\\
&& \nr-6\,(y^2+1)/[(y+1)(y-1)]\,H(1,-1,0;y)\nn\\
&& \nr+6\,(y^2+1)^2/[(y-1)^2(y+1)^2]\,H(0,1,-1,0;y)\nn\\
&& \nr+1/36\,(-217\,y^6+1332\,\zeta(2)\,y^6+282\,y^5+252\,\zeta(2)\,y^5\nn
\\
&& \nr\hspace{1cm}
+1692\,\zeta(2)\,y^4+451\,y^4-348\,y^3+3528\,\zeta(2)\,y^3-467\,y^2\nn
\\
&& \nr\hspace{1cm}
+324\,\zeta(2)\,y^2+66\,y+252\,\zeta(2)\,y-36\,\zeta(2)+233)\nn \\
&& \nr\hspace{2cm} /[(y-1)^3(y+1)^3]\,H(0,0;y)\nn\\
&& \nr-8\,(y^2+1)/[(y-1)^2(y+1)]\,y\,H(0,-1,0;y)\nn\\
&& \nr+2\,(2\,y^4-5\,y^3-14\,y^2-5\,y+6)\nn \\
&& \nr\hspace{1cm} /[(y-1)^2(y+1)^2]\,H(-1,0,0;y)\nn\\
&& \nr+2\,(y^6+5\,y^5+21\,y^4+34\,y^3+17\,y^2+5\,y-3)\nn \\
&& \nr\hspace{1cm} /[(y-1)^3(y+1)^3]\,H(0,-1,0,0;y)\nn\\
&& \nr+1/6\,(89\,y^6+402\,y^5+499\,y^4+354\,y^3+37\,y^2+24\,y+11)\nn \\
&& \nr\hspace{1cm} /[(y-1)^3(y+1)^3]\,H(0,0,0;y)\nn\\
&& \nr-2\,(y^5+5\,y^4+18\,y^3+26\,y^2+y+5)\nn \\
&& \nr\hspace{1cm} /[(y-1)^3(y+1)^2]\,H(-1,0,0,0;y)\nn\\
&& \nr-(12\,y^5+y^4+10\,y^3+28\,y^2-2\,y+1)\nn \\
&& \nr\hspace{1cm} /[(y-1)^3(y+1)^3]\,y\,H(0,0,0,0;y)\nn\\
&& \nr-2\,(y^2+3)\,(y^2+1)/[(y-1)^2(y+1)^2]\,H(1,0,0,0;y)\nn\\
&& \nr+1/3\,(53\,y^4+141\,y^3+228\,y^2+141\,y-17)\nn \\
&& \nr\hspace{1cm} /[(y-1)^2(y+1)^2]\,H(1,0,0;y)\nn\\
&& \nr-2\,(7\,y^6+7\,y^5+33\,y^4+56\,y^3+25\,y^2+7\,y-1)\nn \\
&& \nr\hspace{1cm} /[(y-1)^3(y+1)^3]\,H(0,1,0,0;y)\nn\\
&& \nr+1/9\,(54\,\zeta(2)\,y^4+31\,y^4+105\,y^3+180\,\zeta(2)\,y^2-105\,y\nn
\\
&& \nr\hspace{1cm} -31+126\,\zeta(2))/[(y-1)^2(y+1)^2]\,H(1,0;y)\nn\\
&& \nr-6\,(y^2+1)/[(y+1)(y-1)]\,H(-1,1,0;y)\nn\\
&& \nr+6\,(y^2+1)^2/[(y-1)^2(y+1)^2]\,H(0,-1,1,0;y)\nn\\
&& \nr+2/3\,(34\,y^4-3\,y^3-33\,y^2-3\,y-1)\nn \\
&& \nr\hspace{1cm} /[(y-1)^2(y+1)^2]\,H(0,1,0;y)\nn\\
&& \nr+4\,(y^2+1)^2/[(y-1)^2(y+1)^2]\,H(-1,0,1,0;y)\nn\\
&& \nr-2\,(11\,y^6-5\,y^5-15\,y^4-26\,y^3-27\,y^2-5\,y-1)\nn \\
&& \nr\hspace{1cm} /[(y-1)^3(y+1)^3]\,H(0,0,1,0;y)\nn\\
&& \nr-4\,(y^2+1)^2/[(y-1)^2(y+1)^2]\,H(1,0,1,0;y)\nn\\
&& \nr-10\,(y^2+1)^2/[(y-1)^2(y+1)^2]\,H(0,1,1,0;y)\nn\\
&& \nr-1/3\,(6\,\zeta(3)\,y^4+149\,\zeta(2)\,y^4+342\,\zeta(2)\,y^3+12\,\zeta(3)\,y^2\nn
\\
&& \nr\hspace{1cm}
+576\,\zeta(2)\,y^2+342\,\zeta(2)\,y-77\,\zeta(2)+6\,\zeta(3))\nn \\
&& \nr\hspace{2cm} /[(y-1)^2(y+1)^2]\,H(1;y)\nn\\
&& \nr+\zeta(2)\,(35\,y^6+48\,y^5+227\,y^4+360\,y^3+181\,y^2+48\,y-11)\nn
\\
&& \nr\hspace{1cm} /[(y-1)^3(y+1)^3]\,H(0,1;y)\nn\\
&& \nr+1/540\,(450\,\zeta(3)\,y^6+3240\,\zeta(2)\,\ln(2)\,y^6-7975\,y^6\nn
\\
&& \nr\hspace{1cm}
-13257\,\zeta(2)^2\,y^6+4560\,\zeta(2)\,y^6-3726\,\zeta(2)^2\,y^5\nn
\\
&& \nr\hspace{1cm}
+19440\,\zeta(2)\,\ln(2)\,y^5-46620\,\zeta(2)\,y^5+22680\,\zeta(3)\,y^5\nn
\\
&& \nr\hspace{1cm}
-62700\,\zeta(2)\,y^4+36270\,\zeta(3)\,y^4+29160\,\zeta(2)\,\ln(2)\,y^4\nn
\\
&& \nr\hspace{1cm}
-23193\,\zeta(2)^2\,y^4+23925\,y^4+24120\,\zeta(2)\,y^3\nn \\
&& \nr\hspace{1cm}
-31644\,\zeta(2)^2\,y^3-29160\,\zeta(2)\,\ln(2)\,y^2+68520\,\zeta(2)\,y^2\nn
\\
&& \nr\hspace{1cm}
-30690\,\zeta(3)\,y^2-23925\,y^2-8235\,\zeta(2)^2\,y^2\nn \\
&& \nr\hspace{1cm}
-19440\,\zeta(2)\,\ln(2)\,y-3726\,\zeta(2)^2\,y-22680\,\zeta(3)\,y\nn
\\
&& \nr\hspace{1cm}
+22500\,\zeta(2)\,y+1701\,\zeta(2)^2-6030\,\zeta(3)+7975\nn \\
&& \nr\hspace{1cm} -3240\,\zeta(2)\,\ln(2)-10380\,\zeta(2))\nn \\
&& \nr\hspace{2cm} /[(y-1)^3(y+1)^3]\nn \\
&& \nr+52/3\,(y^2+1)/[(y-1)(y+1)]\,H(1,1,0;y),\nn\\
\Re \, \tilde{c}_{12}  = && \nr 2\,(5\,y^4+10\,y^3+18\,y^2+10\,y+5)\nn \\
&& \nr\hspace{1cm} /[(y+1)^2(y-1)^2]\,H(0,-1,0;y)\nn\\
&& \nr-4\,(3\,y^5-3\,y^4+6\,y^3+2\,y^2-y+1)\nn \\
&& \nr\hspace{1cm} /[(y-1)^3(y+1)^2]\,H(0,0,-1,0;y)\nn\\
&& \nr-1/4\,(272\,\zeta(2)\,y^6-229\,y^6+8\,\zeta(2)\,y^5-266\,y^5-15\,y^4\nn
\\
&& \nr\hspace{1cm}
+200\,\zeta(2)\,y^4+196\,y^3+352\,\zeta(2)\,y^3-168\,\zeta(2)\,y^2+189\,y^2\nn
\\
&& \nr\hspace{1cm} +8\,\zeta(2)\,y+70\,y-96\,\zeta(2)+55)\nn \\
&& \nr\hspace{2cm} /[(y-1)^3(y+1)^3]\,H(0,0;y)\nn\\
&& \nr-2\,(8\,y^4+19\,y^3+46\,y^2+19\,y+8)\nn \\
&& \nr\hspace{1cm} /[(y+1)^2(y-1)^2]\,H(-1,0,0;y)\nn\\
&& \nr+4\,(y^6+y^5+14\,y^4+20\,y^3+14\,y^2+y+1)\nn \\
&& \nr\hspace{1cm} /[(y-1)^3(y+1)^3]\,H(0,-1,0,0;y)\nn\\
&& \nr-2\,(5\,y^6+10\,y^5-60\,y^4-59\,y^3-16\,y^2+7\,y+5)\nn \\
&& \nr\hspace{1cm} /[(y-1)^3(y+1)^3]\,H(0,0,0;y)\nn\\
&& \nr-8\,(y^5+2\,y^4+18\,y^3+14\,y^2+4\,y-1)\nn \\
&& \nr\hspace{1cm} /[(y-1)^3(y+1)^2]\,H(-1,0,0,0;y)\nn\\
&& \nr+(27\,y^6+2\,y^5+27\,y^4+56\,y^3-7\,y^2+2\,y-7)\nn \\
&& \nr\hspace{1cm} /[(y-1)^3(y+1)^3]\,H(0,0,0,0;y)\nn\\
&& \nr+12\,(y^2+1)^2/[(y+1)^2(y-1)^2]\,H(1,0,0,0;y)\nn\\
&& \nr+2\,(3\,y^4+15\,y^3+76\,y^2+15\,y+5)\nn \\
&& \nr\hspace{1cm} /[(y+1)^2(y-1)^2]\,H(1,0,0;y)\nn\\
&& \nr+2\,(7\,y^6-10\,y^5-65\,y^4-104\,y^3-75\,y^2-10\,y-3)\nn \\
&& \nr\hspace{1cm} /[(y-1)^3(y+1)^3]\,H(0,1,0,0;y)\nn\\
&& \nr+16\,(y^2+1)^2/[(y+1)^2(y-1)^2]\,H(1,1,0,0;y)\nn\\
&& \nr-1/2\,(-55\,y^4+80\,\zeta(2)\,y^4-82\,y^3+160\,\zeta(2)\,y^2+82\,y\nn
\\
&& \nr\hspace{1cm} +55+80\,\zeta(2))/[(y+1)^2(y-1)^2]\,H(1,0;y)\nn\\
&& \nr-2\,(8\,y^4+27\,y^3+60\,y^2+27\,y+6)\nn \\
&& \nr\hspace{1cm} /[(y+1)^2(y-1)^2]\,H(0,1,0;y)\nn\\
&& \nr+2\,(11\,y^6+2\,y^5+33\,y^4+56\,y^3+23\,y^2+2\,y+1)\nn \\
&& \nr\hspace{1cm} /[(y-1)^3(y+1)^3]\,H(0,0,1,0;y)\nn\\
&& \nr-4\,\zeta(2)\,(7\,y^4+36\,y^3+132\,y^2+36\,y+11)\nn \\
&& \nr\hspace{1cm} /[(y+1)^2(y-1)^2]\,H(1;y)\nn\\
&& \nr-8\,\zeta(2)\,(4\,y^6-6\,y^5-41\,y^4-72\,y^3-49\,y^2-6\,y-4)\nn \\
&& \nr\hspace{1cm} /[(y-1)^3(y+1)^3]\,H(0,1;y)\nn\\
&& \nr-48\,\zeta(2)\,(y^2+1)^2/[(y+1)^2(y-1)^2]\,H(1,1;y)\nn\\
&& \nr-1/10\,(70\,\zeta(3)\,y^6+120\,\zeta(2)\,\ln(2)\,y^6-115\,y^6-196\,\zeta(2)^2\,y^6\nn
\\
&& \nr\hspace{1cm}
+1580\,\zeta(2)\,y^6+18\,\zeta(2)^2\,y^5+720\,\zeta(2)\,\ln(2)\,y^5\nn
\\
&& \nr\hspace{1cm}
+1880\,\zeta(2)\,y^5-300\,\zeta(3)\,y^5+240\,\zeta(2)\,y^4-1490\,\zeta(3)\,y^4\nn
\\
&& \nr\hspace{1cm}
+1080\,\zeta(2)\,\ln(2)\,y^4-74\,\zeta(2)^2\,y^4+345\,y^4-1720\,\zeta(2)\,y^3\nn
\\
&& \nr\hspace{1cm}
+232\,\zeta(2)^2\,y^3-1080\,\zeta(2)\,\ln(2)\,y^2-1500\,\zeta(2)\,y^2\nn
\\
&& \nr\hspace{1cm}
+1530\,\zeta(3)\,y^2-345\,y^2+258\,\zeta(2)^2\,y^2-720\,\zeta(2)\,\ln(2)\,y\nn
\\
&& \nr\hspace{1cm}
+18\,\zeta(2)^2\,y+300\,\zeta(3)\,y-160\,\zeta(2)\,y+136\,\zeta(2)^2\nn
\\
&& \nr\hspace{1cm}
-110\,\zeta(3)+115-120\,\zeta(2)\,\ln(2)-320\,\zeta(2))\nn \\
&& \nr\hspace{2cm} /[(y-1)^3(y+1)^3]\nn \\
&& \nr+8\,(y^2+1)^2/[(y-1)^2(y+1)^2]\,H(-1,0,-1,0;y)\nn\\
&& \nr-2\,(-24\,\zeta(2)\,y^4+2\,\zeta(3)\,y^4-57\,\zeta(2)\,y^3+4\,\zeta(3)\,y^2\nn
\\
&& \nr\hspace{1cm}
-138\,\zeta(2)\,y^2-57\,\zeta(2)\,y-24\,\zeta(2)+2\,\zeta(3))\nn \\
&& \nr\hspace{2cm} /[(y+1)^2(y-1)^2]\,H(-1;y)\nn\\
&& \nr-8\,(y^2+1)^2/[(y-1)^2(y+1)^2]\,H(-1,0,1,0;y)\nn\\
&& \nr+8\,(y^2+1)^2/[(y-1)^2(y+1)^2]\,H(1,0,1,0;y)\nn\\
&& \nr+4\,(y^2+1)^2/[(y-1)^2(y+1)^2]\,H(0,1,1,0;y)\nn\\
&& \nr+1/8\,(-85\,y^6+104\,\zeta(2)\,y^6+64\,\zeta(3)\,y^6-6\,y^5-160\,\zeta(3)\,y^5\nn
\\
&& \nr\hspace{1cm}
+184\,\zeta(2)\,y^5-1056\,\zeta(3)\,y^4-2312\,\zeta(2)\,y^4+85\,y^4+12\,y^3\nn
\\
&& \nr\hspace{1cm}
-1888\,\zeta(2)\,y^3-1536\,\zeta(3)\,y^3+85\,y^2-1056\,\zeta(3)\,y^2\nn
\\
&& \nr\hspace{1cm}
-152\,\zeta(2)\,y^2-6\,y+360\,\zeta(2)\,y-160\,\zeta(3)\,y+64\,\zeta(3)\nn
\\
&& \nr\hspace{1cm} +248\,\zeta(2)-85)/[(y-1)^3(y+1)^3]\,H(0;y)\nn\\
&& \nr-12\,\zeta(2)\,(y^6+y^5+14\,y^4+20\,y^3+14\,y^2+y+1)\nn \\
&& \nr\hspace{1cm} /[(y-1)^3(y+1)^3]\,H(0,-1;y)\nn\\
&& \nr-4\,(y^2+1)/[(y-1)(y+1)]\,H(1,1,0;y)\nn\\
&& \nr+4\,(-4\,y^5+5\,\zeta(2)\,y^5+2\,y^4+7\,\zeta(2)\,y^4+2\,y^3+74\,\zeta(2)\,y^3+2\,y^2\nn
\\
&& \nr\hspace{1cm}
+54\,\zeta(2)\,y^2+17\,\zeta(2)\,y+2\,y-4-5\,\zeta(2))\nn \\
&& \nr\hspace{1cm} /[(y-1)^3(y+1)^2]\,H(-1,0;y),\nn\\
\eea
and
\bea
\Im \, \tilde{c}_{1}  = && \nr -\frac{4}{11}\, \Im \, \tilde{c}_3
=-\frac{3}{5}\, \Im \, \tilde{c}_5 = 1/3\,(y^2+1)/[(y-1)(y+1)],\\
\Im \, \tilde{c}_{2}  = && \nr \Im \, \tilde{c}_6 = 0,\\
\Im \, \tilde{c}_{4}  = && \nr (y^2+1)^2/[(y-1)^2(y+1)^2]\,H(0;y)\nn\\
&& \nr-(y^2+1)/[(y-1)(y+1)],\\
\Im \, \tilde{c}_{7}  = && \nr -(y^2+1)/[(y-1)(y+1)]\,H(-1;y)\nn\\
&& \nr+(y^2+1)^2/[(y-1)^2(y+1)^2]\,H(0,-1;y)\nn\\
&& \nr+2/[(y-1)(y+1)]\,y^2\,H(0;y)\nn\\
&& \nr-2\,(y^2+1)/[(y-1)^2(y+1)^2]\,y^2\,H(0,0;y)\nn\\
&& \nr+(y^2+1)/[(y-1)(y+1)]\,H(1;y)\nn\\
&& \nr-(y^2+1)^2/[(y-1)^2(y+1)^2]\,H(0,1;y)\nn\\
&& \nr+1/36\,(y^2+1)\,(67\,y^2+18\,\zeta(2)\,y^2+18\,\zeta(2)-67)\nn \\
&& \nr\hspace{1cm} /[(y-1)^2(y+1)^2],\\
\Im \, \tilde{c}_{8}  = && \nr 2\,(y^2+y+2)\,(y^2+1)/[(y+1)^2(y-1)^2]\,H(0;y)\nn\\
&& \nr+3\,(y^2+1)^2/[(y-1)^2(y+1)^2]\,H(0,0;y)\nn\\
&& \nr+4\,(y^2+1)^2/[(y+1)^2(y-1)^2]\,H(1,0;y)\nn\\
&& \nr-2\,(y^2+1)/[(y-1)(y+1)]\,H(1;y)\nn\\
&& \nr+2\,(y^2+1)^2/[(y-1)^2(y+1)^2]\,H(0,1;y)\nn\\
&& \nr-1/2\,(7\,y^4+8\,\zeta(2)\,y^4+2\,y^3+16\,\zeta(2)\,y^2-2\,y-7+8\,\zeta(2))\nn
\\
&& \nr\hspace{1cm} /[(y+1)^2(y-1)^2],\\
\Im \, \tilde{c}_{9}  = && \nr -1/9\,(19\,y^2+6\,y+19)/[(y-1)(y+1)]\,H(0;y)\nn\\
&& \nr-2/3\,(y^2+1)/[(y-1)(y+1)]\,H(0,0;y)\nn\\
&& \nr-4/3\,(y^2+1)/[(y+1)(y-1)]\,H(1,0;y)\nn\\
&& \nr-2/9\,(19\,y^2+6\,y+19)/[(y-1)(y+1)]\,H(1;y)\nn\\
&& \nr-4/3\,(y^2+1)/[(y-1)(y+1)]\,H(0,1;y)\nn\\
&& \nr-8/3\,(y^2+1)/[(y-1)(y+1)]\,H(1,1;y)\nn\\
&& \nr-1/54\,(209\,y^2+78\,y+209)/[(y-1)(y+1)],\\
\Im \, \tilde{c}_{10}  = && \nr 1/9\,(19\,y^4+14\,y^3+14\,y^2+14\,y+19)/(y-1)^4\,H(0;y)\nn\\
&& \nr-2/3\,(y^4-2\,y^3-4\,y^2-2\,y+1)/[(y-1)^3(y+1)]\,H(0,0;y)\nn\\
&& \nr-1/54\,(265\,y^4+208\,y^3+14\,y^2+208\,y+265)\nn \\
&& \nr\hspace{1cm} /[(y-1)^3(y+1)],\\
\Im \, \tilde{c}_{11}  = && \nr 52/3\,(y^2+1)/[(y-1)(y+1)]\,H(1,1;y)\nn\\
&& \nr+2\,(y^2+1)/[(y-1)(y+1)]\,H(-1,-1;y)\nn\\
&& \nr-2\,(y^2+1)^2/[(y-1)^2(y+1)^2]\,H(0,-1,-1;y)\nn\\
&& \nr-(2\,\zeta(2)\,y^4-3\,y^4-2\,y^3+4\,\zeta(2)\,y^2+2\,y+2\,\zeta(2)+3)\nn
\\
&& \nr\hspace{1cm} /[(y-1)^2(y+1)^2]\,H(-1;y)\nn\\
&& \nr-8\,(y^2+1)/[(y-1)^2(y+1)]\,y\,H(0,-1;y)\nn\\
&& \nr-4\,(y^2+1)^2/[(y-1)^2(y+1)^2]\,H(-1,0,-1;y)\nn\\
&& \nr+2\,(7\,y^5-7\,y^4+12\,y^3-4\,y^2+y-1)\nn \\
&& \nr\hspace{1cm} /[(y-1)^3(y+1)^2]\,H(0,0,-1;y)\nn\\
&& \nr+4\,(y^2+1)^2/[(y-1)^2(y+1)^2]\,H(1,0,-1;y)\nn\\
&& \nr-6\,(y^2+1)/[(y-1)(y+1)]\,H(1,-1;y)\nn\\
&& \nr+6\,(y^2+1)^2/[(y-1)^2(y+1)^2]\,H(0,1,-1;y)\nn\\
&& \nr+2\,(y^6+5\,y^5+21\,y^4+34\,y^3+17\,y^2+5\,y-3)\nn \\
&& \nr\hspace{1cm} /[(y-1)^3(y+1)^3]\,H(0,-1,0;y)\nn\\
&& \nr+1/6\,(89\,y^6+402\,y^5+499\,y^4+354\,y^3+37\,y^2+24\,y+11)\nn \\
&& \nr\hspace{1cm} /[(y-1)^3(y+1)^3]\,H(0,0;y)\nn\\
&& \nr-2\,(y^5+5\,y^4+18\,y^3+26\,y^2+y+5)\nn \\
&& \nr\hspace{1cm} /[(y-1)^3(y+1)^2]\,H(-1,0,0;y)\nn\\
&& \nr+1/36\,(468\,\zeta(2)\,y^5-217\,y^5-288\,\zeta(2)\,y^4+499\,y^4-48\,y^3\nn
\\
&& \nr\hspace{1cm}
+1260\,\zeta(2)\,y^3-300\,y^2+252\,\zeta(2)\,y^2+216\,\zeta(2)\,y-167\,y\nn
\\
&& \nr\hspace{1cm} -36\,\zeta(2)+233)/[(y-1)^3(y+1)^2]\,H(0;y)\nn\\
&& \nr+2\,(2\,y^4-5\,y^3-14\,y^2-5\,y+6)/[(y-1)^2(y+1)^2]\,H(-1,0;y)\nn\\
&& \nr-(12\,y^5+y^4+10\,y^3+28\,y^2-2\,y+1)\nn \\
&& \nr\hspace{1cm} /[(y-1)^3(y+1)^3]\,y\,H(0,0,0;y)\nn\\
&& \nr+1/3\,(53\,y^4+141\,y^3+228\,y^2+141\,y-17)\nn \\
&& \nr\hspace{1cm} /[(y-1)^2(y+1)^2]\,H(1,0;y)\nn\\
&& \nr-2\,(7\,y^6+7\,y^5+33\,y^4+56\,y^3+25\,y^2+7\,y-1)\nn \\
&& \nr\hspace{1cm} /[(y-1)^3(y+1)^3]\,H(0,1,0;y)\nn\\
&& \nr+1/9\,(18\,\zeta(2)\,y^4+31\,y^4+105\,y^3+36\,\zeta(2)\,y^2-105\,y\nn
\\
&& \nr\hspace{1cm} +18\,\zeta(2)-31)/[(y-1)^2(y+1)^2]\,H(1;y)\nn\\
&& \nr-6\,(y^2+1)/[(y-1)(y+1)]\,H(-1,1;y)\nn\\
&& \nr+6\,(y^2+1)^2/[(y-1)^2(y+1)^2]\,H(0,-1,1;y)\nn\\
&& \nr+2/3\,(34\,y^4-3\,y^3-33\,y^2-3\,y-1)/[(y-1)^2(y+1)^2]\,H(0,1;y)\nn\\
&& \nr+4\,(y^2+1)^2/[(y-1)^2(y+1)^2]\,H(-1,0,1;y)\nn\\
&& \nr-2\,(y^2+3)\,(y^2+1)/[(y-1)^2(y+1)^2]\,H(1,0,0;y)\nn\\
&& \nr-2\,(11\,y^6-5\,y^5-15\,y^4-26\,y^3-27\,y^2-5\,y-1)\nn \\
&& \nr\hspace{1cm} /[(y-1)^3(y+1)^3]\,H(0,0,1;y)\nn\\
&& \nr-4\,(y^2+1)^2/[(y-1)^2(y+1)^2]\,H(1,0,1;y)\nn\\
&& \nr-10\,(y^2+1)^2/[(y-1)^2(y+1)^2]\,H(0,1,1;y)\nn\\
&& \nr-1/216\,(-2545\,y^6+1620\,\zeta(3)\,y^6+2160\,\zeta(2)\,y^6+3024\,\zeta(3)\,y^5\nn
\\
&& \nr\hspace{1cm}
-570\,y^5+1944\,\zeta(2)\,y^5-432\,\zeta(2)\,y^4+13716\,\zeta(3)\,y^4\nn
\\
&& \nr\hspace{1cm}
+2545\,y^4+1140\,y^3+25920\,\zeta(3)\,y^3+13500\,\zeta(3)\,y^2\nn \\
&& \nr\hspace{1cm}
+2545\,y^2-1728\,\zeta(2)\,y^2-1944\,\zeta(2)\,y+3024\,\zeta(3)\,y-570\,y\nn
\\
&& \nr\hspace{1cm} +1404\,\zeta(3)-2545)/[(y-1)^3(y+1)^3],\\
\Im \, \tilde{c}_{12}  = && \nr
4\,(-4\,y^4+\zeta(2)\,y^4-2\,y^3+2\,\zeta(2)\,y^2+2\,y+\zeta(2)+4)\nn
\\
&& \nr\hspace{1cm} /[(y+1)^2(y-1)^2]\,H(-1;y)\nn\\
&& \nr+2\,(5\,y^4+10\,y^3+18\,y^2+10\,y+5)/[(y+1)^2(y-1)^2]\,H(0,-1;y)\nn\\
&& \nr+8\,(y^2+1)^2/[(y-1)^2(y+1)^2]\,H(-1,0,-1;y)\nn\\
&& \nr-4\,(3\,y^5-3\,y^4+6\,y^3+2\,y^2-y+1)\nn \\
&& \nr\hspace{1cm} /[(y-1)^3(y+1)^2]\,H(0,0,-1;y)\nn\\
&& \nr-1/4\,(-229\,y^5+56\,\zeta(2)\,y^5-64\,\zeta(2)\,y^4-37\,y^4+48\,\zeta(2)\,y^3\nn
\\
&& \nr\hspace{1cm}
+22\,y^3+174\,y^2-144\,\zeta(2)\,y^2+15\,y+32\,\zeta(2)\,y\nn \\
&& \nr\hspace{1cm} -40\,\zeta(2)+55)/[(y-1)^3(y+1)^2]\,H(0;y)\nn\\
&& \nr-2\,(8\,y^4+19\,y^3+46\,y^2+19\,y+8)/[(y+1)^2(y-1)^2]\,H(-1,0;y)\nn\\
&& \nr+4\,(y^6+y^5+14\,y^4+20\,y^3+14\,y^2+y+1)\nn \\
&& \nr\hspace{1cm} /[(y-1)^3(y+1)^3]\,H(0,-1,0;y)\nn\\
&& \nr-2\,(5\,y^6+10\,y^5-60\,y^4-59\,y^3-16\,y^2+7\,y+5)\nn \\
&& \nr\hspace{1cm} /[(y-1)^3(y+1)^3]\,H(0,0;y)\nn\\
&& \nr-8\,(y^5+2\,y^4+18\,y^3+14\,y^2+4\,y-1)\nn \\
&& \nr\hspace{1cm} /[(y-1)^3(y+1)^2]\,H(-1,0,0;y)\nn\\
&& \nr+(27\,y^6+2\,y^5+27\,y^4+56\,y^3-7\,y^2+2\,y-7)\nn \\
&& \nr\hspace{1cm} /[(y-1)^3(y+1)^3]\,H(0,0,0;y)\nn\\
&& \nr+12\,(y^2+1)^2/[(y+1)^2(y-1)^2]\,H(1,0,0;y)\nn\\
&& \nr+2\,(3\,y^4+15\,y^3+76\,y^2+15\,y+5)/[(y+1)^2(y-1)^2]\,H(1,0;y)\nn\\
&& \nr+2\,(7\,y^6-10\,y^5-65\,y^4-104\,y^3-75\,y^2-10\,y-3)\nn \\
&& \nr\hspace{1cm} /[(y-1)^3(y+1)^3]\,H(0,1,0;y)\nn\\
&& \nr+16\,(y^2+1)^2/[(y+1)^2(y-1)^2]\,H(1,1,0;y)\nn\\
&& \nr-1/2\,(32\,\zeta(2)\,y^4-55\,y^4-82\,y^3+64\,\zeta(2)\,y^2+82\,y\nn
\\
&& \nr\hspace{1cm} +32\,\zeta(2)+55)/[(y+1)^2(y-1)^2]\,H(1;y)\nn\\
&& \nr-2\,(8\,y^4+27\,y^3+60\,y^2+27\,y+6)/[(y+1)^2(y-1)^2]\,H(0,1;y)\nn\\
&& \nr-8\,(y^2+1)^2/[(y-1)^2(y+1)^2]\,H(-1,0,1;y)\nn\\
&& \nr+2\,(11\,y^6+2\,y^5+33\,y^4+56\,y^3+23\,y^2+2\,y+1)\nn \\
&& \nr\hspace{1cm} /[(y-1)^3(y+1)^3]\,H(0,0,1;y)\nn\\
&& \nr+8\,(y^2+1)^2/[(y-1)^2(y+1)^2]\,H(1,0,1;y)\nn\\
&& \nr-4\,(y^2+1)/[(y-1)(y+1)]\,H(1,1;y)\nn\\
&& \nr+4\,(y^2+1)^2/[(y-1)^2(y+1)^2]\,H(0,1,1;y)\nn\\
&& \nr+1/8\,(-56\,\zeta(2)\,y^6+64\,\zeta(3)\,y^6-85\,y^6-6\,y^5-160\,\zeta(3)\,y^5\nn
\\
&& \nr\hspace{1cm}
-136\,\zeta(2)\,y^5+85\,y^4-392\,\zeta(2)\,y^4-1056\,\zeta(3)\,y^4\nn
\\
&& \nr\hspace{1cm}
-1536\,\zeta(3)\,y^3+12\,y^3-1056\,\zeta(3)\,y^2+360\,\zeta(2)\,y^2+85\,y^2\nn
\\
&& \nr\hspace{1cm}
+136\,\zeta(2)\,y-6\,y-160\,\zeta(3)\,y+64\,\zeta(3)-85+88\,\zeta(2))\nn
\\
&& \nr\hspace{1cm} /[(y-1)^3(y+1)^3], 
\eea
respectively
\bea
\Re \, \tilde{d}_j  = && \nr 0 \quad \mbox{for}\quad j=1\ldots 7 , \\
\Re \, \tilde{d}_{8}  =&& \nr 2\,(5\,y^2+2\,y+5)/[(y-1)(y+1)^3]\,y\,H(0;y)\nn\\
&& \nr-4\,(3\,y^2+2\,y+3)\,(y^2+1)/[(y-1)^2(y+1)^4]\,y\,H(0,0;y)\nn\\
&& \nr+4\,(-y^4+9\,\zeta(2)\,y^4+6\,\zeta(2)\,y^3+18\,\zeta(2)\,y^2+2\,y^2+6\,\zeta(2)\,y\nn
\\
&& \nr\hspace{1cm} -1+9\,\zeta(2))/[(y-1)^2(y+1)^4]\,y,\\
\Re \, \tilde{d}_{9}  = && \nr 2/9\,(51\,y^2+26\,y+51)/[(y-1)(y+1)^3]\,y\,H(0;y)\nn\\
&& \nr+4/3\,(3\,y^2+2\,y+3)/[(y-1)(y+1)^3]\,y\,H(0,0;y)\nn\\
&& \nr+8/3\,(3\,y^2+2\,y+3)/[(y-1)(y+1)^3]\,y\,H(1,0;y)\nn\\
&& \nr-4/9\,(36\,\zeta(2)\,y^2+19\,y^2+24\,\zeta(2)\,y-19+36\,\zeta(2))\nn
\\
&& \nr\hspace{1cm} /[(y-1)(y+1)^3]\,y,\\
\Re \, \tilde{d}_{10}  = && \nr
2/9\,(51\,y^4+176\,y^3+144\,\zeta(2)\,y^2-70\,y^2+176\,y+51)\nn \\
&& \nr\hspace{1cm} /[(y-1)^3(y+1)^3]\,y\,H(0;y)\nn\\
&& \nr-4/3\,(3\,y^4+14\,y^3-2\,y^2+14\,y+3)\nn \\
&& \nr\hspace{1cm} /[(y-1)^4(y+1)^2]\,y\,H(0,0;y)\nn\\
&& \nr-16/[(y-1)^3(y+1)^3]\,y^3\,H(0,0,0;y)\nn\\
&& \nr-8/9\,(-4\,y^2+9\,\zeta(2)\,y^2+32\,y-4+9\,\zeta(2))\nn \\
&& \nr\hspace{1cm} /[(y-1)^2(y+1)^2]\,y,\\
\Re \, \tilde{d}_{11}  = && \nr -72\,\zeta(2)\,(3\,y^2+2\,y+3)/[(y-1)^2(y+1)^4]\,y^2\,H(-1;y)\nn\\
&& \nr+24\,\zeta(2)\,(5\,y^4+14\,y^3+42\,y^2+14\,y+5)\nn \\
&& \nr\hspace{1cm} /[(y-1)^3(y+1)^5]\,y^2\,H(0,-1;y)\nn\\
&& \nr+1/18\,(648\,\zeta(2)\,y^6-687\,y^6+26\,y^5+1008\,\zeta(3)\,y^5+8712\,\zeta(2)\,y^5\nn
\\
&& \nr\hspace{1cm}
+11808\,\zeta(2)\,y^4+4752\,\zeta(3)\,y^4+687\,y^4+9216\,\zeta(3)\,y^3\nn
\\
&& \nr\hspace{1cm}
-52\,y^3+14112\,\zeta(2)\,y^3+687\,y^2+4752\,\zeta(3)\,y^2\nn \\
&& \nr\hspace{1cm}
-648\,\zeta(2)\,y^2+1008\,\zeta(3)\,y+26\,y-648\,\zeta(2)\,y-687)\nn
\\
&& \nr\hspace{2cm} /[(y-1)^3(y+1)^5]\,y\,H(0;y)\nn\\
&& \nr-4\,(y^4+24\,\zeta(2)\,y^3+64\,\zeta(2)\,y^2-2\,y^2+24\,\zeta(2)\,y+1)\nn
\\
&& \nr\hspace{1cm} /[(y-1)^3(y+1)^3]\,y\,H(-1,0;y)\nn\\
&& \nr+16/[(y-1)^2(y+1)^2]\,y^2\,H(0,-1,0;y)\nn\\
&& \nr-32/[(y-1)^3(y+1)^3]\,y^3\,H(0,0,-1,0;y)\nn\\
&& \nr-1/3\,(-39\,y^6+184\,y^5+84\,\zeta(2)\,y^5+87\,y^4+432\,\zeta(2)\,y^4\nn
\\
&& \nr\hspace{1cm}
+1416\,\zeta(2)\,y^3-152\,y^3+432\,\zeta(2)\,y^2-81\,y^2-32\,y\nn \\
&& \nr\hspace{1cm} +84\,\zeta(2)\,y+33)/[(y-1)^3(y+1)^5]\,y\,H(0,0;y)\nn\\
&& \nr+24\,(3\,y^2+2\,y+3)/[(y-1)^2(y+1)^4]\,y^2\,H(-1,0,0;y)\nn\\
&& \nr-8\,(5\,y^4+14\,y^3+42\,y^2+14\,y+5)\nn \\
&& \nr\hspace{1cm} /[(y-1)^3(y+1)^5]\,y^2\,H(0,-1,0,0;y)\nn\\
&& \nr-2\,(9\,y^4+112\,y^3+146\,y^2+196\,y+9)\nn \\
&& \nr\hspace{1cm} /[(y-1)^3(y+1)^5]\,y^3\,H(0,0,0;y)\nn\\
&& \nr+16\,(3\,y^2+8\,y+3)/[(y-1)^3(y+1)^3]\,y^2\,H(-1,0,0,0;y)\nn\\
&& \nr+4\,(y^4+5\,y^3+38\,y^2+5\,y+1)\nn \\
&& \nr\hspace{1cm} /[(y-1)^3(y+1)^5]\,y^2\,H(0,0,0,0;y)\nn\\
&& \nr-2\,(9\,y^4+92\,y^3+130\,y^2+92\,y+9)\nn \\
&& \nr\hspace{1cm} /[(y-1)^2(y+1)^4]\,y\,H(1,0,0;y)\nn\\
&& \nr+8\,(7\,y^4+29\,y^3+62\,y^2+29\,y+7)\nn \\
&& \nr\hspace{1cm} /[(y-1)^3(y+1)^5]\,y^2\,H(0,1,0,0;y)\nn\\
&& \nr-2/3\,(3\,y^2+88\,y+3)/[(y-1)(y+1)^3]\,y\,H(1,0;y)\nn\\
&& \nr+48\,(y^2+y+1)/[(y-1)^2(y+1)^4]\,y^2\,H(0,1,0;y)\nn\\
&& \nr-8\,(5\,y^4+14\,y^3+30\,y^2+14\,y+5)\nn \\
&& \nr\hspace{1cm} /[(y-1)^3(y+1)^5]\,y^2\,H(0,0,1,0;y)\nn\\
&& \nr+12\,\zeta(2)\,(3\,y^4+40\,y^3+62\,y^2+40\,y+3)\nn \\
&& \nr\hspace{1cm} /[(y-1)^2(y+1)^4]\,y\,H(1;y)\nn\\
&& \nr-48\,\zeta(2)\,(4\,y^4+17\,y^3+32\,y^2+17\,y+4)\nn \\
&& \nr\hspace{1cm} /[(y-1)^3(y+1)^5]\,y^2\,H(0,1;y)\nn\\
&& \nr+2/45\,(605\,y^6-540\,\zeta(2)\,\ln(2)\,y^6-270\,\zeta(3)\,y^6-180\,\zeta(2)\,y^6\nn
\\
&& \nr\hspace{1cm}
+621\,\zeta(2)^2\,y^5-3690\,\zeta(3)\,y^5-2160\,\zeta(2)\,\ln(2)\,y^5\nn
\\
&& \nr\hspace{1cm}
+9390\,\zeta(2)\,y^5-1815\,y^4+5670\,\zeta(2)\,y^4+3906\,\zeta(2)^2\,y^4\nn
\\
&& \nr\hspace{1cm}
-2700\,\zeta(2)\,\ln(2)\,y^4-5490\,\zeta(3)\,y^4-5100\,\zeta(2)\,y^3\nn
\\
&& \nr\hspace{1cm}
+4626\,\zeta(2)^2\,y^3-5760\,\zeta(2)\,y^2+2700\,\zeta(2)\,\ln(2)\,y^2\nn
\\
&& \nr\hspace{1cm}
+5490\,\zeta(3)\,y^2+3906\,\zeta(2)^2\,y^2+1815\,y^2+621\,\zeta(2)^2\,y\nn
\\
&& \nr\hspace{1cm}
+2160\,\zeta(2)\,\ln(2)\,y-4290\,\zeta(2)\,y+3690\,\zeta(3)\,y\nn \\
&& \nr\hspace{1cm}
+540\,\zeta(2)\,\ln(2)+270\,\zeta(2)+270\,\zeta(3)-605)\nn \\
&& \nr\hspace{2cm} /[(y-1)^3(y+1)^5]\,y,\\
\Re \, \tilde{d}_{12}  = && \nr
-24\,\zeta(2)\,(3\,y^4+7\,y^3+64\,y^2+7\,y+3)\nn \\
&& \nr\hspace{1cm} /[(y-1)^2(y+1)^4]\,y\,H(-1;y)\nn\\
&& \nr-48\,\zeta(2)\,(y^4-19\,y^3-16\,y^2-19\,y+1)\nn \\
&& \nr\hspace{1cm} /[(y-1)^3(y+1)^5]\,y^2\,H(0,-1;y)\nn\\
&& \nr+1/2\,(61\,y^6+96\,\zeta(2)\,y^6-120\,\zeta(2)\,y^5+22\,y^5+32\,\zeta(3)\,y^5\nn
\\
&& \nr\hspace{1cm}
-61\,y^4+1056\,\zeta(3)\,y^4+3168\,\zeta(2)\,y^4-44\,y^3+1664\,\zeta(3)\,y^3\nn
\\
&& \nr\hspace{1cm}
+896\,\zeta(2)\,y^3-144\,\zeta(2)\,y^2+1056\,\zeta(3)\,y^2-61\,y^2\nn
\\
&& \nr\hspace{1cm}
+32\,\zeta(3)\,y+22\,y-200\,\zeta(2)\,y+61-240\,\zeta(2))\nn \\
&& \nr\hspace{1cm} /[(y-1)^3(y+1)^5]\,y\,H(0;y)\nn\\
&& \nr-32\,(-y^4+2\,\zeta(2)\,y^3+y^3+34\,\zeta(2)\,y^2+2\,\zeta(2)\,y+y-1)\nn
\\
&& \nr\hspace{1cm} /[(y-1)^3(y+1)^3]\,y\,H(-1,0;y)\nn\\
&& \nr-8\,(3\,y^4+6\,y^3+14\,y^2+6\,y+3)/[(y-1)^2(y+1)^4]\,y\nn \\
&& \nr\hspace{1cm} \,H(0,-1,0;y)\nn\\
&& \nr+64/[(y-1)^3(y+1)^3]\,y^3\,H(0,0,-1,0;y)\nn\\
&& \nr+(-125\,y^6+24\,\zeta(2)\,y^5-230\,y^5+24\,\zeta(2)\,y^4-187\,y^4\nn
\\
&& \nr\hspace{1cm}
+480\,\zeta(2)\,y^3+204\,y^3+24\,\zeta(2)\,y^2+269\,y^2+26\,y\nn \\
&& \nr\hspace{1cm} +24\,\zeta(2)\,y+43)/[(y-1)^3(y+1)^5]\,y\,H(0,0;y)\nn\\
&& \nr+8\,(3\,y^4+7\,y^3+64\,y^2+7\,y+3)\nn \\
&& \nr\hspace{1cm} /[(y-1)^2(y+1)^4]\,y\,H(-1,0,0;y)\nn\\
&& \nr+16\,(y^4-19\,y^3-16\,y^2-19\,y+1)\nn \\
&& \nr\hspace{1cm} /[(y-1)^3(y+1)^5]\,y^2\,H(0,-1,0,0;y)\nn\\
&& \nr+4\,(14\,y^5-175\,y^4-56\,y^3-14\,y^2+6\,y+9)\nn \\
&& \nr\hspace{1cm} /[(y-1)^3(y+1)^5]\,y\,H(0,0,0;y)\nn\\
&& \nr+32\,(y^2+17\,y+1)/[(y-1)^3(y+1)^3]\,y^2\,H(-1,0,0,0;y)\nn\\
&& \nr-8\,(y^4+5\,y^3+38\,y^2+5\,y+1)\nn \\
&& \nr\hspace{1cm} /[(y-1)^3(y+1)^5]\,y^2\,H(0,0,0,0;y)\nn\\
&& \nr-4\,(15\,y^4+16\,y^3+198\,y^2+16\,y+15)\nn \\
&& \nr\hspace{1cm} /[(y-1)^2(y+1)^4]\,y\,H(1,0,0;y)\nn\\
&& \nr+16\,(y^4+37\,y^3+54\,y^2+37\,y+1)\nn \\
&& \nr\hspace{1cm} /[(y-1)^3(y+1)^5]\,y^2\,H(0,1,0,0;y)\nn\\
&& \nr-2\,(25\,y^2+86\,y+25)/[(y+1)^3(y-1)]\,y\,H(1,0;y)\nn\\
&& \nr+8\,(3\,y^4+13\,y^3+64\,y^2+13\,y+3)\nn \\
&& \nr\hspace{1cm} /[(y-1)^2(y+1)^4]\,y\,H(0,1,0;y)\nn\\
&& \nr+16\,(y^4-19\,y^3-28\,y^2-19\,y+1)\nn \\
&& \nr\hspace{1cm} /[(y-1)^3(y+1)^5]\,y^2\,H(0,0,1,0;y)\nn\\
&& \nr+24\,\zeta(2)\,(9\,y^4+14\,y^3+102\,y^2+14\,y+9)\nn \\
&& \nr\hspace{1cm} /[(y-1)^2(y+1)^4]\,y\,H(1;y)\nn\\
&& \nr-192\,\zeta(2)\,(8\,y^2+13\,y+8)/[(y-1)^3(y+1)^5]\,y^3\,H(0,1;y)\nn\\
&& \nr-1/5\,(85\,y^6-240\,\zeta(2)\,\ln(2)\,y^6+60\,\zeta(3)\,y^6-1640\,\zeta(2)\,y^6\nn
\\
&& \nr\hspace{1cm}
+68\,\zeta(2)^2\,y^5+320\,\zeta(3)\,y^5-960\,\zeta(2)\,\ln(2)\,y^5\nn
\\
&& \nr\hspace{1cm}
-3480\,\zeta(2)\,y^5-255\,y^4-4400\,\zeta(2)\,y^4+20\,\zeta(2)^2\,y^4\nn
\\
&& \nr\hspace{1cm}
-1200\,\zeta(2)\,\ln(2)\,y^4+3340\,\zeta(3)\,y^4+3600\,\zeta(2)\,y^3\nn
\\
&& \nr\hspace{1cm}
-960\,\zeta(2)^2\,y^3+5560\,\zeta(2)\,y^2+1200\,\zeta(2)\,\ln(2)\,y^2\nn
\\
&& \nr\hspace{1cm}
-3340\,\zeta(3)\,y^2+20\,\zeta(2)^2\,y^2+255\,y^2+68\,\zeta(2)^2\,y\nn
\\
&& \nr\hspace{1cm}
+960\,\zeta(2)\,\ln(2)\,y-120\,\zeta(2)\,y-320\,\zeta(3)\,y+240\,\zeta(2)\,\ln(2)\nn
\\
&& \nr\hspace{1cm} +480\,\zeta(2)-60\,\zeta(3)-85)/[(y-1)^3(y+1)^5]\,y,
\eea
and
\bea
\Im \, \tilde{d}_j  = && \nr 0 \quad \mbox{for} \quad j = 1 \ldots 7,\\
\Im \, \tilde{d}_{8}  = && \nr -4\,(3\,y^2+2\,y+3)\,(y^2+1)/[(y-1)^2(y+1)^4]\,y\,H(0;y)\nn\\
&& \nr+2\,(5\,y^2+2\,y+5)/[(y-1)(y+1)^3]\,y,\\
\Im \, \tilde{d}_{9}  = && \nr 4/3\,(3\,y^2+2\,y+3)/[(y-1)(y+1)^3]\,y\,H(0;y)\nn\\
&& \nr+8/3\,(3\,y^2+2\,y+3)/[(y-1)(y+1)^3]\,y\,H(1;y)\nn\\
&& \nr+2/9\,(51\,y^2+26\,y+51)/[(y-1)(y+1)^3]\,y,\\
\Im \, \tilde{d}_{10}  = && \nr -4/3\,(3\,y^4+14\,y^3-2\,y^2+14\,y+3)/[(y-1)^4(y+1)^2]\,y\,H(0;y)\nn\\
&& \nr-16/[(y-1)^3(y+1)^3]\,y^3\,H(0,0;y)\nn\\
&& \nr+2/9\,(51\,y^4+176\,y^3-70\,y^2+176\,y+51)/[(y-1)^3(y+1)^3]\,y,\\
\Im \, \tilde{d}_{11}  = && \nr -4/[(y-1)(y+1)]\,y\,H(-1;y)\nn\\
&& \nr+16/[(y-1)^2(y+1)^2]\,y^2\,H(0,-1;y)\nn\\
&& \nr-32/[(y-1)^3(y+1)^3]\,y^3\,H(0,0,-1;y)\nn\\
&& \nr+1/3\,(39\,y^5-223\,y^4-60\,\zeta(2)\,y^4-252\,\zeta(2)\,y^3+136\,y^3\nn
\\
&& \nr\hspace{1cm}
-252\,\zeta(2)\,y^2+16\,y^2-60\,\zeta(2)\,y+65\,y-33)\nn \\
&& \nr\hspace{2cm} /[(y-1)^3(y+1)^4]\,y\,H(0;y)\nn\\
&& \nr+24\,(3\,y^2+2\,y+3)/[(y-1)^2(y+1)^4]\,y^2\,H(-1,0;y)\nn\\
&& \nr-8\,(5\,y^4+14\,y^3+42\,y^2+14\,y+5)\nn \\
&& \nr\hspace{1cm} /[(y-1)^3(y+1)^5]\,y^2\,H(0,-1,0;y)\nn\\
&& \nr-2\,(9\,y^4+112\,y^3+146\,y^2+196\,y+9)\nn \\
&& \nr\hspace{1cm} /[(y-1)^3(y+1)^5]\,y^3\,H(0,0;y)\nn\\
&& \nr+16\,(3\,y^2+8\,y+3)/[(y-1)^3(y+1)^3]\,y^2\,H(-1,0,0;y)\nn\\
&& \nr+4\,(y^4+5\,y^3+38\,y^2+5\,y+1)/[(y-1)^3(y+1)^5]\,y^2\,H(0,0,0;y)\nn\\
&& \nr-2\,(9\,y^4+92\,y^3+130\,y^2+92\,y+9)\nn \\
&& \nr\hspace{1cm} /[(y-1)^2(y+1)^4]\,y\,H(1,0;y)\nn\\
&& \nr+8\,(7\,y^4+29\,y^3+62\,y^2+29\,y+7)\nn \\
&& \nr\hspace{1cm} /[(y-1)^3(y+1)^5]\,y^2\,H(0,1,0;y)\nn\\
&& \nr-2/3\,(3\,y^2+88\,y+3)/[(y-1)(y+1)^3]\,y\,H(1;y)\nn\\
&& \nr+48\,(y^2+y+1)/[(y-1)^2(y+1)^4]\,y^2\,H(0,1;y)\nn\\
&& \nr-8\,(5\,y^4+14\,y^3+30\,y^2+14\,y+5)\nn \\
&& \nr\hspace{1cm} /[(y-1)^3(y+1)^5]\,y^2\,H(0,0,1;y)\nn\\
&& \nr-1/18\,(687\,y^6-26\,y^5-648\,\zeta(2)\,y^5-1008\,\zeta(3)\,y^5-4752\,\zeta(3)\,y^4\nn
\\
&& \nr\hspace{1cm}
-1296\,\zeta(2)\,y^4-687\,y^4-9216\,\zeta(3)\,y^3+52\,y^3\nn \\
&& \nr\hspace{1cm}
+1296\,\zeta(2)\,y^2-4752\,\zeta(3)\,y^2-687\,y^2+648\,\zeta(2)\,y\nn
\\
&& \nr\hspace{1cm} -1008\,\zeta(3)\,y-26\,y+687)/[(y-1)^3(y+1)^5]\,y,\\
\Im \, \tilde{d}_{12}  = && \nr 32\,(y^2+y+1)/[(y+1)^3(y-1)]\,y\,H(-1;y)\nn\\
&& \nr-8\,(3\,y^4+6\,y^3+14\,y^2+6\,y+3)/[(y-1)^2(y+1)^4]\,y\,H(0,-1;y)\nn\\
&& \nr+64/[(y-1)^3(y+1)^3]\,y^3\,H(0,0,-1;y)\nn\\
&& \nr-(125\,y^5-8\,\zeta(2)\,y^4+105\,y^4+64\,\zeta(2)\,y^3+82\,y^3-286\,y^2\nn
\\
&& \nr\hspace{1cm} +64\,\zeta(2)\,y^2+17\,y-8\,\zeta(2)\,y-43)\nn \\
&& \nr\hspace{2cm} /[(y-1)^3(y+1)^4]\,y\,H(0;y)\nn\\
&& \nr+8\,(3\,y^4+7\,y^3+64\,y^2+7\,y+3)/[(y-1)^2(y+1)^4]\,y\,H(-1,0;y)\nn\\
&& \nr+16\,(y^4-19\,y^3-16\,y^2-19\,y+1)\nn \\
&& \nr\hspace{1cm} /[(y-1)^3(y+1)^5]\,y^2\,H(0,-1,0;y)\nn\\
&& \nr+4\,(14\,y^5-175\,y^4-56\,y^3-14\,y^2+6\,y+9)\nn \\
&& \nr\hspace{1cm} /[(y-1)^3(y+1)^5]\,y\,H(0,0;y)\nn\\
&& \nr+32\,(y^2+17\,y+1)/[(y-1)^3(y+1)^3]\,y^2\,H(-1,0,0;y)\nn\\
&& \nr-8\,(y^4+5\,y^3+38\,y^2+5\,y+1)/[(y-1)^3(y+1)^5]\,y^2\,H(0,0,0;y)\nn\\
&& \nr-4\,(15\,y^4+16\,y^3+198\,y^2+16\,y+15)\nn \\
&& \nr\hspace{1cm} /[(y-1)^2(y+1)^4]\,y\,H(1,0;y)\nn\\
&& \nr+16\,(y^4+37\,y^3+54\,y^2+37\,y+1)\nn \\
&& \nr\hspace{1cm} /[(y-1)^3(y+1)^5]\,y^2\,H(0,1,0;y)\nn\\
&& \nr-2\,(25\,y^2+86\,y+25)/[(y+1)^3(y-1)]\,y\,H(1;y)\nn\\
&& \nr+8\,(3\,y^4+13\,y^3+64\,y^2+13\,y+3)/[(y-1)^2(y+1)^4]\,y\,H(0,1;y)\nn\\
&& \nr+16\,(y^4-19\,y^3-28\,y^2-19\,y+1)\nn \\
&& \nr\hspace{1cm} /[(y-1)^3(y+1)^5]\,y^2\,H(0,0,1;y)\nn\\
&& \nr+1/2\,(61\,y^6+96\,\zeta(2)\,y^6+104\,\zeta(2)\,y^5+32\,\zeta(3)\,y^5+22\,y^5\nn
\\
&& \nr\hspace{1cm}
+368\,\zeta(2)\,y^4-61\,y^4+1056\,\zeta(3)\,y^4+1664\,\zeta(3)\,y^3-44\,y^3\nn
\\
&& \nr\hspace{1cm}
-368\,\zeta(2)\,y^2+1056\,\zeta(3)\,y^2-61\,y^2-104\,\zeta(2)\,y+22\,y\nn
\\
&& \nr\hspace{1cm}
+32\,\zeta(3)\,y-96\,\zeta(2)+61)/[(y-1)^3(y+1)^5]\,y. 
\eea
%
\begin{boldmath}
\subsection{Threshold Expansions \label{subsec_threshold_expansion}}
\end{boldmath}
In this Section we  expand the 
one and 
two loop form factors near threshold $S\sim4m^2$, i.e. $y \rightarrow 1$ in powers of
\be
\beta = \sqrt{1-\frac{4m^2}{S}} \, .
\ee
Up to and including terms
of order $\beta^0$ 
the non-vanishing real and imaginary parts of the coefficents $\tilde{a}_i$, $\tilde{b}_i$, $\tilde{c}_j$ and $\tilde{d}_j$
($i=1\dots 3$, $j=1 \dots 12$) are:
\bea
\Re \, \tilde{a}_2    &=& \frac{3\,\zeta (2)}{\beta}\, \,-\,2
,\\
\Re \, \tilde{a}_3    &=&  -\frac{6\,\zeta (2)}{\beta}\, \left( -1+\ln(\beta) +\ln(2)  \right)
,\\
\Im \, \tilde{a}_1    &=& -{\frac {1}{2\,\beta}},\\
\Im \, \tilde{a}_2    &=&   \frac{1}{\beta}\,\left( -1+\ln(\beta) +\ln(2)  \right)
,\\
\Im \, \tilde{a}_3    &=&   \frac{1}{\beta}\,\Big( - \ln^2 (\beta)-2+2\,
\ln(\beta) -\ln^2 (2) \nn \\
&& -2\,\ln(2) \ln(\beta) +2\,\ln(2) +\zeta (2)  \Big)
,\\
\Re \, \tilde{b}_2    &=& -1,\\
\Re \, \tilde{b}_3    &=& -\frac{6\,\zeta (2)}{\beta}\, +2
,\\
\Im \, \tilde{b}_2    &=& \frac{1}{\beta,}\\
\Im \, \tilde{b}_3    &=& \frac{1}{\beta}\,\left( 2-2\,\ln(\beta) -2\,\ln(2)  \right)
,\\
\Re \, \tilde{c}_4    &=& -\frac{3\,\zeta (2)}{4\, \beta^2}\, -\frac{3}{2}\,\zeta (2) 
,\\
\Re \, \tilde{c}_8    &=& \frac{3\,\zeta (2)}{\beta^2}  \left(
  -1+\ln(\beta) +\ln(2)  \right) \nn \\
&& +3\,\zeta (2)  \left( 2\,\ln(\beta) -1+2\,\ln(2)  \right),\\
\Re \, \tilde{c}_9    &=& \frac{2 \, \zeta (2)}{3\,\beta} \left( -11+6\,\ln(\beta) +6\,\ln(2)  \right) +{\frac {14}{9}}
,\\
\Re \, \tilde{c}_{10} &=& {\frac {80}{9}}-\frac{16}{3}\,\zeta (2) 
,\\
\Re \, \tilde{c}_{11} &=& - \frac{\zeta (2)}{6\,\beta}  \left(
  66\,\ln(\beta) -97+66\,\ln(2)  \right)-\frac{9}{2}\,\zeta
\left( 3 \right) -{\frac {101}{18}}\nn\\
&&-18\,\zeta (2) \ln(2)+{\frac {89}{6}}\,\zeta (2) -4\,\zeta (2) \ln(\beta)  
,\\
\Re \, \tilde{c}_{12} &=& \frac{3\,\zeta (2)}{2\,\beta^2} \Big( -4\,
  \ln^2 (2)+3\,\zeta (2) +8\,\ln(2) -8\,\ln(2) \ln(\beta)\nn \\
&&  -4+8\,\ln(\beta) -4\, \ln^2 (\beta) \Big) - \frac{6\,\zeta
  (2)}{\beta}-12\,\zeta (2)  \ln^2 (2)\nn \\
&& -24\,\zeta (2)\ln(2) \ln(\beta) -12\,\zeta ( 2)  \ln^2 (\beta) +{\frac {23}{6}}\nn \\
&& +9\, \zeta^2 (2) -8\,\zeta (2) -{\frac {27}{4}}\,\zeta (3)
+2\,\zeta (2) \ln(\beta) + 21\,\zeta (2) \ln(2) 
,\\
\Im \, \tilde{c}_1    &=& -\frac{1}{6\,\beta},\\
\Im \, \tilde{c}_3    &=& \frac{11}{24\,\beta},\\
\Im \, \tilde{c}_5    &=& \frac{5}{18\,\beta},\\
\Im \, \tilde{c}_7    &=& -\frac{31}{72\,\beta},\\
\Im \, \tilde{c}_8    &=& -\frac{3\,\zeta (2)}{2\,\beta^2}+\frac{1}{\beta}-3\,\zeta (2) 
,\\
\Im \, \tilde{c}_9    &=& \frac{1}{\beta}\Big( {\frac
    {62}{27}}-{\frac {22}{9}}\,\ln(2) +\frac{2}{3}\, \ln^2
  (2)-{\frac {22}{9}}\,\ln(\beta) \nn \\
&& +\frac{2}{3}\, \ln^2 (\beta)+\frac{4}{3}\,\ln(2) \ln(\beta)  \Big) 
,\\
\Im \, \tilde{c}_{11} &=& \frac{1}{\beta}\,\Big( {\frac {97}{18}}\,\ln(\beta) -{\frac
    {239}{54}}-{\frac {11}{6}}\, \ln^2 (\beta)-{\frac {11}{6}}\, \ln^2
  2\nn \\
&& +{\frac {97}{18}}\,\ln(2) -\frac{11}{3}\,\ln(2) \ln(\beta)  \Big)+2\,\zeta (2) 
,\\
\Im \, \tilde{c}_{12} &=& \frac{6\,\zeta (2)}{\beta^2}
\left(-1+\ln(\beta) +\ln(2)  \right) + \frac{1}{\beta}\, \left( 2-2\,\ln(\beta)
  -2\,\ln(2)  \right) \nn \\
&& +\zeta (2)  \left( 12\,\ln(2) +12\,\ln(\beta) -1 \right) 
,\\
\Re \, \tilde{d}_8    &=&  \frac{3\,\zeta (2)}{\beta^2}+\frac{3}{2}\,\zeta (2) 
,\\
\Re \, \tilde{d}_9    &=&  \frac{4\,\zeta (2)}{\beta}-{\frac {11}{9}}
,\\
\Re \, \tilde{d}_{10} &=& -\frac{8}{3}\,\zeta (2) +{\frac {49}{9}}
,\\
\Re \, \tilde{d}_{11} &=& -\frac{11\,\zeta (2)}{\beta} + {\frac {26}{3}}\,\zeta (2) -18\,\zeta
\left( 2 \right) \ln(2) \nn \\
&&-8\,\zeta (2) \ln(\beta) -\frac{15}{2}\,\zeta (3) +{\frac {151}{36}}
,\\
\Re \, \tilde{d}_{12} &=& -\frac{6\,\zeta (2)}{\beta^2}  \left( 2\,\ln(\beta) -1+2\,\ln(2)  \right)- \frac{3\,\zeta (2)}{\beta}-15\,\zeta (2) \ln(2) \nn \\
&& -\frac{9}{2}\,\zeta (2) -{\frac {45}{4}}\,\zeta (3) -20\,\zeta (2) \ln(\beta) +{\frac {41}{12}}
,\\
\Im \, \tilde{d}_8    &=& \frac{1}{2\,\beta},\\
\Im \, \tilde{d}_9    &=&  \frac{1}{\beta}\,\left( \frac{4}{3}\,\ln(2) +\frac{4}{3}\,\ln(\beta) -{\frac {16}{9}} \right)
,\\
\Im \, \tilde{d}_{11} &=&   \frac{1}{\beta}\,\left( -\frac{11}{3}\,\ln(2) +{\frac {32}{9}}-\frac{11}{3}\,\ln(\beta)  \right)+4\,\zeta (2) 
,\\
\Im \, \tilde{d}_{12} &=&  \frac{6\,\zeta (2)}{\beta^2}+ \frac{1}{\beta}\,\left( -3-\ln(2) -\ln(\beta)  \right) +10\,\zeta (2) 
.
\eea
These expansions are relevant, for instance,  for studies of $Q\bar{Q}$ production near
threshold. For a study using the one-loop axial form factors see \cite{KuhnTeubner}.
\begin{boldmath}
\subsection{Asymptotic Expansions \label{subsec_asymptotic_expansion}}
\end{boldmath}
In this section, we write down the expansions of the form factors in
the limit $S \gg m^2$, i.e. $y \rightarrow 0$. Putting
$r=\frac{S}{m^2}$ and $L=\ln r$, 
and keeping terms up to $r^{-2}$ we find for the coefficients that
are non-vanishing to that order:
\bea
\Re \, \tilde{a}_1    &=&  \frac{1}{r^2}\,\left( -3+2\,L \right) -\frac{2}{r}-1+L,\\
\Re \, \tilde{a}_2    &=& \frac{1}{r^2}\,\left( -8+13\,L+8\,\zeta (2)
  -{L}^{2} \right) + \frac{1}{r}\,\left( -1+5\,L \right)
\nn \\
&& -2+\frac{3}{2}\,L-\frac{1}{2}\,{L}^{2}+4\,\zeta (2) 
,\\
\Re \, \tilde{a}_3    &=&  \frac{1}{r^2}\,\left( 4\,\zeta (3)
  +52\,\zeta (2)
  +\frac{1}{3}\,{L}^{3}-\frac{13}{2}\,{L}^{2}-5+32\,L-8\,L\zeta (2)
\right) \nn \\
&& + \frac{1}{r}\,\left( 20\,\zeta (2) -\frac{5}{2}\,{L}^{2}-3+6\,L \right)
\nn \\
&& +6\,\zeta (2) -4+2\,\zeta (3) -\frac{3}{4}\,{L}^{2}+4\,L+\frac{1}{6}\,{L}^{3}-4\,L\zeta (2) 
,\\
\Im \, \tilde{a}_1    &=& -\frac{2}{r^2}-1
,\\
\Im \, \tilde{a}_2    &=&  \frac{1}{r^2}\,\left( -13+2\,L \right) -\frac{5}{r}-{\frac {3}{2}}+L
,\\
\Im \, \tilde{a}_3    &=&  \frac{1}{r^2}\, \left( 13\,L-32-{L}^{2}+4\,\zeta \left( 2
  \right)  \right)+ \frac{1}{r}\,\left( -6+5\,L \right)
\nn \\
&& +2\,\zeta (2) -\frac{1}{2}\,{L}^{2}-4+\frac{3}{2}\,L
,\\
\Re \, \tilde{b}_2    &=&  \frac{1}{r^2}\,\left( 12-4\,L \right) +  \frac{1}{r}\,\left( 4-6\,L \right)
,\\
\Re \, \tilde{b}_3    &=&  \frac{1}{r^2}\,\left( 12-32\,L+2\,{L}^{2}-16\,\zeta (2)  \right) + \frac{1}{r}\left( 8-12\,L+3\,{L}^{2}-24\,\zeta (2)  \right) 
,\\
\Im \, \tilde{b}_2    &=& \frac{4}{r^2}+\frac{6}{r}
\\
\Im \, \tilde{b}_3    &=& \frac{1}{r^2}\,\left( 32-4\,L \right) +  \frac{1}{r}\,\left( -6\,L+12 \right)
,\\
\Re \, \tilde{c}_1    &=&  \frac{1}{r^2}\,\left( -1+\frac{2}{3}\,L \right)-\frac{2}{3\,r}-{\frac {1}{3}}+\frac{1}{3}\,L
,\\
\Re \, \tilde{c}_3    &=&  \frac{1}{r^2}\,\left( {\frac {11}{4}}-{\frac {11}{6}}\,L \right) +\frac{11}{6\,r}+{\frac {11}{12}}-{\frac {11}{12}}\,L
,\\
\Re \, \tilde{c}_4    &=&  \frac{1}{r^2}\,\left( 5+2\,{L}^{2}-12\,\zeta \left( 2
  \right) -5\,L \right) + 
\frac{1}{r}\,\left( -2\,L+2 \right)\nn \\
&& +\frac{1}{2}\,{L}^{2}-3\,\zeta (2) +{\frac {1}{2}}-L
,\\
\Re \, \tilde{c}_5    &=&  \frac{1}{r^2}\,\left( {\frac {5}{3}}-{\frac {10}{9}}\,L \right) +\frac{10}{9\,r}+{\frac {5}{9}}-\frac{5}{9}\,L
,\\
\Re \, \tilde{c}_7    &=&  \frac{1}{r^2}\,\left( \frac{13}{2}\,\zeta (2)
  -2\,\zeta (3) -{L}^{2}+\frac{1}{3}\,{L}^{3}+{\frac{47}{9}}\,L-6\,L\zeta (2) -{\frac {55}{12}} \right)\nn \\
&& + \frac{1}{r}\,\left( -{\frac {67}{18}}+\zeta (2)  \right) +\frac{1}{2}\,\zeta (2) -{\frac {49}{36}}-\frac{1}{2}\,L\zeta (2) +{\frac {67}{36}}\,L-\frac{1}{2}\,\zeta (3) 
,\\
\Re \, \tilde{c}_8    &=&  \frac{1}{r^2}\,\left( -2\,{L}^{3}+16-116\,\zeta \left( 2
  \right) +{\frac {37}{2}}\,{L}^{2}+40\,L\zeta (2)
  -{\frac {79}{2}}\,L \right) \nn \\
&& + \frac{1}{r}\,\left( -38\,\zeta (2) +6\,{L}^{2}-9\,L+5 \right)
\nn \\
&& +2-\frac{1}{2}\,{L}^{3}-\frac{7}{2}\,L+2\,{L}^{2}-13\,\zeta (2) +10\,L\zeta (2) 
,\\
\Re \, \tilde{c}_9    &=&  \frac{1}{r^2}\,\left( -{\frac {356}{9}}\,\zeta \left( 2
  \right) -\frac{8}{3}\,\zeta (3) -\frac{2}{9}\,{L}^{3}+\frac{8}{3}\,L\zeta
  \left( 2 \right) +{\frac {49}{9}}\,{L}^{2}+{\frac {23}{2}}-{\frac
    {959}{27}}\,L \right) \nn \\
&& + \frac{1}{r}\,\left( -{\frac {89}{9}}\,L+\frac{5}{3}\,{L}^{2}+{\frac {59}{27}}-{\frac
    {32}{3}}\,\zeta (2)  \right) \nn \\
&& -\frac{1}{9}\,{L}^{3}+{\frac {19}{18}}\,{L}^{2}-{\frac {209}{54}}\,L-{\frac {64}{9}}\,\zeta (2) -\frac{4}{3}\,\zeta (3) +{\frac {106}{27}}+\frac{4}{3}\,L\zeta (2) 
,\\
\Re \, \tilde{c}_{10} &=&   \frac{1}{r^2}\,\left( -\frac{16}{3}\,L\zeta (2)
  +{\frac {1153}{18}}+{\frac {229}{9}}\,{L}^{2}+12\,\zeta (2) +\frac{4}{9}\,{L}^{3}-{\frac {2230}{27}}\,L \right)\nn
\\
&& + \frac{1}{r}\,\left( {\frac {385}{27}}-{\frac {161}{9}}\,L+{\frac
    {17}{3}}\,{L}^{2}+\frac{4}{3}\,\zeta (2)  \right)
\nn \\
&& +{\frac {19}{18}}\,{L}^{2}-{\frac {22}{3}}\,\zeta (2) -{\frac {265}{54}}\,L+{\frac {383}{27}}-\frac{1}{9}\,{L}^{3}+\frac{4}{3}\,L\zeta (2) 
,\\
\Re \, \tilde{c}_{11} &=&   \frac{1}{r^2}\,\left( {\frac {241}{4}}+{\frac {98}{5}}\,
  \zeta^2(2)+144\,\zeta \left( 2
  \right) \ln(2)+{\frac {59}{3}}\,L\zeta (2) +{\frac
    {173}{18}}\,{L}^{2}+{\frac {17}{18}}\,{L}^{3}\right.\nn \\
&& \left.-{\frac
    {4111}{18}}\,\zeta (2) -110\,L\zeta (3)
  +{\frac {1319}{6}}\,\zeta (3) -10\,{L}^{2}\zeta \left(
    2 \right) +{\frac {15619}{108}}\,L \right)\nn \\
&& +  \frac{1}{r}\,\left( \frac{1}{24}\,{L}^{4}+{\frac {69}{10}}\, \zeta^2(2)-14\,L\zeta (3) +{\frac
    {1189}{36}}\,L+{\frac {5}{12}}\,{L}^{2}-{\frac {137}{3}}\,\zeta
  \left( 2 \right)  \right. \nn \\
&& \left.+55\,\zeta (3)+36\,\zeta \left( 2
  \right) \ln(2)+7\,L\zeta (2) -\frac{7}{2}\,{L}^{2}\zeta (2) -{\frac {157}{108}} \right)\nn \\
&& -{\frac {1595}{108}}-\frac{11}{3}\,L\zeta (2) -{\frac
  {233}{72}}\,{L}^{2}+{\frac {11}{36}}\,{L}^{3}+{\frac {67}{6}}\,\zeta(3) -{\frac {63}{20}}\,  \zeta^2 (2)\nn \\
&& +\frac{1}{2}\,{L}^{2}\zeta (2) +{\frac {173}{9}}\,\zeta (2) -\frac{13}{2}\,L\zeta (3) +6\,\zeta (2) \ln(2)+{\frac {2545}{216}}\,L
,\\
\Re \, \tilde{c}_{12} &=&  \frac{1}{r^2}\,\left( {\frac {841}{4}}+{\frac {351}{5}}\,
  \zeta^2(2)-288\,\zeta \left( 2
  \right) \ln(2)+{L}^{4}+350\,L\zeta (2) -625\,\zeta
  \left( 2 \right) \right. \nn \\
&& \left.-148\,L\zeta (3) +95\,{L}^{2}-{\frac
    {85}{6}}\,{L}^{3}+190\,\zeta (3) -55\,{L}^{2}\zeta
  \left( 2 \right) +{\frac {37}{2}}\,L \right) \nn \\
&& +  \frac{1}{r}\,\left( -\frac{16}{3}\,{L}^{3}-74\,\zeta (2)
  -\frac{1}{12}\,{L}^{4}+\frac{9}{5}\, \zeta^2 (2)-20\,L\zeta (3) -{\frac {91}{4}}\,L\right. \nn \\
&& \left. +{\frac
    {59}{4}}\,{L}^{2}+{L}^{2}\zeta (2) +10\,\zeta \left(
    3 \right) -72\,\zeta (2) \ln(2)+113\,L\zeta \left( 2
  \right) +{\frac {107}{4}} \right)\nn \\
&& +{\frac {23}{2}}-\frac{5}{3}\,{L}^{3}+31\,L\zeta (2)
+{\frac {55}{8}}\,{L}^{2}-32\,\zeta (2) -11\,\zeta (3) +{\frac {68}{5}}\,  \zeta^2 (2)\nn \\
&& -12\,{L}^{2}\zeta (2) +{\frac {7}{24}}\,{L}^{4}+8\,L\zeta (3) -12\,\zeta (2) \ln(2)-{\frac {85}{8}}\,L
,\\
\Im \, \tilde{c}_1    &=& -\frac{2}{3\,r^2}-{\frac {1}{3}}
,\\
\Im \, \tilde{c}_3    &=& \frac{11}{6\,r^2}+{\frac {11}{12}}
,\\
\Im \, \tilde{c}_4    &=&  \frac{1}{r^2}\,\left( 5-4\,L \right)+\frac{2}{r}+1-L
,\\
\Im \, \tilde{c}_5    &=& \frac{10}{9\,r^2}+{\frac {5}{9}}
,\\
\Im \, \tilde{c}_7    &=&  \frac{1}{r^2}\,\left( 2\,\zeta (2) -{\frac {47}{9}}+2\,L-{L}^{2} \right) +\frac{1}{2}\,\zeta (2) -{\frac {67}{36}}
,\\
\Im \, \tilde{c}_8    &=&   \frac{1}{r^2}\,\left( -16\,\zeta (2) +6\,{L}^{2}+{\frac {79}{2}}-37\,L \right)+  \frac{1}{r}\,\left(
  9-12\,L \right)\nn \\
&& +{\frac {7}{2}}-4\,L-4\,\zeta (2) +\frac{3}{2}\,{L}^{2}
,\\
\Im \, \tilde{c}_9    &=&  \frac{1}{r^2}\,\left( {\frac
    {959}{27}}+\frac{2}{3}\,{L}^{2}-{\frac {98}{9}}\,L \right)
+ \frac{1}{r}\left( -\frac{10}{3}\,L+{\frac {89}{9}} \right)\nn \\
&& -{\frac {19}{9}}\,L+{\frac {209}{54}}+\frac{1}{3}\,{L}^{2},\\
\Im \, \tilde{c}_{10} &=&   \frac{1}{r^2}\,\left( -\frac{4}{3}\,{L}^{2}-{\frac
    {458}{9}}\,L+{\frac {2230}{27}} \right)+  \frac{1}{r}\,\left(
  {\frac {161}{9}}-{\frac {34}{3}}\,L \right)\nn \\
&& +\frac{1}{3}\,{L}^{2}-{\frac {19}{9}}\,L+{\frac {265}{54}}
,\\
\Im \, \tilde{c}_{11} &=&  \frac{1}{r^2}\,\left( 110\,\zeta (3) -{\frac
    {15619}{108}}-{\frac {173}{9}}\,L+20\,L\zeta (2)
  -31\,\zeta (2) -{\frac {17}{6}}\,{L}^{2} \right)
\nn \\
&& +  \frac{1}{r}\,\left( -\frac{1}{6}\,{L}^{3}-7\,\zeta (2) +5\,L\zeta \left(
    2 \right) -{\frac {1189}{36}}+14\,\zeta (3) -\frac{5}{6}\,L
\right)\nn \\
&& -{\frac {11}{12}}\,{L}^{2}+{\frac {233}{36}}\,L+\frac{13}{2}\,\zeta (3) -L\zeta (2) -{\frac {2545}{216}}
,\\
\Im \, \tilde{c}_{12} &=&  
\frac{1}{r^2}\,\left( 62\,L\zeta (2) +{\frac{85}{2}}\,{L}^{2}-{\frac {37}{2}}+148\,\zeta (3)
  -4\,{L}^{3}-180\,\zeta (2) -190\,L \right)\nn \\
&& +  \frac{1}{r}\,\left( 20\,\zeta (3) -49\,\zeta (2)
  +16\,{L}^{2}+\frac{1}{3}\,{L}^{3}+2\,L\zeta \left(2 \right) +{\frac
    {91}{4}}-{\frac {59}{2}}\,L \right)\nn \\
&& +{\frac{85}{8}}-8\,\zeta (3) +5\,{L}^{2}-{\frac {55}{4}}\,L-\frac{7}{6}\,{L}^{3}+10\,L\zeta (2) -11\,\zeta (2) 
,\\
\Re \, \tilde{d}_8    &=&  \frac{1}{r^2}\, \left( -20-4\,{L}^{2}+24\,\zeta \left( 2
  \right) +28\,L \right)\nn \\
&& +  \frac{1}{r}\,\left( -6\,{L}^{2}+10\,L-4+36\,\zeta (2)  \right)
,\\
\Re \, \tilde{d}_9    &=&   \frac{1}{r^2}\,\left( {\frac {196}{9}}\,L+{\frac{32}{3}}
  \,\zeta (2) - {\frac{44}{3}}-\frac{4}{3}\,{L}^{2} \right)\nn \\
&& +  \frac{1}{r}\,\left( -{\frac {76}{9}}+16\,\zeta (2) -2\,{L}^{2}+{\frac {34}{3}}\,L \right)
,\\
\Re \, \tilde{d}_{10} &=&  \frac{1}{r^2}\, \left( -44-{\frac
    {52}{3}}\,{L}^{2}-16\,\zeta (2) +{\frac {628}{9}}\,L
\right)\nn \\
&& +  \frac{1}{r}\left( -8\,\zeta (2) +{\frac {32}{9}}-2\,{L}^{2}+{\frac {34}{3}}\,L \right)
,\\
\Re \, \tilde{d}_{11} &=&   \frac{1}{r^2}\,\left( -96\,\zeta (2) \ln
  2+{\frac {157}{3}}-{\frac {43}{3}}\,{L}^{2}-{\frac
    {401}{9}}\,L-36\,L\zeta (2) -{\frac {138}{5}}\,
  \zeta^2(2)\right. \nn \\
&& \left.-\frac{1}{6}\,{L}^{4}+56\,L\zeta
  \left( 3 \right) +14\,{L}^{2}\zeta (2) +{\frac
    {680}{3}}\,\zeta (2) -164\,\zeta (3)
\right)\nn \\
&& +  \frac{1}{r}\,\Big( -12\,\zeta (3) -{\frac
    {229}{6}}\,L-12\,\zeta (2) \nn \\
&& -24\,\zeta (2) \ln(2)+\frac{11}{2}\,{L}^{2}+{\frac {242}{9}} \Big)
,\\
\Re \, \tilde{d}_{12} &=&   \frac{1}{r^2}\,\bigg( -115+43\,L+360\,\zeta \left( 2
  \right) -100\,L\zeta (2) +16\,L\zeta (3)
  -12\,{L}^{2}\zeta (2)  \nn \\
&& +192\,\zeta (2) \ln
  2+{\frac {68}{5}}\, \zeta^2 (2)+\frac{1}{3}\,{L}^{4}-64\,\zeta (3) +4\,{L}^{3}-67\,{L}^{2}
\bigg)\nn \\
&& +  \frac{1}{r}\,\left( 96\,\zeta (2) +48\,\zeta (2) \ln
  2+{\frac{61}{2}}\,L+6\,{L}^{3}-17\right. \nn \\
&& \left. -12\,\zeta (3) -{\frac {43}{2}}\,{L}^{2}-120\,L\zeta (2)  \right)
,\\
\Im \, \tilde{d}_8    &=&   \frac{1}{r^2}\,\left( -28+8\,L \right)+  \frac{1}{r}\,\left( -10+12\,L \right)
,\\
\Im \, \tilde{d}_9    &=&  \frac{1}{r^2}\,\left( -{\frac {196}{9}}+\frac{8}{3}\,L \right) +  \frac{1}{r}\,\left( -{\frac {34}{3}}+4\,L \right)
,\\
\Im \, \tilde{d}_{10} &=&   \frac{1}{r^2}\,\left( -{\frac {628}{9}}+{\frac{104}{3}}\,L \right) +  \frac{1}{r}\,\left( -{\frac {34}{3}}+4\,L \right)
,\\
\Im \, \tilde{d}_{11} &=&   \frac{1}{r^2}\,\left( {\frac{401}{9}}+\frac{2}{3}\,{L}^{3}+{\frac {86}{3}}\,L-20\,L\zeta
  \left( 2 \right) +36\,\zeta (2) -56\,\zeta \left( 3
  \right)  \right)\nn \\
&& + \frac{1}{r}\,\left( {\frac {229}{6}}-11\,L \right) 
,\\
\Im \, \tilde{d}_{12} &=&   \frac{1}{r^2}\,\left( -43-16\,\zeta (3)
  +52\,\zeta (2) -\frac{4}{3}\,{L}^{3}+8\,L\zeta (2)
  -12\,{L}^{2}+134\,L \right)\nn \\
&& +  \frac{1}{r}\,\left( 43\,L-{\frac {61}{2}}+48\,\zeta (2) -18\,{L}^{2} \right)
.
\eea
The limit $r \to \infty$ corresponds to the massless limit $m \to 0$.
Therefore the chirality-flipping form factors $F_2$ and $G_2$ must
be of order $1/r$, and those terms in the chirality-conserving
form factor $F_1$ that survive the limit  $r \to \infty$ must be equal
to the corresponding terms in $G_1$. The asymptotic expansions given
above and in \cite{bbghlmr} satisfy these constraints.

All the results of this Section can be obtained in an electronic
form by downloading the source of this manuscript from
http://www.arxiv.org.
\section{Summary}

In this paper we calculated the axial vector form factors $G_1$ and
$G_2$ to second order in the QCD coupling, excluding the contributions
from Fig.~\ref{triangle-graph}~(a) and (b). The results for the two
form factors were obtained keeping the full dependence  on the mass of
the heavy quark, as well as on the arbitrary momentum transfer. \par
The renormalized
form factors are expressed in terms of the ${\rm \overline{MS}}$ coupling
$\alpha_s$ for $(N_f +1)$ quark flavors and of the on-shell mass of the
heavy quark.
The expressions for the unsubtracted as well as the UV-renormalized form
factors are given in  closed analytic form as a Laurent expansion in
$\epsilon=(4-D)/2$. The coefficients of this expansion have a suitable
representation in terms of 1-dimensional harmonic polylogarithms.
Poles in $\epsilon$, which correspond to soft and collinear divergences,
are still present  in $G_{1,2,R}$,  like in $F_{1,2,R}$.
Depending on the
observable considered, these divergences must cancel either among
each other or against  the divergences arising from the real radiation,
which in this paper was not  taken into account.
\par
As already discussed in the introduction and in
our preceeding paper \cite{bbghlmr} our results for the vector and axial 
vector form
factors are part of the order $\alpha_S^2 $ QCD corrections
to the differential electroproduction cross sections of heavy quarks 
and, moreover,
have a number of immediate  applications. These include studies of the 
extrapolation of the continuum amplitudes
to the ${\bar Q} Q$ production threshold and studies concerning the 
generic singularity
structure of QCD amplitudes with massive quarks.
Of general interest is the computation of the
NNLO QCD corrections to the forward-backward asymmetry of heavy quarks.
We shall report on this in a future publication.

\section*{Acknowledgement}

We are grateful to J.~Vermaseren for his kind assistance in the use
of the algebra manipulating program {\tt FORM}~\cite{FORM}, by which
many of our calculations were carried out.
This work was partially supported by the European Union under
contract HPRN-CT-2000-00149, by Deutsche Forschungsgemeinschaft (DFG), 
SFB/TR9, by DFG-Graduiertenkolleg RWTH Aachen, by the Swiss 
National Funds (SNF) under contract 200021-101874,  and by the USA DoE under 
the grant DE-FG03-91ER40662, Task J.


\end{fmffile}


\begin{thebibliography}{99} 
\def    \np     #1#2#3{{\it Nucl. Phys.} {\bf #1} (19#2) #3}
\def    \nptwoth     #1#2#3{{\it Nucl. Phys.} {\bf #1} (20#2) #3}
\def    \nptwothnz     #1#2#3{{\it Nucl. Phys.} {\bf #1} (19#2) #3}
\def    \prep   #1#2#3{{\it Phys. Rep.} {\bf #1}  (19#2) #3}
\def    \pl     #1#2#3{{\it Phys. Lett.} {\bf #1} (19#2) #3}
\def    \pltwoth     #1#2#3{{\it Phys. Lett.} {\bf #1} (20#2) #3}
\def    \plold  #1#2#3{{\it Phys. Lett.} {\bf #1B} (19#2) #3}
\def    \prl    #1#2#3{{\it Phys. Rev. Lett.} {\bf #1}  (19#2) #3} 
\def    \pr     #1#2#3{{\it Phys. Rev.} {\bf #1}  (19#2) #3}
\def    \prd    #1#2#3{{\it Phys. Rev.} {\bf D#1}  (19#2) #3} 
\def    \zeit   #1#2#3{{\it Z. Phys.} {\bf C#1}  (19#2) #3}
\def    \cmp    #1#2#3{{\it Comm. Math. Phys.} {\bf #1}  (19#2) #3}
\def    \ibid   #1#2#3{{\it ibid.} {\bf #1} (19#2) #3}
\def    \nc     #1#2#3{{\it Nuovo Cim.} {\bf #1} (19#2) #3}
\def    \acta   #1#2#3{{\it Acta Phys. Polon.} {\bf #1} (19#2) #3}
\def    \tmp    #1#2#3{{\it Theor. Math. Phys.} {\bf #1} (19#2) #3}
\def    \eur    #1#2#3{{\it Eur. Phys.  J.} {\bf C#1} (19#2) #3}
\def    \comp    #1#2#3{{\it Comput. Phys. Commun.} {\bf #1} (20#2) #3}
\def    \hepph  #1 {{\tt hep-ph/#1}}
\def    \hepex  #1 {{\tt hep-ex/#1}}
\def    \mathph  #1 {{\tt math-ph/#1}}
\parskip 0pt
\itemsep=0pt


%
\bibitem{bbghlmr}
W.~Bernreuther, R.~Bonciani, T.~Gehrmann, R.~Heinesch, T.~Leineweber, P.~Mastrolia and E.~Remiddi,
\nptwoth{B706}{05}{245} 
(\hepph{0406046}).


\bibitem{EWWorkingGroup}
 The LEP collaborations ALEPH, DELPHI, L3, OPAL, the LEP Electroweak
 Working Group, and the SLD Heavy Flavour Group, {\it A Combination of
 Preliminary Electroweak Measurements and Constraints on the Standard
 Model}, report LEPEWWG/2003-01, April 2003.

\bibitem{Freitas:2004mn}
A.~Freitas and K.~Moenig,
(\hepph{0411304}).

\bibitem{tesla}
 J.~A.~Aguilar-Saavedra {\it et al.}  [ECFA/DESY LC Physics Working Group
 Collaboration], ``TESLA Technical Design Report Part III: Physics at an
 $e^+e^-$ Linear Collider'', DESY-report 2001-011 (\hepph{0106315}).

\bibitem{triangle}
W.~Bernreuther, R.~Bonciani, T.~Gehrmann,
R.~Heinesch, T.~Leineweber and
E.~Remiddi, in preparation.

\bibitem{HV}
G.~'t Hooft and M.~J.~G.~Veltman,
\nptwothnz{B44}{72}{189}

\bibitem{Larin}
S.~A.~Larin,
\pl{B303}{93}{113}
(\hepph{9302240}).


\bibitem{Polylog}
  E. Remiddi and J. A. M. Vermaseren, {\it Int. J. Mod. Phys.} 
  {\bf A15} (2000) 725 ({\tt hep-ph/9905237}). 

\bibitem{Polylog3}
  T. Gehrmann and E. Remiddi, \comp{141}{01}{296} 
  ({\tt hep-ph/0107173}).


\bibitem{Broadhurst}
D.~J.~Broadhurst, N.~Gray and K.~Schilcher,
\zeit{52}{91}{111}

\bibitem{Melnikov}
K.~Melnikov and T.~van Ritbergen, \nptwoth{B591}{00}{515}
(\hepph{0005131}).


\bibitem{RoPieRem1}
  R. Bonciani, P. Mastrolia and E. Remiddi, \nptwoth{B661}{03}{289} 
[Erratum-ibid. \nptwoth{B702}{04}{259}] (\hepph{0301170}).

\bibitem{RoPieRem2}
  R. Bonciani, P. Mastrolia and E. Remiddi, \nptwoth{B676}{04}{399}
  (\hepph{0307295}).

\bibitem{RoPieRem3}
  R. Bonciani, P. Mastrolia and E. Remiddi, \nptwoth{B690}{04}{138}
  (\hepph{0311145}).


\bibitem{Lap}
  S. Laporta and E. Remiddi, \pl{B379}{96}{283} (\hepph{9602417}).\\
  S.~Laporta, {\it Int.\ J.\ Mod.\ Phys.}  {\bf A 15} (2000) 5087
 (\hepph{0102033}).






\bibitem{Chet}
  F.V. Tkachov, \pl{B100}{81}{65}.\\
  K.G. Chetyrkin and F.V. Tkachov, \np{B192}{81}{159}.



\bibitem{Rem3}
  T. Gehrmann and E. Remiddi, \nptwoth{B580}{00}{485} 
  ({\tt hep-ph/9912329}). 
  


\bibitem{Kot}
  A. V. Kotikov, \pl{B254}{91}{158}.

\bibitem{Rem1} 
  E. Remiddi, \nc{110A}{97}{1435} ({\tt hep-th/9711188}).

\bibitem{Rem2}
  M. Caffo, H. Czy\.{z}, S. Laporta and E. Remiddi, 
  \acta{B29}{98}{2627} ({\tt hep-ph/9807119}).\\
  M. Caffo, H. Czy\.{z}, S. Laporta and E. Remiddi, 
  \nc{A111}{98}{365} ({\tt hep-ph/9805118}).

\bibitem{Jersak:1981sp}
J.~Jersak, E.~Laermann and P.~M.~Zerwas,
\prd{25}{82}{1218}
[Erratum-ibid. \prd{36}{87}{310}].

\bibitem{KuhnTeubner}
J.~H.~K\"uhn and T.~Teubner,
\eur{9}{99}{221}
(\hepph{9903322}).



\bibitem{FORM} J.A.M.\ Vermaseren, Symbolic Manipulation with
               {\tt FORM}, Version 2, CAN, Amsterdam, 1991; 
               New features of {\tt FORM}, (\mathph{0010025}).


\end{thebibliography}
\end{document}